\documentclass[%
 reprint,
superscriptaddress,
 amsmath,amssymb,
 aps,
pra,
]{revtex4-2}

\usepackage[english]{babel}

\usepackage{graphicx}
\usepackage{dcolumn}
\usepackage{bm}

\usepackage[super]{nth}
\usepackage{xspace}
\usepackage{siunitx}
\usepackage{tabularx} 
\usepackage{graphicx} 
\usepackage[margin=1in,letterpaper]{geometry} 

\usepackage[final]{hyperref} 
\hypersetup{
	colorlinks=true,       
	linkcolor=blue,        
	citecolor=blue,        
	filecolor=magenta,     
	urlcolor=blue         
}
\usepackage[toc,acronyms,nonumberlist]{glossaries}

\newcommand{\fm}[1]{\ifmmode#1\else$#1$\fi}
\newcommand{\caP}{\fm{{\text{Ca}^{+}}}\xspace}
\newcommand{\Ca}{{\fm{\text{Ca}^{+}}}\xspace}
\newcommand{\alP}{\fm{{\text{Al}^{+}}}\xspace}
\newcommand{\Al}{\fm{{\text{Al}^{+}}}\xspace}
\newcommand{\CaP}{\fm{{^{40}\caP}}\xspace}
\newcommand{\AlP}{\fm{{^{27}\alP}}\xspace}
\newcommand{\Sr}{\fm{\text{Sr}}\xspace}
\newcommand{\Sres}{\fm{{^{87}\Sr}}\xspace}

\newcommand{\wrf}{\fm{\omega_\mathrm{rf}}\xspace}
\newcommand{\Erf}{\fm{E_\mathrm{rf}}\xspace}
\def\ssz{\fm{{}^1\mathrm{S}_0}\xspace}
\def\tpz{\fm{{}^3\mathrm{P}_0}\xspace}
\def\tpo{\fm{{}^3\mathrm{P}_1}\xspace}
\def\dsoh{\fm{{}^2\mathrm{S}_{1/2}}\xspace}
\def\dpoh{\fm{{}^2\mathrm{P}_{1/2}}\xspace}

\def\ssztpz{\ssz\fm{\leftrightarrow} \tpz}
\def\ssztpo{\ssz\fm{\leftrightarrow} \tpo}
\def\dsoh{\fm{{}^2\mathrm{S}_{1/2}\xspace}}
\def\ddfh{\fm{{}^2\mathrm{D}_{5/2}\xspace}}
\def\dsohddfh{\dsoh \fm{\leftrightarrow} \ddfh \xspace}
\def\mP{\fm{m_\mathrm{P}}\xspace}
\def\mS{\fm{m_\mathrm{S}}\xspace}
\def\DS{\fm{\Delta_\mathrm{S}}\xspace}
\def\DP{\fm{\Delta_\mathrm{P}}\xspace}

\usepackage{booktabs}
\usepackage{braket}

\begin{document}

\preprint{APS/123-QED}
    
\title{An \Al clock with $\num{1.6e-18}$ systematic uncertainty and its frequency ratios}

\author{Fabian Dawel}%
\email{Fabian.Dawel@ptb.de}
 \affiliation{Physikalisch-Technische Bundesanstalt, Bundesallee 100, 38116 Braunschweig, Germany}
 \affiliation{Institut für Quantenoptik, Leibniz Universität Hannover, Welfengarten 1, 30167 Hannover, Germany}
\author{Johannes Kramer}%
\affiliation{Institut für Quantenoptik, Leibniz Universität Hannover, Welfengarten 1, 30167 Hannover, Germany}
\author{Derwell Drapier}%
\affiliation{Physikalisch-Technische Bundesanstalt, Bundesallee 100, 38116 Braunschweig, Germany}
\author{Lennart Pelzer}%
\affiliation{Physikalisch-Technische Bundesanstalt, Bundesallee 100, 38116 Braunschweig, Germany}
\author{Kai Dietze}%
 \affiliation{Physikalisch-Technische Bundesanstalt, Bundesallee 100, 38116 Braunschweig, Germany}
 \affiliation{Institut für Quantenoptik, Leibniz Universität Hannover, Welfengarten 1, 30167 Hannover, Germany}
\author{Mirza A. Ali}
\affiliation{Physikalisch-Technische Bundesanstalt, Bundesallee 100, 38116 Braunschweig, Germany}
\affiliation{Institut für Quantenoptik, Leibniz Universität Hannover, Welfengarten 1, 30167 Hannover, Germany}
\author{Marek Hild}%
 \affiliation{Physikalisch-Technische Bundesanstalt, Bundesallee 100, 38116 Braunschweig, Germany}
\author{Vincent Barbé}%
 \affiliation{Physikalisch-Technische Bundesanstalt, Bundesallee 100, 38116 Braunschweig, Germany}
 \affiliation{current address: Laboratoire Kastler Brossel, Sorbonne Université, CNRS, ENS-PSL Research University, Collège de France, 4 place Jussieu, 75005 Paris, France}
\author{Steven A.~King}%
 \affiliation{Physikalisch-Technische Bundesanstalt, Bundesallee 100, 38116 Braunschweig, Germany}
 \author{Joshua Klose}%
 \affiliation{Physikalisch-Technische Bundesanstalt, Bundesallee 100, 38116 Braunschweig, Germany}
\author{Kilian Stahl}%
 \affiliation{Physikalisch-Technische Bundesanstalt, Bundesallee 100, 38116 Braunschweig, Germany}
 \author{Johannes Rahm}%
 \affiliation{Physikalisch-Technische Bundesanstalt, Bundesallee 100, 38116 Braunschweig, Germany}
  \author{Navraj Poudel}%
 \affiliation{Physikalisch-Technische Bundesanstalt, Bundesallee 100, 38116 Braunschweig, Germany}
\author{Sören Dörscher}%
 \affiliation{Physikalisch-Technische Bundesanstalt, Bundesallee 100, 38116 Braunschweig, Germany}
\author{Stefan Weyers}%
 \affiliation{Physikalisch-Technische Bundesanstalt, Bundesallee 100, 38116 Braunschweig, Germany}
\author{Erik Benkler}%
 \affiliation{Physikalisch-Technische Bundesanstalt, Bundesallee 100, 38116 Braunschweig, Germany}
\author{Christian Lisdat}%
 \affiliation{Physikalisch-Technische Bundesanstalt, Bundesallee 100, 38116 Braunschweig, Germany}
\author{Piet O.~Schmidt}%
 \affiliation{Physikalisch-Technische Bundesanstalt, Bundesallee 100, 38116 Braunschweig, Germany}
 \affiliation{Institut für Quantenoptik, Leibniz Universität Hannover, Welfengarten 1, 30167 Hannover, Germany}
	
\date{\today}

\begin{abstract}
Advances in optical clocks motivate a redefinition of the second, requiring rigorous evaluations of systematic uncertainties and robust consistency among the clocks.
Here, we report the full evaluation of the systematic frequency shifts of an \AlP single-ion clock, and the measurement of its absolute frequency and frequency ratio with a \Sres optical lattice clock at PTB. The evaluated total systematic fractional frequency uncertainty is \num{1.6e-18}, mainly limited by the accuracy of the relevant atomic coefficients and by background gas collisions. The absolute frequency of the clock has been measured to be $\nu_{\Al}=\SI{1 121 015 393 207 859.19(24)}{\hertz}$, obtained by comparison with two primary caesium fountain clocks at PTB. The frequency ratio between the \Al and \Sr optical clocks has been determined to be $\nu_{\Al}/\nu_{\Sres}=\num{2.611 701 431 781 462 668(36)}$, limited by the accuracy of the \Sr clock. This ratio differs by $8.6\sigma$ and $1.2\sigma$ from the 2021 and 2025 frequency ratio published by the BACON collaboration, respectively. These results represent an important contribution toward a future redefinition of the second using optical clocks, and underscore the importance of independent measurements of clock-candidate frequency ratios across different institutions.
\end{abstract}
	
\maketitle
\textit{Introduction}---
Optical clocks based on trapped ions and neutral atoms have surpassed microwave clocks in stability and accuracy by more than two orders of magnitude, reaching estimated relative systematic frequency uncertainties of a few \num{e-18} and below \cite{marshall_high-stability_2025, ma_quantum-logic-based_2024, zhang_liquid-nitrogen-cooled_2026, hausser_115in-172yb_2025, arnold_optical_2025_2, lindvall_88sr_2025, tofful_171yb_2024_2, sanner_optical_2019, jia_improved_2026_2, lu_ntsc_2025_2, aeppli_clock_2024, ohmae_transportable_2021_2, mcgrew_atomic_2018, nosske_transportable_2025_2}. These record uncertainties enable many applications ranging from tests of fundamental physics, such as the search for dark matter, variation of fundamental constants and tests of relativity \cite{beloy_frequency_2021, filzinger_improved_2023, lange_improved_2021, sanner_optical_2019, takamoto_test_2020, kobayashi_search_2022, sherrill_analysis_2023, roberts_search_2020, banerjee_oscillating_2025, arakawa_probing_2026_2, filzinger_ultralight_2025, dreissen_improved_2022}, to centimeter-level height difference measurements in relativistic geodesy \cite{mehlstaubler_atomic_2018, grotti_geodesy_2018, bjerhammar_relativistic_1985, takano_geopotential_2016}. Validation of the estimated uncertainties requires frequency ratio measurements between different realizations of the same or different clock transitions. This is also a mandatory requirement for the envisioned redefinition of the second \cite{dimarcq_roadmap_2024}.

In contrast to frequency ratio measurements between two clocks realizing the same clock transitions, the ratio between clocks utilizing different transitions is \textit{a priori} unknown. Frequency ratio measurements between optical clocks at a level of \num{e-17} and below with the same \cite{arnold_optical_2025_2, sanner_optical_2019, ushijima_cryogenic_2015, takamoto_test_2020, takano_geopotential_2016, mcgrew_atomic_2018} and different \cite{aeppli_atomic_2025_2, beloy_frequency_2021, hausser_115in-172yb_2025} species have been performed previously. However, unresolved discrepancies have been identified at a level of \num{e-17} \cite{aeppli_atomic_2025_2, beloy_frequency_2021, amy-klein_international_2024_2}, corresponding to many times the estimated systematic uncertainty of the involved clocks. This underscores the need for independent validation of optical clocks via frequency ratio measurements by different institutions.

Here, we present the full evaluation of an aluminum ion clock at PTB using quantum logic spectroscopy with a co-trapped calcium ion. We compare this clock with PTB's strontium lattice clock Sr3 \cite{klo26} and find a discrepancy in the frequency ratio that differs by $\SI{8.6}{\sigma}$ and $\SI{1.2}{\sigma}$ from two previously published ratios by the BACON collaboration \cite{beloy_frequency_2021, aeppli_atomic_2025_2}. Furthermore, we measure the absolute frequency of our \Al clock through a direct comparison with PTB's primary caesium fountain clocks CSF1 and CSF2 \cite{weyers_advances_2018} and using the measured absolute frequency of Sr3 \cite{klo26}.

\textit{Setup}---
In the following, we provide a summary of the experimental setup. More details can be found in \cite{kramer_aluminum_2023, dawel_high-stability_2025_2, supplement}.
We prepare an \AlP/\CaP ion crystal in a linear Paul trap loaded by ablation from solid-state targets, followed by 2-step photoionization. A trapped single \Ca ion has secular motional frequencies of $(\omega_z,\omega_x,\omega_y)=(1.16,2.03,2.05)$\,MHz. 
The trap is located in an ultra-high-vacuum stainless-steel chamber with a pressure of \SI{11(6)}{\nano\pascal}. 
The residual background pressure was determined from position-swap events of an \Al/\Ca crystal recorded with an EMCCD camera using the model of Ref.~\cite{hankin_systematic_2019}, yielding a mean event time of \SI{212}{\second}.
A magnetic bias field of \SI{0.15}{\milli\tesla} is generated by coils along the axial direction of the trap. It is stabilized to \SI{<2}{\nano\tesla} for \SI{1}{\second} to \SI{100}{\second} using a fluxgate sensor near the vacuum chamber and an additional set of coils.

\begin{figure}[ht] 
	\centering \includegraphics[width=1\columnwidth]{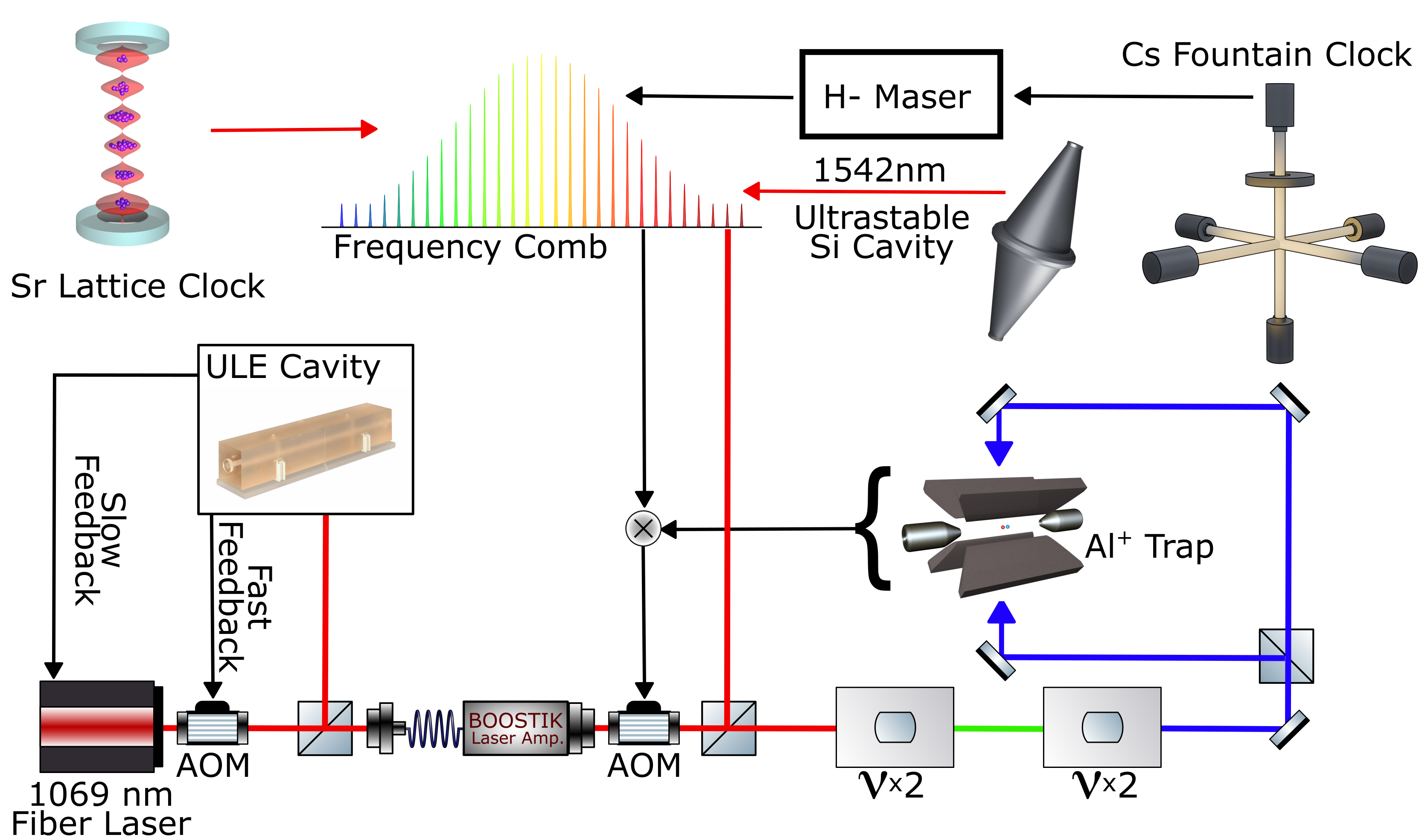}
	\caption{Simplified sketch of the frequency ratio measurement setup of the Sr and \Al clocks. 
    $\nu\times 2$  indicate second harmonic generation stages. AOM: acousto-optic modulator. Symbols from \cite{complib}.}
	\label{fig:clock_laser_path} 
\end{figure}

Fig.~\ref{fig:clock_laser_path} shows the clock laser and frequency comparison setup. 
Frequency stabilized \SI{1069}{\nano\meter} light \cite{amairi_reducing_2013, amairi_long_2014} is guided through a $\SI{100}{\meter}$-long fiber to a fiber amplifier in the lab. The major part of the amplified light is frequency quadrupled to \SI{267}{\nano\meter}, the \Al clock transition wavelength, in two second-harmonic generation stages with piecewise optical pathlength stabilization \cite{kraus_phase-stabilized_2022}, where the last reference mirror is less than \SI{23}{\centi\meter} away from the vacuum chamber. The remaining part of the amplified light is sent to an optical frequency comb, where it is used to measure the frequency of the clock and to transfer lock \cite{scharnhorst_high-bandwidth_2015} the \SI{1069}{\nano\meter} laser to an ultrastable Si cavity with a stability of \num{4e-17} at \SI{1}{\second} \cite{matei_1.5_2017}. Each optical clock measures its ratio against the same silicon cavity's resonance frequency via optical frequency combs, in this way the cavities common-mode noise is rejected. All electronic components are frequency referenced to the same hydrogen maser and the frequency measurements are synchronized.

\textit{Clock operation}---
The clock is operated on the \ssztpz transition of \Al, which has \SI{8}{\milli\hertz} natural linewidth and a hyperfine-induced ($I=5/2$) Zeeman splitting in both states. 
At the start of each clock cycle, \Al is prepared in one of the stretched states of the ground state \ssz via frequency-addressed optical pumping on the \ssztpo transition, taking advantage of the \SI{300}{\micro\second}-short excited-state lifetime. Doppler cooling on \Ca is applied in each clock interrogation cycle for at least \SI{2.5}{\milli\second}.
Subsequent electromagnetically induced transparency (EIT) cooling \cite{roos_experimental_2000, lechner_electromagnetically-induced-transparency_2016, scharnhorst_experimental_2017_2} for \SI{10}{\milli\second} prepares all motional modes near the motional ground-state. 

The clock transition is probed in a Ramsey sequence with \SI{250}{\milli\second} dark time between \SI{25}{\milli\second} long $\pi/2$ pulses. During the entire Ramsey sequence, the \Ca ion is continuously EIT cooled to reduce the 2nd-order Doppler shift (also known as time dilation shift, TDS) \cite{dawel_high-stability_2025_2}. The \Al clock state is read out using a quantum nondemolition measurement with 10 repetitions on the \ssztpo transition for higher state detection fidelity \cite{hume_high-fidelity_2007}. 

The \ssztpz transition is driven on a \mbox{$\Delta m=0$} transition starting in randomized-order in one of the stretched states, to eliminate first-order Zeeman shifts.  Additionally, the clock transition is driven from two opposing directions to reduce first-order Doppler shifts. An error signal is generated from the resulting excitation probabilities, which steers the laser to each of the transition resonances. A typical clock cycle takes \SI{19.1}{\second} with an average duty cycle of \num{56}{\%} when including service measurements to monitor and to correct for systematic shifts (see below).

In the following, we discuss all the relevant systematic shifts and their uncertainty, summarized in Table~\ref{tab:error_budget}. Additional details are provided in the Supplemental Material \cite{supplement}.

\textit{Dc magnetic fields}---
The first-order dc magnetic Zeeman shift is suppressed by averaging the frequencies of the $\Delta m_F = 0$ transitions of the two stretched states $m_F = \pm 5/2$. The difference between these transition frequencies is proportional to the magnetic field at the position of the \Al ion. The magnetic field uncertainty is mainly limited by the uncertainty of the $g$-factor difference $g_P-g_S=\num{-0.00118437(8)}$ \cite{rosenband_observation_2007}. 
The magnetic field is used for the calculation of the second-order dc magnetic Zeeman shift $\Delta\nu/\nu=-C_2 \langle B^2\rangle$ with $C_2=\SI{71.944(24)}{\mega\hertz\per\tesla^2}$ \cite{brewer_measurements_2019}, where the uncertainty of the shift is mainly limited by the $C_2$ coefficient. For the employed magnetic field of $\SI{0.15}{\milli\tesla}$, the uncertainty is \num{5.4e-19}.

\textit{Ac magnetic fields}---
We measured the ac magnetic field arising from the rf  trap drive using the Autler-Townes splitting technique \cite{gan_oscillating-magnetic-field_2018} on the single trapped \Ca ion.
Additional permanent magnets are employed to generate a ground-state Zeeman level splitting of \Ca  matching the trap drive frequency at $\wrf\approx 2\pi\times \SI{28}{\mega\hertz}$. Near this resonance, each Zeeman level is split by the Rabi frequency of the magnetic coupling strength, as observed by probing the \dsohddfh transition. The ac magnetic field amplitude perpendicular to the dc bias field is directly proportional to the splitting frequency of the Autler-Townes pair. By measuring along different directions, the amplitude of the trap drive's magnetic field is determined to be $\sqrt{\langle B_{\text{ac}}^2\rangle}=\SI{16.01(15)}{\micro\tesla}$, which results in a relative clock frequency shift of \num{-164.6(3.1)e-19}.

\textit{First-order Doppler shift}---
The ions are confined in a three-dimensional near-harmonic potential deep in the recoil-free Lamb-Dicke regime, in which the spatial extend of the ion is much smaller than the laser wavelength. Nevertheless, there can be a mean velocity over the probe duration due to changes in the trapping potential, e.g. from surface charging by laser beams, thermal expansions, or vibration of the trap or the optical setup.
The first-order Doppler shift is determined by taking the frequency difference of two counterpropagating probe beams \cite{brewer_27al+_2019} to \num{-31(28)e-19}, limited by the statistical uncertainty of the acquired data. The average transition frequency from both directions reduces the measured shift depending on the angular mismatch between the two beams. For an angular mismatch of \SI{6}{\milli\radian} estimated from measuring the coupling efficiency into the optical fibers delivering the respective other beam, the uncertainty of the average is bounded below \num{2.3e-19}.

\textit{Time dilation shift (TDS)}---
Residual kinetic energy causes a TDS, which can be described by the motional mode frequency and mean occupation, including the zero-point energy. EIT cooling is used during interrogation which results in a thermal motional state. This allows us to use sideband thermometry \cite{monroe_resolved-sideband_1995} to determine the mean motional occupation. The steady state motional mode occupation results in a TDS from residual thermal motion of \num{-16.9(2.0)e-19}, which is analyzed in more detail in Ref.~\cite{dawel_high-stability_2025_2}.

\begin{figure}[ht] 
	\centering \includegraphics[width=1\columnwidth]{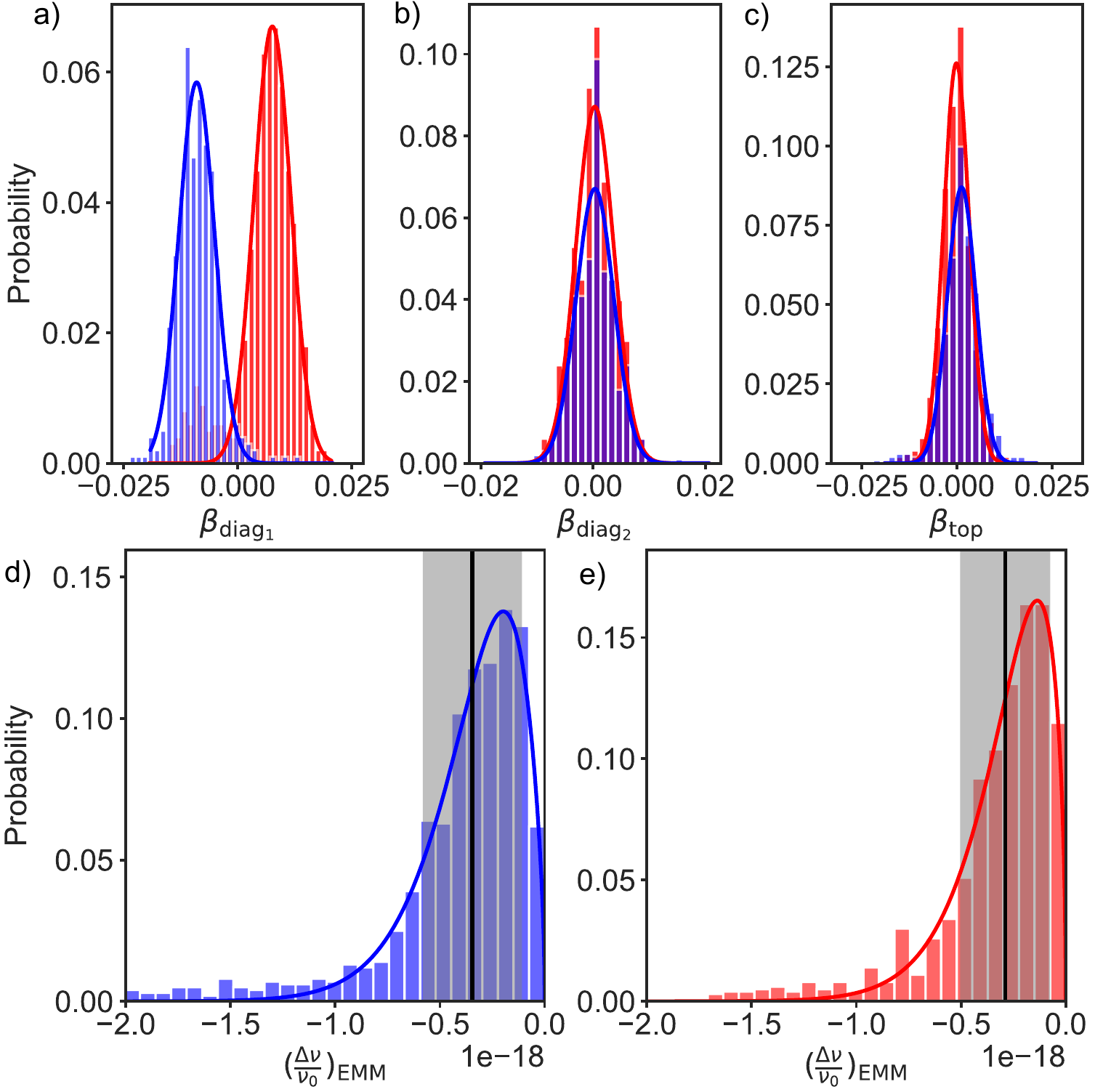}
    \caption{a)-c) Modulation index measured from three directions during the clock measurement campaign. The micromotion is amplitude controlled via compensation electrodes for signals of the directions "diag2" and "top". The "diag1" direction was kept at a non-zero but symmetric level for both crystal configurations. All directions show a gaussian distribution. The different colors indicate the orientation of the crystal \AlP - \CaP (blue) and \CaP - \AlP (red). d) and e) show the relative frequency shift of each measurement for the \AlP - \CaP and \CaP - \AlP orientation respectively. The shift is nearly independent of the orientation of the crystal.}
	\label{fig:MM_plot} 
\end{figure}

\textit{Excess micromotion shift}---
 Static electric fields push the ion into the trap rf field, leading to excess micromotion (EMM). The kinetic energy of this driven motion gives rise to an additional TDS \cite{keller_precise_2015}. EMM is monitored during the clock lock using interleaved service measurements after every \nth{5} clock feedback cycle and is compensated after every \nth{10} cycle employing phase modulated sideband spectroscopy \cite{arnold_enhanced_2024} on \Ca. The phase sensitivity of this methods enables measurement and minimization of micromotion using two-point sampling. The micromotion is adjusted such that both ions experience nearly the same amount of stray electric field independent of the crystal orientation. The shift is estimated by:
\begin{equation}
    \left(\frac{\Delta\nu}{\nu}\right)_{\mathrm{EMM}}=-\left(\frac{\wrf}{2c_0k_{729}}\right)^2\sum_{i=1}^3 \beta_i^2
\end{equation}
with $\wrf$ the angular trap drive frequency, $c_0$ the speed of light, $k_{729}$ the wave vector of the probing laser, and $\beta_i$ the modulation index of the micromotion direction obtained from the measurement. As the $\beta_i$ are normally distributed, the shift follows a chi-square distribution (see Fig.~\ref{fig:MM_plot}). The measurement of \Ca is transferred to \Al by measuring the distributions of $\beta$ for both \Ca orientations in a two-ion crystal. 
The shift on \Al is determined to be \num{-3.5(3.8)e-19} by fitting the noncentral $\chi^2$ distribution to the sum over $\beta_i^2$. 
EMM also causes an ac-Stark shift \cite{keller_precise_2015} of \num{-0.13(10)e-20} which has a marginal contribution due to the small static differential polarizability of \Al.

\textit{Clock laser ac-Stark shift}---
Off-resonant coupling of the probed clock states to other states by the clock laser results in an ac-Stark shift of the clock transition, proportional to the square of the probe Rabi frequency. This shift is estimated by rescaling the result reported in Ref.~\cite{chou_frequency_2010} to our experimental parameters. 
Adjusting for the Ramsey pulse sequence \cite{taichenachev_compensation_2010}, we obtain a clock laser light shift of \num{0(20)e-20}. The impact of polarization mismatches of the clock laser \cite{yudin_probe-field-ellipticity-induced_2023, king_optical_2022} are below $10^{-20}$.

\textit{EIT cooling laser ac-Stark shift}--
Cooling during the clock interrogation reduces the TDS, but introduces an ac-Stark shift $\Delta \nu_\mathrm{ac,L} = -\frac{1}{4h} \mathrm{E^2_L}\Delta\alpha(\lambda_\mathrm{L})$ on the clock transition via its differential polarizability at the cooling wavelength, $\Delta\alpha(\lambda_\mathrm{L})$. The electric field strength $\mathrm{E_L}$ of the cooling laser is determined via the frequency shift of the \SI{729}{\nano\meter} transition on \Ca. As \Ca and \Al are separated only by a few micrometers, the electric field measured on \Ca can be transferred to \Al. The ac-Stark shift for all cooling lasers is at \num{-93(11)e-19} level, where the main uncertainty comes from the differential polarizability of \Al at \SI{397}{\nano\meter} \cite{dawel_high-stability_2025_2}.

\textit{Blackbody radiation (BBR) shift}---
Thermal radiation from the environment shifts the clock transition via its differential polarizability $\Delta\alpha$, which is small for \Al across the BBR spectrum.
Nevertheless, \Al is exposed to blackbody radiation from chamber and trap parts. The temperature around the vacuum chamber (typically $\SI{299.3}{\kelvin}$) and the temperature on one of the trap's isolating sapphire discs are monitored using PT$100$ sensors. When operating the trap, the temperature of the latter rises by \SI{2.7}{\kelvin}, which translates to an effective temperature rise seen by the ion of \SI{1.0(5)}{\kelvin} \cite{dolezal_analysis_2015}. Including this temperature rise leads to a conservative temperature estimate of $T=\SI{300.3(3.0)}{\kelvin}$ at the ion position, a clock frequency shift of \num{-32.1(1.7)e-19} is estimated.

\textit{Collisions}---
Polarization of neutral background-gas particles close to the ions leads to an attractive potential that depends on the ion-atom distance \cite{meir_dynamics_2016}. 
At small distances, spiraling (Langevin) collision take place which cause phase and motional shifts. At large distances, purely elastic collisions cause motional shifts (TDS) through transfer of kinetic energy \cite{hankin_systematic_2019, barrett_analysis_2025_2}. For large energy transfers, interaction of the clock light with the ion is suppressed by a reduction of coupling strength (motional Debye-Waller factors). For the phase shift from Langevin collisions, a frequency shift of \num{0.2(2.8)e-19} is obtained using the analytical equations from Ref.~\cite{davis_improved_2019}. 
The increase in motional energy due to collisions is not captured by sideband thermometry of the ion crystal \cite{chen_sympathetic_2017, rasmusson_optimized_2021}, but needs careful modeling of the dynamics.
The model of Ref.~\cite{hankin_systematic_2019} is employed and scaled to our experimental parameters. As cooling during the interrogation reduces the motional shift, the effect of the Debye-Waller suppression is lowered. Taking both effects into account, a collisional shift of \num{-5.4(6.5)e-19} is estimated, limited by the accuracy of the model and pressure in the chamber.

\textit{AOM phase chirp}---
The pulses used for the Ramsey interrogation change the rf power in the acousto-optic modulators (AOMs) used for switching. This can cause phase chirps due to thermal and transient electronic effects in the AOMs \cite{degenhardt_influence_2005, rosenband_frequency_2008}.
An interferometric setup containing the AOM in one arm \cite{kazda_phase_2016} is used to measure the AOM phase chirp. From the measured phase excursion, a fractional frequency shift of \num{0(30)e-20} is estimated for a Ramsey experiment.

\textit{Servo error}---
The clock laser drifts during the interrogation due to length changes of the reference cavity. This results in a shift that can be reduced by the use of a double integrator \cite{peik_laser_2006} and  depends linearly on the drift rate of the laser \cite{yuan_suppression_2021_2}. The frequency shift and uncertainty of the servo is estimated by a numerical Monte Carlo simulation incorporating the laser drift and lock gain parameters. For each lock run, a linear slope is fitted to each measurement interval. An average drift rate of \SI{-87}{\micro\hertz\per\second} over all measurement days at 1069\,nm is extracted from the clock laser corrections, ranging from \SI{-180}{\micro\hertz\per\second} to \SI{28}{\micro\hertz\per\second}. For optimal parameters, the frequency uncertainty is \num{2e-19}.

\textit{Electric quadrupole shift}---
Electric field gradients couple to electric quadrupole moments (QPM) of the clock states, resulting in a frequency shift of \num{-0.23(8)e-19}. This effect is small in \Al due to the low QPM in the \ssz and \tpz states \cite{beloy_hyperfine-mediated_2017}.

\begin{table}[h!]
\caption{\label{tab:error_budget} The total estimated systematic uncertainty budget \Al clock in $10^{-19}$}
\label{tab:4s_state_Ca}
\begin{ruledtabular}
\begin{tabular}{l S[table-format=7.3] S[table-format=6.3]}
Effect & {Shift}  & {Uncertainty}  \\
\hline
Cooling laser light shift & -93 & 11 \\
Collisions & -5.4 & 6.5 \\
Quadratic Zeeman - dc & -14900.3 & 5.4 \\
Excess micromotion & -3.5 & 3.8 \\
Quadratic Zeeman - ac & -164.6 & 3.1 \\
Phase chirp & 0.0 & 3.0 \\
First-order Doppler shift & 0.0& <2.3 \\
Clock laser light shift & 0.0 & 2.0 \\
Time dilation shift & -16.9 & 2.0 \\
Servo error   & 0.0 & <2.0 \\
Black-body radiation & -32.1 & 1.7\\
Electric quadrupole shift & -0.23 & 0.08 \\
Ac-Stark shift trap & -0.013 & 0.010 \\
\hline
\textbf{Total} & -15216 & 16 \\
\end{tabular}
\end{ruledtabular}
\end{table}

\begin{figure}[ht] 
	\centering \includegraphics[width=0.95\columnwidth]{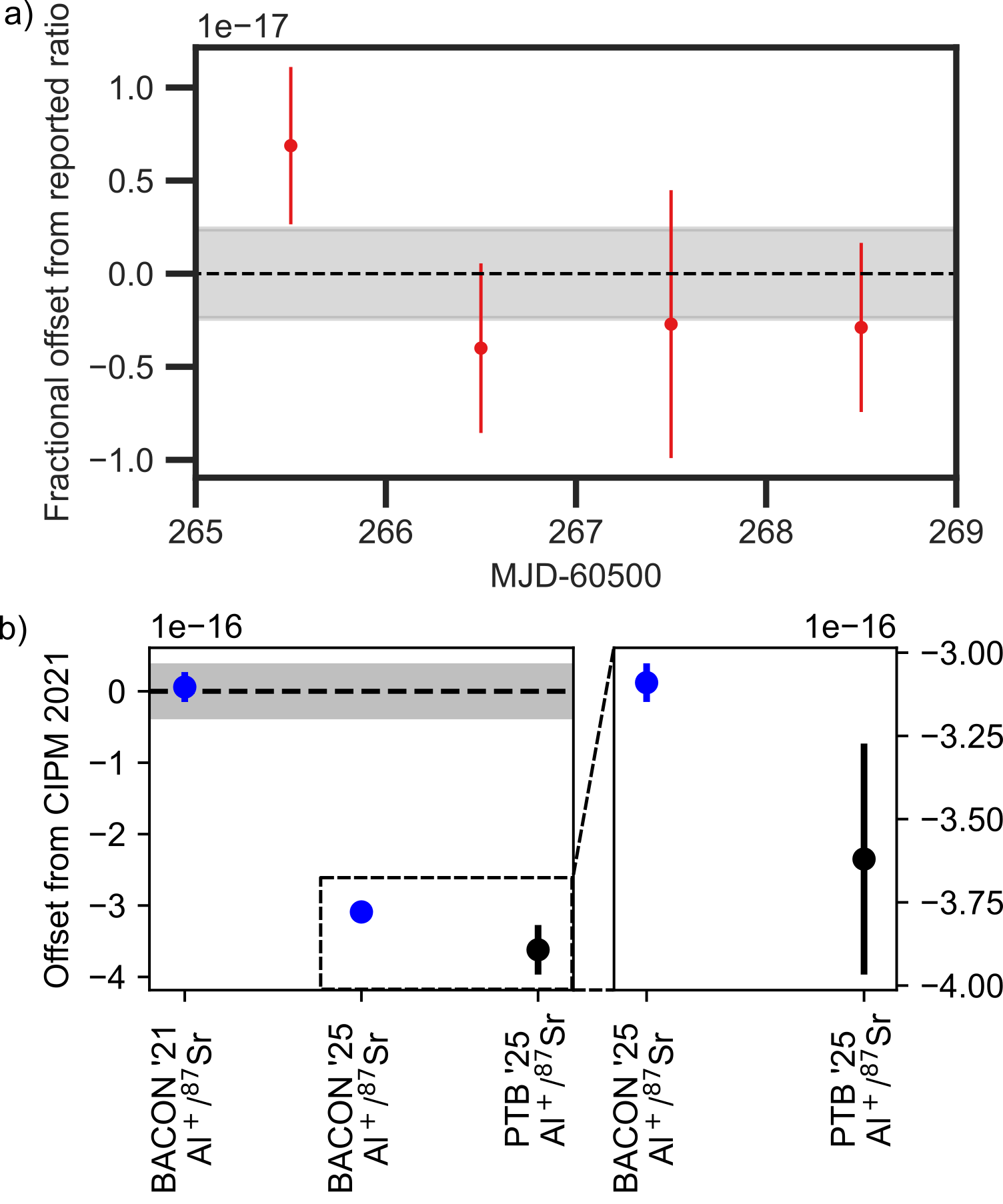}
    \caption{a) Shows the daily mean of the frequency ratio measured between \Al and Sr at each modified Julian date (MJD). The error bars assume white frequency noise behavior and are extrapolated to the measurement time of each day. The gray area shows the statistical uncertainty of all measurements. b) Comparisons of measurements of the \Al Sr frequency ratio. 
    }
	\label{fig:Frequency_ratio} 
\end{figure}

\textit{Frequency ratio measurement}---
The \Al clock is compared to the Sr3 optical lattice clock at PTB \cite{klo26} with an instability of \num{1.1e-16}$\sqrt{\SI{1}{\second}/\tau}$ and a fractional systematic uncertainty of \num{1.3e-17} during these measurements. The \Al clock used non-optimal servo parameters, which increase its systematic uncertainty to \num{3.5e-18} for the frequency ratio measurement.
The evaluation of the systematic frequency shifts in Sr3 is described in detail in Ref.\ \cite{klo26}.
At the time of the measurements reported here, Sr3 was subject to a frequency offset caused by the clock laser source, which is corrected retrospectively and increases the fractional systematic uncertainty from few \num{e-18} to \num{1.3e-17} \cite{klo26}.

Relative to \Al the gravitational redshift due to the different elevations is \num{-13761.9(6.3)e-19} for \Sr, \num{-13948.7(7.7)e-19} for CSF1, and \num{-14064.4(7.7)e-19} for CSF2. It has been determined through local leveling combined with a local measurement of Earth's gravity acceleration \cite{denker_geodetic_2018, riedel_direct_2020}.

The frequency ratio of the two clocks was measured between Modified Julian Dates (MJDs) 60765 and 60678. The mean result of each measurement is shown in Fig. \ref{fig:Frequency_ratio}. Overall \SI{60483}{\second} of data with a stability of $\num{5.9e-16} \sqrt{\SI{1}{\second}/\tau}$ \cite{dawel_high-stability_2025_2} are used. The reduced chi-squared value shows values of the daily measurements of $\chi^2 _{\mathrm{red}}=1.33$, which is consistent with the uncertainties. The measured frequency ratio is
\begin{equation}
    \nu_{\mathrm{Al}^+}/\nu_{^{87}\mathrm{Sr}}=\num{2.611 701 431 781 462 668(36)}
\end{equation}
This differs by $\SI{8.6}{\sigma}$ from the value reported in Ref.~\cite{beloy_frequency_2021} but only by $\SI{1.2}{\sigma}$ from that reported recently in Ref.~\cite{aeppli_atomic_2025_2} (see also Fig.~\ref{fig:Frequency_ratio}). It marks one of the first inter-species frequency ratio measurements at the low \num{1e-17} level conducted at two different institutions.

We also compared the \Al clock transition with two caesium fountain clocks at PTB \cite{weyers_advances_2018}, to measure its absolute frequency. For this, both the \Al clock laser and the Cs fountain clocks are synchronously compared against the same hydrogen maser for \SI{310}{\hour} (see Fig. \ref{fig:clock_laser_path}), such that the maser noise is common-mode rejected, at least on the relevant long averaging times. Following the method described in Refs.~\cite{schwarz_long_2020, grebing_realization_2016} yields an absolute frequency of $\nu_{\mathrm{Al}^+}=$\SI{1\,121\,015\,393\,207\,859.19(24)}{\hertz}, which corresponds to a relative uncertainty of \num{2.1e-16}. Using the absolute frequency of Sr3 \cite{klo26} measured via PTB's Cs fountain clocks, an absolute frequency of $\nu_{\mathrm{Al}^+}=$\SI{1 121 015 393 207 859.20(18)}{\hertz} is measured. Both absolute frequencies agree within their uncertainties with each other and with the recommended value of BIPM \cite{BIPM_recommended} and the one reported by NIST \cite{leopardi_measurement_2021}. 

\textit{Conclusion}---
The full evaluation of an \Al optical clock and measurement of its absolute frequency presented in this letter mark an important progress towards fulfilling the criteria for a redefinition of the SI second \cite{dimarcq_roadmap_2024} including \Al as a candidate system. 
However, the observed deviations in optical frequency ratios at a level of \num{1e-17} measured by different institutions illustrates the continued need for the validation of estimated systematic uncertainty budgets through inter-institutional comparisons of the frequency ratios.

\section*{Acknowledgements}
We thank Daniele Nicolodi, Thomas Legero, and Uwe Sterr for providing the Si cavity-stabilized laser. We also thank Burghard Lipphardt to provide values of the ratio between the maser and Si laser light. The project was supported by the Max-Planck–Riken–PTB–Center for Time, Constants and Fundamental Symmetries, and the Deutsche Forschungsgemeinschaft (DFG, German Research Foundation) under Germany’s Excellence Strategy – EXC-2123 QuantumFrontiers – 390837967 and Project-ID 274200144 – SFB 1227 DQ-mat, projects B03 and B02. This project also received funding from the European Metrology Programme for Innovation and Research (EMPIR) and the European Partnership on Metrology, cofinanced by the five participating States and the European Union’s Horizon 2020 and Horizon Europe research and innovation programmes (Projects No. 20FUN01 TSCAC 23FUN02 CoCoRICO 23FUN03 HIOC). This project further received funding from the European Research Council (ERC) under the European Union’s Horizon 2020 research and innovation programme (grant agreement No 101019987). This project has been supported by the German Federal Ministry of Education and Research within the funding program "Clusters4Future", contract number 03ZU1209EE (QVLS-iLabs) and by the State of Lower Saxony, Hannover, Germany, through Niedersächsisches Vorab (project QVLS-Q1).

\bibliographystyle{unsrt}
\bibliography{EQM_Master,supplement} 

\onecolumngrid
\clearpage

\appendix
\section{Preparation of the \texorpdfstring{\AlP/\CaP}{} crystal} 

We load about 5 to 10 \Ca ions into the trap using pulsed laser ablation of a calcium target near the trap and photoionization with cw lasers at \SI{375}{\nano\meter} and \SI{422}{\nano\meter} wavelength. This is followed by pulsed laser ablation of \Al from an aluminum target next to the calcium target and photoionization using a cw laser at \SI{394}{\nano\meter} wavelength. Ablating aluminum is typically accompanied by a decrystallization of the \Ca ions. After recrystallization by lowering the rf trapping potential, successful loading of \Al is indicated by a dark ion in the \Ca crystal as observed by an EMCCD camera. If recrystallization is not achieved within \SI{30}{\second}, ablation loading is repeated until success.
After a dark ion is crystallized, the excess \Ca ions are released from the trap by lowering the rf trapping potential until a single \Ca is trapped together with the dark ion. We identify the dark ion as \Al through motional sideband spectroscopy on \Ca.
The typical preparation time of a single \Ca--\Al ion crystal in our system ranges between \SI{5}{\minute} to \SI{25}{\minute}.

\begin{figure}[thbp]
    \centering
    \includegraphics[width=12.5cm]{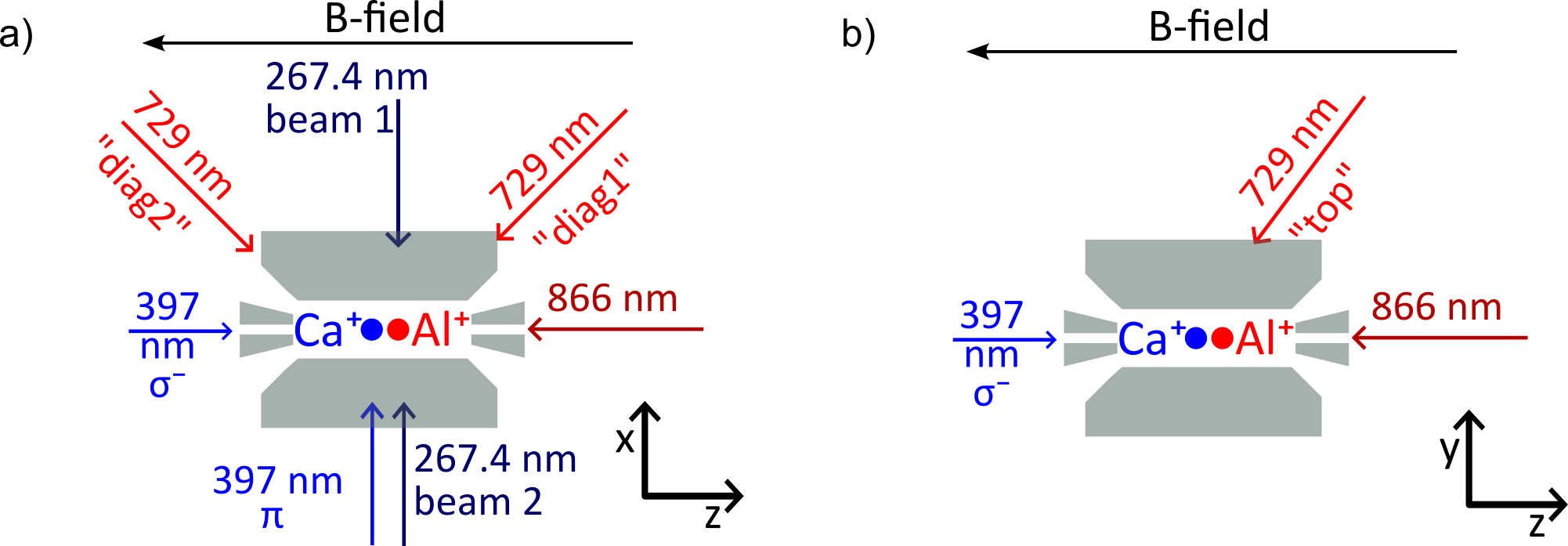}
    \caption{Laser setup used in the experiment. a) shows the projection on the xz plane, while b) shows the projection on the yz plane.}
    \label{fig:Simple_laser_setup}
\end{figure}

\section{Motional shifts}
\label{sec:Motional_shifts}
Laser light which interrogates an atom with a velocity component parallel to the laser's wavevector $v_\parallel=\hat{k}\cdot\vec{v}$ experiences a Doppler shift (DS). Additionally, any velocity of the atom introduces a time dilation shift, which further changes the laser frequency $\nu$ in the moving frame of the atom. Therefore, the atom is interrogated at a frequency $\nu'$ \cite{chou_optical_2010}:

\begin{equation}
    \nu'=\nu\gamma\left(1-\frac{v_{\parallel}}{c_0}\right)
\end{equation}
where $\gamma=1/\sqrt{1-(v/c_0)^2}$ is the Lorentz factor. To excite the atomic transition the frequency at the atom must fulfill $\langle\nu'\rangle=\nu_0$. 
We can calculate the frequency difference $\delta\nu$ as \cite{chou_optical_2010,ludlow_optical_2015}:
\begin{equation}
    \frac{\delta \nu}{\nu_0}=\underbrace{\frac{\langle v_{\parallel}\rangle}{c_0}}_\textrm{first-order DS} - \underbrace{\frac{\langle v^2 \rangle}{2c_0^2}}_\textrm{TDS}+\underbrace{\frac{\langle v_{\parallel}\rangle^2}{c_0^2}}_\textrm{second-order DS} + O((v/c_0)^3).
\end{equation}
where $O((v/c_0)^3)$ are higher order terms. The first and third term are the non-relativistic Doppler shift terms up to second-order. The second term is due to the time-dilation shift (TDS) from the expansion of the Lorentz factor. 

\subsection{First-order Doppler shift}
\label{sec:first_order_Doppler}

\begin{figure}[thbp]
    \centering
    \includegraphics[width=5cm]{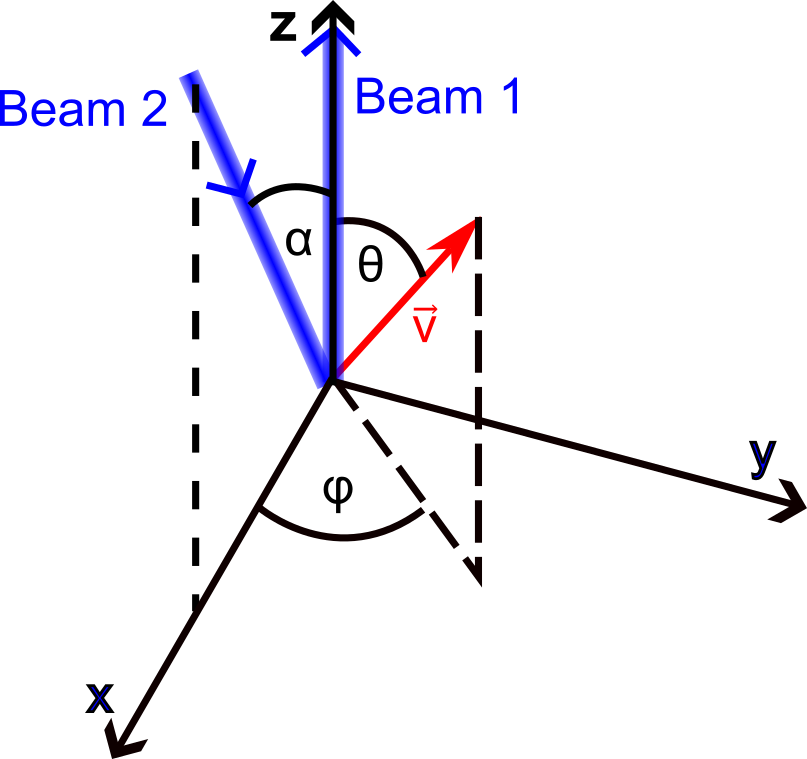}
    \caption[Different angles first-order Doppler]{Beam 1 is propagating along the $z$-axis. For beam 2 the coordinate system is rotated, such that it is in the $x-z$ plane. The ion's velocity direction is described by two angles $\theta$ and $\varphi$.  Note the change in coordinate system compared to Fig.~\ref{fig:Simple_laser_setup}.}
    \label{fig:angles_first_order_doppler}
\end{figure}

The ions are confined in the electromagnetic potential of an ion trap. Therefore, the average velocity should be zero for times exceeding the periodic motion \cite{leibfried_quantum_2003, chou_optical_2010, ludlow_optical_2015}. Nevertheless, first-order Doppler shifts can arise from slow continuous movements of the ion against the laser beam \cite{ludlow_optical_2015}. The ion can move because of rf or dc voltage changes due to settling effects of power supplies and the response characteristics of low pass filters. Additionally, UV light can charge dielectric parts of the trap or produce patch potentials, which may also contribute to a drift in the position of the ion. Moreover, optical path length drifts of the laser beam with respect to the ion trap from, e.g thermal expansion of the optical breadboard or the ion trap, as well as slow vibrations could lead to first-order Doppler shifts. If the ion starts to drift and is pushed into the rf field of the trap, this motion is also visible as an increase in excess micromotion \cite{brewer_27al+_2019}.

Here, we follow the approach of Ref.~\cite{brewer_27al+_2019} to estimate the first-order Doppler shift and reduce its uncertainty using two counterpropagating probe beam directions. The coordinate system can be chosen such (see also Fig.~\ref{fig:angles_first_order_doppler}) that one clock beam propagates along the $z$-axis (note the change in coordinate system compared to Fig.~\ref{fig:Simple_laser_setup}) and that the second probe beam is counterpropagating with angle mismatch $\alpha$:
 \begin{equation}
     \hat{k}_1=\hat{z} \textrm{ and } \hat{k}_2=-\sin(\alpha)\hat{x} - \cos(\alpha)\hat{z}
 \end{equation}
with $\hat{k}_{1,2}$ the normalized wavevector of the first and second probe beam, respectively, and $\hat{x},\hat{y},\hat{z}$ the normalized coordinate vectors. The ion itself will have a drift velocity of $\vec{v}$, which can be described in this coordinate system as:

\begin{equation}
    \vec{v}=v\sin(\Theta)\cos(\varphi)\hat{x}+v\sin(\Theta)\sin(\varphi)\hat{y}+v\cos(\Theta)\hat{z},
\end{equation}

where $\Theta \in [0,\pi]$ and $\varphi \in [0,2\pi)$ follow the definition of spherical coordinates. The difference frequency $\Delta \nu_{\mathrm{dif}}$ between the two-probe laser can be measured and depends on the first-order Doppler shifted frequencies $\Delta \nu_{1,2}$ of each beam:

\begin{equation}
    \frac{\Delta \nu_{\mathrm{dif}}}{\nu_0} = \frac{1}{2} \left(\frac{\Delta \nu_1}{\nu_0} -\frac{\Delta \nu_2}{\nu_0}\right) = \frac{v}{2 c_0}(\cos(\Theta)+\cos(\alpha) \cos(\Theta) + \sin(\alpha) \sin(\Theta) \cos(\varphi))
\end{equation}

with $\nu_0$ the unperturbed frequency of the clock transition.
We can further simplify the result if we assume that the angle difference $\alpha$ between both beams is small. Thus, we get
\begin{equation}
    \frac{\Delta \nu_{\mathrm{dif}}}{\nu_0} \approx \frac{v}{2 c_0}(2\cos(\Theta)  + \alpha \sin(\Theta)\cos(\varphi)).
\end{equation}
Here, it is evident that $\Delta\nu_{\mathrm{dif}}$ becomes large when the velocity is high along $z$  ($\Theta\approx0$). The difference due to motion in the other directions is suppressed by $\alpha$. For $2\cos(\Theta)=-\alpha\sin(\Theta)\cos(\varphi)$, the sensitivity for the velocity is lost.

To calculate the effect of the velocity on the frequency shift we need to take the average of $\Delta \nu_{1,2}$:
\begin{equation}
    \frac{\Delta \nu_{av}}{\nu_0}=\frac{1}{2}\left(\frac{\Delta \nu_1}{\nu_0}+ \frac{\Delta \nu_2}{\nu_0}\right)\approx-\frac{v}{2 c_0}(\alpha \sin(\Theta)\cos(\varphi)),
\end{equation}
where we again have used the small angle approximation. The frequency shift averages to zero for perfectly counterpropagating probe beams. In case of $\alpha \neq 0$, a residual shift occurs. The relevant velocity can be determined from the difference frequency measurement: 
\begin{equation}
    \frac{\Delta \nu_{av}}{\nu_0}=-\frac{\Delta \nu_{dif}}{\nu_0}\frac{\alpha \sin(\Theta)\cos(\varphi)}{2 \cos(\Theta) + \alpha \sin(\Theta)\cos(\varphi)} 
    \label{eq:average_shift_first_order_doppler}
\end{equation}
\begin{figure}[htp]
    \centering
    \includegraphics[width=0.6\linewidth]{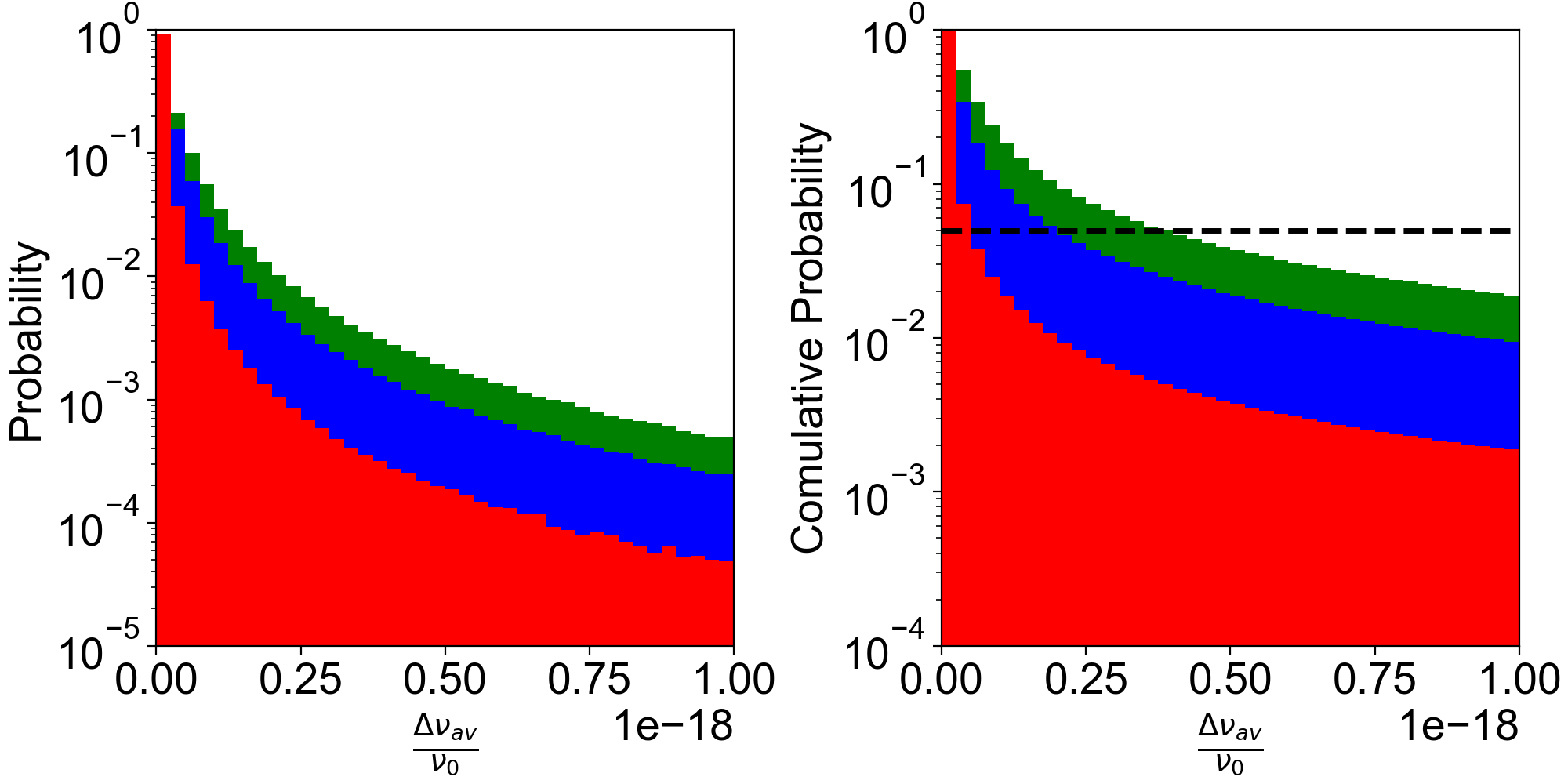}
    \caption[First-order Doppler uncertainty dependence on the beam angle between the counterpropagating beams]{Probability to see a certain first-order Doppler shift depending on the direction of the ion velocity. In this figure different angles between the counterpropagating beams $\alpha=$\numlist{0.01;0.006;0.001}\SI{}{\radian} are plotted (green, blue,red). The left figure shows the probability of finding an ion with a certain frequency shift. The right figure shows the cumulative probability to find the ion with a first-order Doppler shift of $\frac{\Delta \nu_{av}}{\nu_0}$ or higher. The black dashed line marks $5\,$\%.}
    \label{fig:First_order_doppler_residual}
\end{figure}
The equations show that the first-order Doppler shift of two nearly counterpropagating beam still depends on the angles of the ion's velocity to the interrogation beams. 
To estimate the influence of different ion velocity directions on a measured frequency difference $\Delta \nu_{\mathrm{dif}}$ we use a Monte-Carlo simulation to predict the probability of having a certain first-order Doppler shift. This also captures the loss of sensitivity for the condition stated above. For this we generated a uniformly distributed set of angles $\theta, \varphi$ that produce a frequency difference of $\Delta\nu_\mathrm{dif}/\nu_0=\num{5.9e-18}$, which corresponds to the magnitude of the measured frequency difference in a $1\sigma$ band (see below). For different angles $\alpha$,  the ion's velocity and resulting frequency shift can be calculated using Eq.~\eqref{eq:average_shift_first_order_doppler}. The probability to find a certain first-order Doppler shift for a certain $\alpha$ is shown in Fig.~\ref{fig:First_order_doppler_residual}, assuming isotropic probabilities of the angle distribution.   

\begin{figure}[htp]
    \centering
    \includegraphics[width=0.6\linewidth]{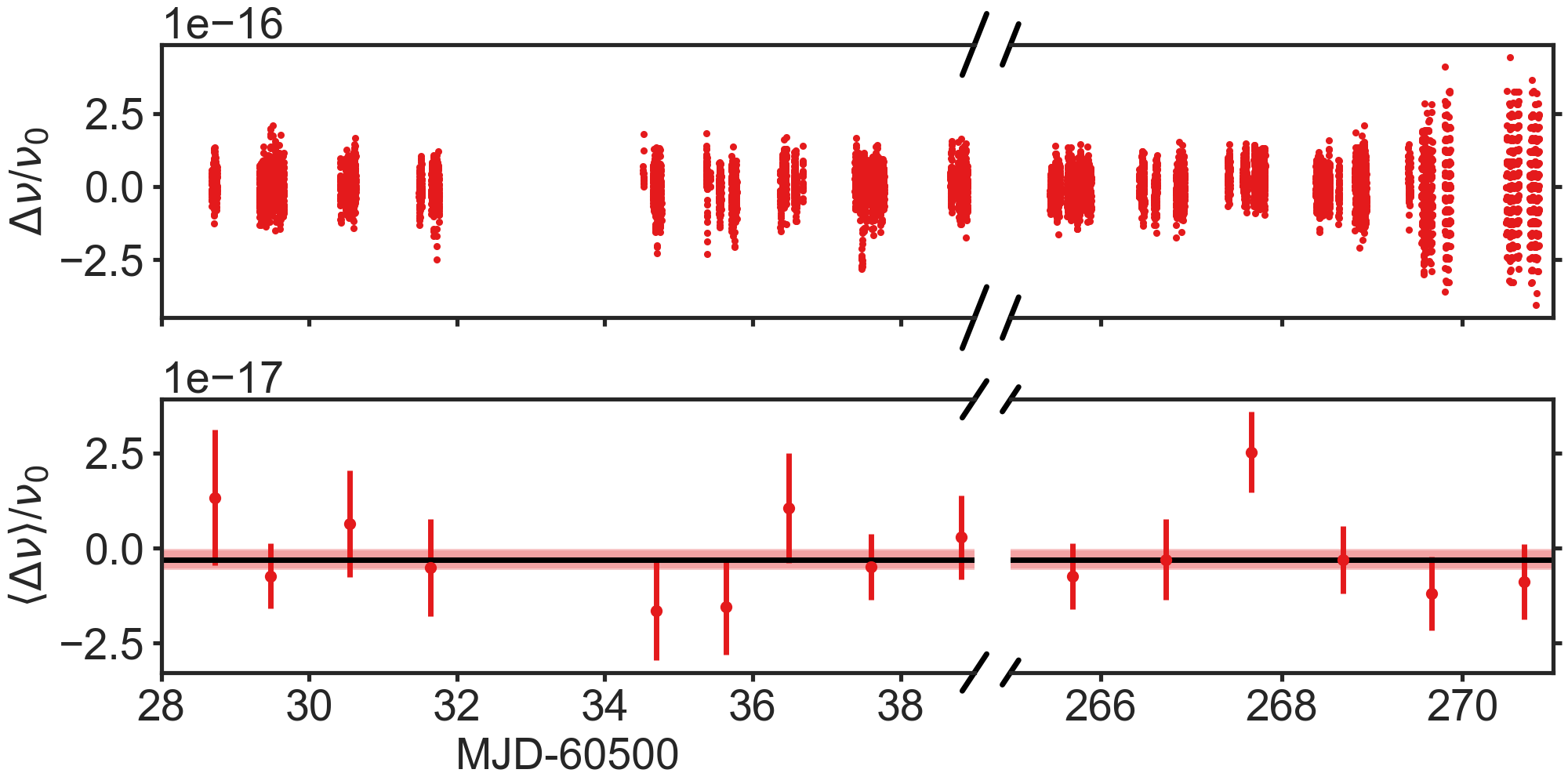}
    \caption[Measurement first-order Doppler shift]{Measurement of the first-order Doppler shift. The upper plot shows all frequency measurement points. The lower plot shows a daily average of the upper points. The black solid line depicts the mean and the red area the lowest point of the Allan deviation between the frequency difference of the two probe beam directions.}
    \label{fig:first_order_Doppler_iqloc}
\end{figure}

We ensure a counterpropagating alignment of the two probe beams by coupling into the fiber of the other beam. The distance from the ion to each fiber coupler is around \SI{0.9}{\meter} and the coupling efficiency is \SI{10}{\percent}. Using the overlap integral on the lens, this leads to an angle difference of not more than \SI{6}{\milli\radian}. Our measurement results of the first-order Doppler shift during a clock run in August 2024 and April 2025 are shown in Fig.~\ref{fig:first_order_Doppler_iqloc}.

The average frequency shift measured is at \num{-3.1(2.8)e-18}, which corresponds to a drift of \SI{0.9(0.8)}{\nano\meter\per\second} (for $\alpha=0$). There is no visible change between modified Julian dates (MJD) 60528 to 60539 and the ones at 60766 to 60771, which shows a good reproducibility of the small, but statistically insignificant shift. Additional measurements are required to reduce the statistical uncertainty.

Nevertheless, some shift may remain, but for an angle mismatch of around \SI{6}{\milli\radian}, the probability of the shift being \num{>2.3e-19} is \SI{<5}{\percent} ($2\sigma$). Therefore, we estimate the first-order Doppler shift uncertainty at \num{<2.3e-19}.  This is mainly limited by the measurement of the frequency difference and the unknown ion velocity direction.

\subsection{Time dilation shift (TDS)}
\label{sec:second_order_Doppler}
The time dilation shift depends on the kinetic energy of the ion in the trap, assuming that all other motional kinetic energy is small and can be neglected. The ion's motion in the trap can be described via the harmonic oscillator states, which we will focus on in the following. The time dilation shift can be calculated by \cite{berkeland_minimization_1998}
\begin{equation}
    \frac{\Delta \nu}{\nu_0}=-\frac{\langle E_{\text{kin}}\rangle}{mc^2}=-\frac{\langle v^2 \rangle}{2c^2}
\end{equation}
with $\langle E_{\text{kin}}\rangle$ the average kinetic energy over a time period T ($\langle\cdot\rangle$ indicates a time average) and $m,v$ the mass and velocity of the clock ion, respectively. The motional energy of the trapped particle is described by \cite{wubbena_sympathetic_2012}
\begin{equation}
    \langle E_{\text{kin}} \rangle= \frac{1}{2}\hbar \sum_i b_i^2 \omega_{S,i}\left(\langle \bar{n}_i \rangle + \frac{1}{2}\right)\left(1+S_i\right),
\end{equation}
where $b_i$ is the normalized eigenvector of the $i$-th mode, $\bar{n}_i$ is the mean motional state of the $i$-th mode, and $S_i$ is an additional factor for the contribution of the intrinsic micromotion, which also depends on the chosen mode (see \cite{wubbena_sympathetic_2012}). This factor vanishes in the axial direction [$S_{\textrm{IP},z}=S_{\textrm{OP},z}=0$(IP=in-phase and OP=out-of-phase)], as there is only a negligible rf field. The other parameters $S_i$ are defined as:
\begin{equation}
    S_{\textrm{IP},i}=\frac{2\epsilon^2/\mu}{2\epsilon^2/\mu-2\alpha-(1-\sqrt{\mu}b_1/b_2)}
\end{equation}
\begin{equation}
     S_{\textrm{OP},i}=\frac{2\epsilon^2/\mu}{2\epsilon^2/\mu-2\alpha-(1+\sqrt{\mu}b_2/b_1)},   
\end{equation}
in this formulas, $\epsilon=\sqrt{\omega_x^2/\omega_z^2+1/2}$ (single \Ca secular frequencies), $\mu=m_{Al}/m_{Ca}$ is the mass ratio, and the symmetry of the radial modes is given via $\alpha\approx0.5$. In our case, $S_{\textrm{IP},i}=1.22$ and $S_{\textrm{OP},i}=3.18$.
Our modes are symmetric, thus $S_{IP,x}=S_{IP,y}$ and $S_{OP,x}=S_{OP,y}$.
One important factor for the motion of ions is the heating rate. Over time, the motional energy of the ion will increase because of electric field noise \cite{brownnutt_ion-trap_2015}. The mean time averaged motional state is then given by:
\begin{equation}
        \langle \bar{n}\rangle = \frac{1}{T}\int^T _0 \bar{n}_0+\dot{\bar{n}}t\,dt=\bar{n}_0+\frac{\dot{\bar{n}}T}{2}.
    \label{eq:heating_rate}
\end{equation}
Here, $T$ is the time of the clock interrogation, $\dot{\bar{n}}$ is the heating rate and $\bar{n}_0$ is the mean motional state before the clock interrogation. Eq.~\eqref{eq:heating_rate} shows that the TDS increases for longer interrogation times. However, this can be mitigated by cooling during the clock interrogation. In this case, the time dependence vanishes, and the ion's motional state depends on the steady state of the applied cooling method. We apply EIT cooling during the clock interrogation, which cools the ions close to the motional ground-state but introduces a light shift \cite{dawel_high-stability_2025_2}. The light shift is evaluated in Sec.~\ref{sec:light_shifts_cooling}.

The TDS depends on the motional frequency and the mean motional state of each mode. The motional frequency can be determined in the resolved sideband regime by comparing the transition frequency difference between the carrier transition and a blue or red sideband. The main uncertainty of the motional frequencies is due to the mode stability. In Ref.~\cite{kramer_aluminum_2023} a Ramsey motion technique \cite{wineland_experimental_1998} was used to measure the stability of single \Ca sideband frequencies in our setup. The radial mode stability is best at \SI{20}{\second} with a frequency uncertainty of \SI{100}{\hertz}. The axial mode has the highest frequency stability at \SI{200}{\second} with \SI{1.5}{\hertz} uncertainty. The stability of the radial modes are worse because of thermal fluctuations of the helical resonator and the rf amplifier. Since the motional modes are not tracked during the clock interrogation, the fluctuations of the radial modes are taken with an uncertainty of \SI{6}{\kilo\hertz} as in Ref.~\cite{kramer_aluminum_2023}.

For the measurement of the motional state of the ion, we use the sideband thermometry method. This method relies on the fact that the excitation ratio between the blue and red sideband just depends on the mean motional state. The excitation probability of a first-order red sideband (RSB) for an ion in a thermal state has the form \cite{rasmusson_optimized_2021}:

\begin{equation}
    P_{\mathrm{RSB}}(t)=\sum_{m=1}^\infty\left(\frac{\bar{n}}{\bar{n}+1}\right)^m\sin^2\left(\frac{\Omega_{m,m-1}t}{2}\right)=\left(\frac{\bar{n}}{\bar{n}+1}\right)P_{\mathrm{BSB}}(t),
\end{equation}

which can be related to the blue sideband (BSB) excitation probability $P_{\mathrm{BSB}}$.
Therefore, we can calculate the mean motional state by:
\begin{equation}
    \bar{n}=\frac{R}{1-R} \hspace{1cm} \mathrm{with}\hspace{1cm} R=\frac{P_{\mathrm{RSB}}}{P_{\mathrm{BSB}}},
\end{equation}
where $R$ is the ratio of the RSB and BSB excitation probability. This method gives only reliable results if the motional state is a thermal state. The motional state distribution depends on the cooling process, e.g. Doppler cooling shows a thermal state distribution \cite{stenholm_semiclassical_1986}, while sideband cooling shows a non-thermal distribution \cite{rasmusson_optimized_2021} and can be modeled by double-thermal \cite{chen_sympathetic_2017} or a three-component thermal distribution \cite{brewer_27al+_2019}. In our experiment, we mainly rely on EIT cooling for motional ground-state preparation. For EIT the two \dsoh  Zeeman states and one \dpoh state of \Ca are used in a $\Lambda$ configuration, where the states are coupled with a strong $\sigma$ beam and a weak $\pi$ beam \cite{morigi_ground_2000}. This cooling was studied with a chain of \Ca ions, which showed thermal distributions \cite{lechner_multi-mode_2016}. For now, we will assume that after EIT cooling the ions are in a thermal state and use sideband thermometry to measure the mean motional state. 

To start the clock interrogation at a low motional state, we apply EIT cooling for \SI{10}{\milli\second} before the first clock pulse. This assures that the ions are thermalized at the EIT cooling limit. Most of the modes are cooled in \SI{1}{\milli\second} to \SI{2}{\milli\second}, just the radial IP modes need longer cooling times due to their small coupling strengths. After precooling, we keep the EIT beam on for the entire interrogation. The $\sigma^-$ beam causes a light shift of \SI{2.3}{\mega\hertz}. This is set a bit above the motional mode center frequency of \SI{2.08}{\mega\hertz} to have better cooling of the radial IP motional modes.

\begin{figure}[]
    \centering
    \includegraphics[width=0.6\linewidth]{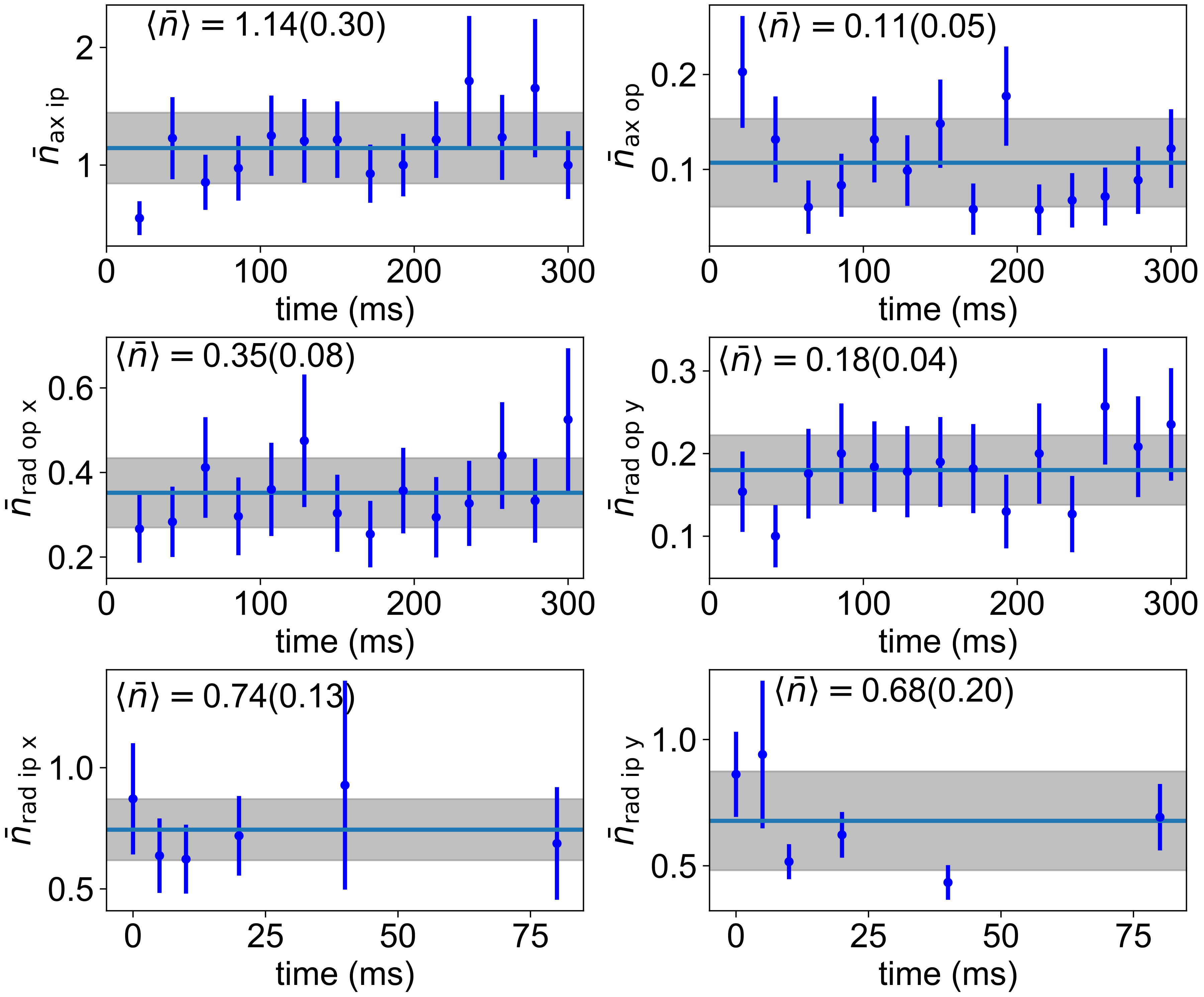}
    \caption[Cooling limits EIT cooling]{Achieved mean motional state occupation of all modes during EIT cooling. The length of the cooling was varied to confirm that the motional state is constant over time. The axial and radial OP modes, which are closer to the EIT cooling optimum, reach lower populations than the IP modes. The solid line shows the mean value, while the gray area depicts the uncertainty. The shown $\bar{n}$ for the radial IP modes are taken with slightly different EIT parameters for the measurement in 2024. Nevertheless, it shows that the $\bar{n}$ is constant during the entire interrogation, but the $\bar{n}$ can be different.}
    \label{fig:EIT_cooling limit}
\end{figure}

In Fig.~\ref{fig:EIT_cooling limit} the cooling limits of EIT cooling are shown for long times (after the \SI{10}{\milli\second} of initial cooling). We measured the excitation ratio between the RSB and BSB of the axial modes and radial OP modes with the \SI{729}{\nano\meter} transition of calcium. For the radial IP modes, we did frequency scans and used the fitted amplitudes to obtain the temperature of the modes. \Ca has very small participation in the radial IP modes, thus, probe times of \SI{400}{\micro\second} are used. Due to the long probe times, the drift of the sideband adds some noise.

\begin{table*}[htp]
\caption[Summary of all important parameters for the second-order Doppler shift]{\label{tab:secularfreq} Motional mode parameters of an \Al-\Ca crystal. Here $\omega$ is the motional mode frequency, $b_m$ are the normalized eigenvector components of \Al, $\eta_{729}$ the Lamb-Dicke parameter of the \SI{729}{\nano\meter} laser on \Ca (which includes the projection of the laser on the mode), $S$ is the multiplicative contribution due to intrinsic micromotion. The mean motional state after Doppler and EIT cooling are given by $\bar{n}_\text{Dop}$ and $\bar{n}_\text{EIT}$, respectively. The heating rate is $\dot{\bar{n}}$. The TDS with cooling column shows the resulting frequency shift of each mode for the measured $\bar{n}_\text{EIT}$. The radial (rad) IP mode and the axial (ax) OP mode have the largest weight for the secular motional shift on \Al, which is shown in the next-to-last column. The last column shows the shift per time when no cooling is applied during the interrogation. In the last two columns we neglected the zero-point energy. This table is taken from Ref.~\cite{dawel_high-stability_2025_2}.}
\resizebox{\columnwidth}{!}{
\begin{tabular}{l|cccccccccc}
Mode & $\omega$  & $b_m$ & $S$ &$\eta_{729}$ & $\bar{n}_\text{Dop}$ & $\bar{n}_\text{EIT}$ & TDS with &$\dot{\bar{n}}$ & Shift/quantum  & Shift/time \\
 & (MHz) &  & & &  &  &  cooling $(10^{-18})$& (quantum/s) & ($10^{-19}$) & ($10^{-18}$t/s)\\
\hline
ax IP & 1.25& 0.562 & 0 & 0.051 & 9.9 & 1.1(4) & 0.054(15) &56(6) & 0.33& 0.97(10)\\
ax OP & 2.26& 0.827 & 0 & 0.026 & 5.3 & 0.11(5) & 0.077(6)&3.7(1.0)& 1.27 & 0.31(6)\\
rad x OP & 1.759& 0.157& 3.17 & 0.039 & 6.7 & 0.35(8) & 0.00125(12)&37(4)& 0.15& 0.28(3)\\
rad y OP & 1.766& 0.157 & 3.17 & 0.030 & 6.7 & 0.18(4) &0.00101(7)&7.8(1.3)& 0.15& 0.07(1)\\
rad x IP & 2.868& 0.988 & 1.21 & 0.005& 4.0 & 0.78(19) &0.66(10) &53(8)& 5.2& 14.4(2.1)\\
rad y IP & 2.912& 0.988 & 1.21 & 0.004& 4.0 & 1.18(34) & 0.88(18)&11(5)& 5.2 & 3.7(1.3)
 
\end{tabular}
}
\end{table*}

The results of the motional states and heating rates are summarized in Tab.~\ref{tab:secularfreq}, which show that the largest impact is due to the radial IP modes. In these eigenmodes \Al oscillates much stronger than \Ca. Compared to this, the radial out-of-phase modes have no significant impact on the overall TDS of \Al, where the motion is much stronger on the \Ca ion. The motional state occupation of the axial in-phase mode is large, but it has only a small impact on the overall shift due to its low frequency and its nonexistent micromotion contribution. The shift/time column shows the effect of long interrogation times on the motional shift, where a probe time of \SI{0.3}{\second} will cause a frequency shift of \num{-71.2(7.7)e-19} without cooling. Applying cooling during the interrogation we arrive at a final motional shift of \num{-16.9(2.0)e-19}. The shift is primarily limited by the large shifts of the radial in-phase modes.

The shift can potentially be reduced by increasing the cooling for the radial in-phase modes. A set of EIT cooling parameters, optimized for the radial IP mode, could bring the modes closer to the motional ground-state, thereby reducing the overall shift and uncertainty. While this is possible, a higher frequency shift of the $\sigma$ beam causes a larger light shift, which is the main source of uncertainty so far (see Sec.~\ref{sec:light_shifts_cooling}). 

\subsection{Excess micromotion}
\label{sec:excess_micromotion}
In the last section, frequency shifts were caused by the secular motion and intrinsic micromotion of the ion in the trap, but there is also an additional motion of the ion caused by stray electric fields. The stray electric fields can arise from imperfect electrode geometry, from patch potentials on charged surfaces, e.g. from UV light exposure. Also a phase shift between the two rf blades can cause an additional micromotion. This effect will be neglected because so far it was not resolved in any measurements in our setup. The equation of motion is influenced by adding a static electric field in the following way \cite{berkeland_minimization_1998}:
\begin{equation}
    x(t)\approx [x_0+x_1\cos(\omega_0 t)]\left(1+\frac{q_x}{2}\cos(\wrf t) \right),
\end{equation}
where $x_0$ is a position offset due to the stray electric field and $x_1$ is the amplitude of the secular motion. The term $x_0(1+q_x\cos(\wrf t)/2)$ consists of a static offset and an oscillating term. This shows that a static electric field will move the ion by $x_0$ into the rf field where it will start an additional oscillation with an amplitude $x_0q_x/2$ at the trap drive frequency. This oscillating motion is called excess micromotion (EMM). 

EMM can be minimized by applying electric fields counteracting the stray field through compensation electrodes.
It can be characterized by measuring the phase modulation on the laser interaction. The ion is moving in a periodic motion $\vec{v}=\vec{v}_0\sin(\wrf t)$ modulated by the trap drive frequency. Thus, the spatial phase of the laser light gets modulated by the motion of the ion \cite{keller_precise_2015}
\begin{equation}
    \vec{k}\cdot \vec{r} = \frac{\vec{k} \vec{v_0}}{\wrf}\cos(\wrf t) = \beta \cos(\wrf t),
\label{eq:Modulationindex_EMM}
\end{equation}
where $\beta$ is the modulation index. Using this phase modulation one can write the electric field of the laser as:
\begin{equation}
    E\propto \exp(i (\omega_L t + \vec{k}\cdot\vec{r})) = \sum_{n=-\infty}^{n=\infty} J_n(\beta)\exp(i(\omega_L-n\wrf)t)
    \label{eq:electric_field_MM_modulation}
\end{equation}
where $J_n(\beta)$ are the Bessel functions of the first kind. This shows that the light gets additional resonant components at $\pm n\wrf$. Since in our case $\beta\ll 1$, we can neglect all higher-order terms ($n>1$). Measuring the coupling strength of a first-order modulation sideband ($\omega_L=\omega\pm\wrf$) and the carrier transition ($\omega_L=\omega$) the modulation index can be determined through
\begin{equation}\label{eq:rfmodindex}
    \frac{\Omega_{\pm1}}{\Omega_0}=\frac{J_1(\beta)}{J_0(\beta)}\approx \frac{\beta}{2}=\frac{\vec{k} \vec{v_0}}{2\wrf}.
\end{equation} 
We can use this relation to calculate the relative frequency shift caused by micromotion \cite{keller_precise_2015}
\begin{equation}
     \left\langle\frac{\Delta \nu}{\nu} \right\rangle = -\frac{1}{2 c^2}\left\langle v^2 \right\rangle = -\frac{\wrf^2}{4 k^2 c^2}\beta^2=-\frac{\wrf^2}{ k^2 c^2}\left(\frac{\Omega_{\pm1}}{\Omega_0}\right)^2.
     \label{eq:MM_2nd_order_shift}
\end{equation}

The equation of motion of a single ion is given by \cite{keller_precise_2015}
\begin{equation}
    m\dot{\vec{v}}= Q \Erf \cos(\wrf t)
\end{equation}
where $Q$ is the electric charge and $\dot{\vec{v}}$ represents the time derivative of the velocity (acceleration). Through integration and comparison to Eq.~\ref{eq:rfmodindex}, we obtain $\Erf=\beta m\wrf^2/kQ$. Using this result, we can rewrite Eq.~\eqref{eq:MM_2nd_order_shift} to express the EMM-induced TDS as a function of the excess rf electric field \Erf \cite{keller_precise_2015}
\begin{equation}
    \left\langle\frac{\Delta \nu}{\nu} \right\rangle = -\frac{1}{2 c^2}\left\langle v^2 \right\rangle =-\left(\frac{v_0}{2c}\right)^2= -\left(\frac{Q}{2c \wrf m }\Erf\right)^2 
\end{equation}
Following \cite{kramer_aluminum_2023}, we can translate the frequency shift of \Ca to \Al assuming the electric field \Erf is the same for both ions and get:
\begin{equation}
    \left\langle\frac{\Delta \nu}{\nu} \right\rangle_\mathrm{Al} =  \left\langle\frac{\Delta \nu}{\nu} \right\rangle_\mathrm{Ca} \frac{m_\mathrm{Ca}^2}{m_\mathrm{Al}^2}.
\end{equation}
Here, the index clarifies whether the shift is associated with \Al or \Ca.  

Measuring the sideband modulation using one laser captures the projection of the wavevector $\vec{k}$ onto the excess micromotion $\vec{v}_0$ of the ion (Eq.~\eqref{eq:Modulationindex_EMM}). Hence, to determine $\vec{v_0}$, one must assess three linear independent directions. In Ref.~\cite{kramer_aluminum_2023} the micromotion of a single \Ca ion was measured on the $S_{1/2} \leftrightarrow D_{5/2}$ transition for different ion positions by changing the voltage of the micromotion compensation electrodes. The point with the lowest modulation index for all laser directions is the position with the lowest micromotion. Ref.~\cite{kramer_aluminum_2023} achieves modulation indices of $\beta\lesssim0.01$ with a total relative shift of $-1.70(25)\times10^{-18}$, assuming a symmetric positioning of the \Al and \Ca around the minimum in axial direction. 

\begin{figure}[htp]
    \centering
    \includegraphics[width=0.6\linewidth]{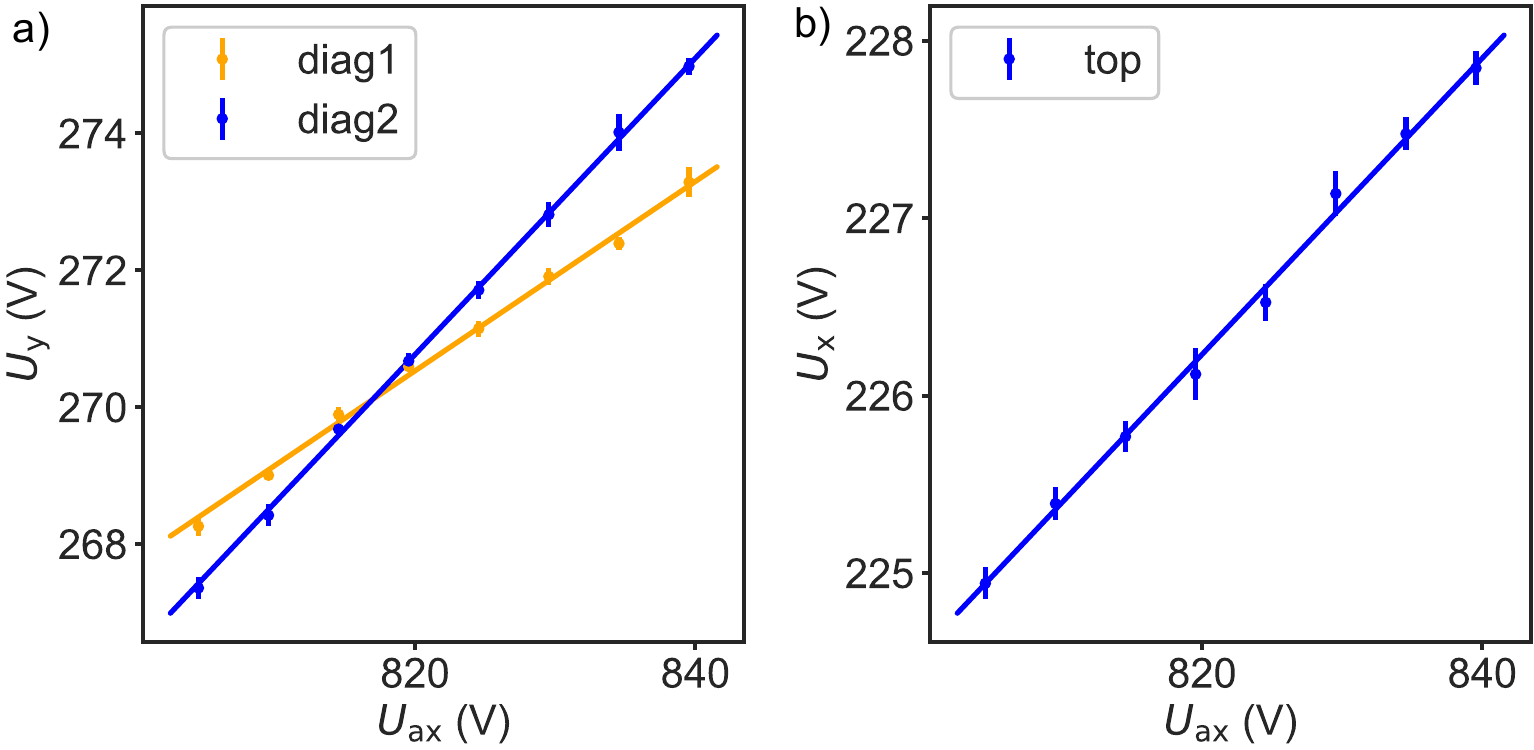}
    \caption[Single \Ca compensation voltage for different axial voltages]{Single \Ca compensation voltage for different axial voltages. The compensation voltage minima depend on the axial trapping voltage. a) For the $y$ direction the crossing point marks the spot of the smallest axial micromotion and the typical operation point. Diagonal beam~1 scales with $0.138(5)U_{ax}+157(4)$ and diagonal beam~2 scales with $0.217(2)U_{ax}+93.1(1.9)$. b) The compensation voltage in $x$ direction is measured using the top beam and scales with $0.0835(16)U_{ax}+157.7(1.4)$. }
    \label{fig:single_Ca_MM_compensation_point}
\end{figure}

We remeasured the compensation voltages of the trap by examining the radial micromotion minimum at different axial voltages. This is illustrated in Fig.~\ref{fig:single_Ca_MM_compensation_point}, where a linear dependence between the compensation voltages to the axial voltage is observed, indicating that the axial potential pushes the ion into the rf potential. 

\begin{figure}[htp]
    \centering
    \includegraphics[width=0.8\linewidth]{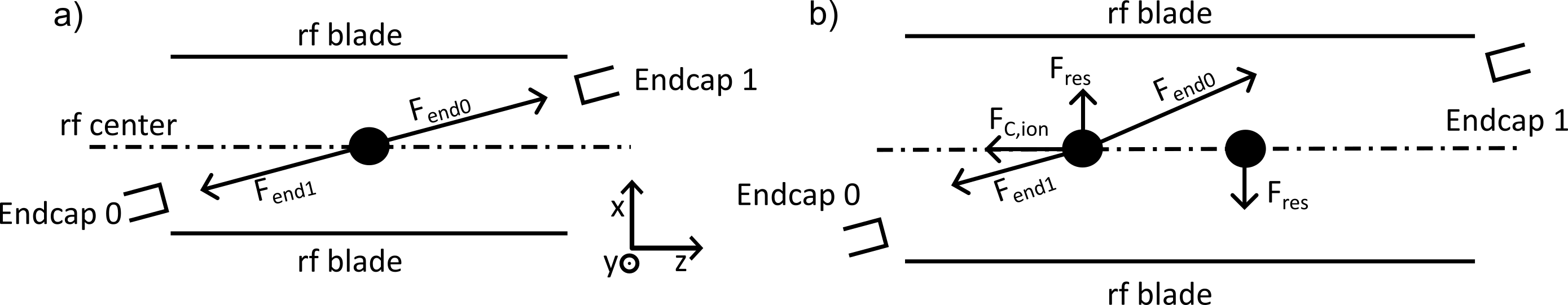}
    \caption[Micromotion change when going from one ion to two-ions]{Micromotion change when dc and rf quadrupole fields are not aligned. a) If a single ion is shifted in position towards one of the endcaps by changing the axial confinement provided by the dc potential applied to them, it moves out of the rf center (the ion moves along the directions of the arrows). This can be compensated by applying a compensation voltage along the $x-y$ direction to shift the ion back into the rf center. b) For a two-ion crystal, the Coulomb interaction will lead to a tilt of the crystal.}
    \label{fig:MM_two_ion_explain}
\end{figure}

For a single ion, a compensated point can be found by changing the two compensation electrodes. This changes for an ion string, as illustrated in Fig.~\ref{fig:MM_two_ion_explain}. It shows that the Coulomb force of the ion-ion interaction introduces an effective force, which is not aligned with the dc quadrupole field of the endcaps. This will move both ions away from each other and out of the rf center, thus tilting the crystal. Since both ions are moving in opposite directions, this shift cannot be compensated by two compensation electrodes producing a cylindrically symmetric field aligned with the axial direction of the trap. However, in a trap with multiple segments it might be possible to adjust the axial dc confinement so that it aligns with the rf confinement.

\begin{figure}[htp]
    \centering
    \includegraphics[width=0.6\linewidth]{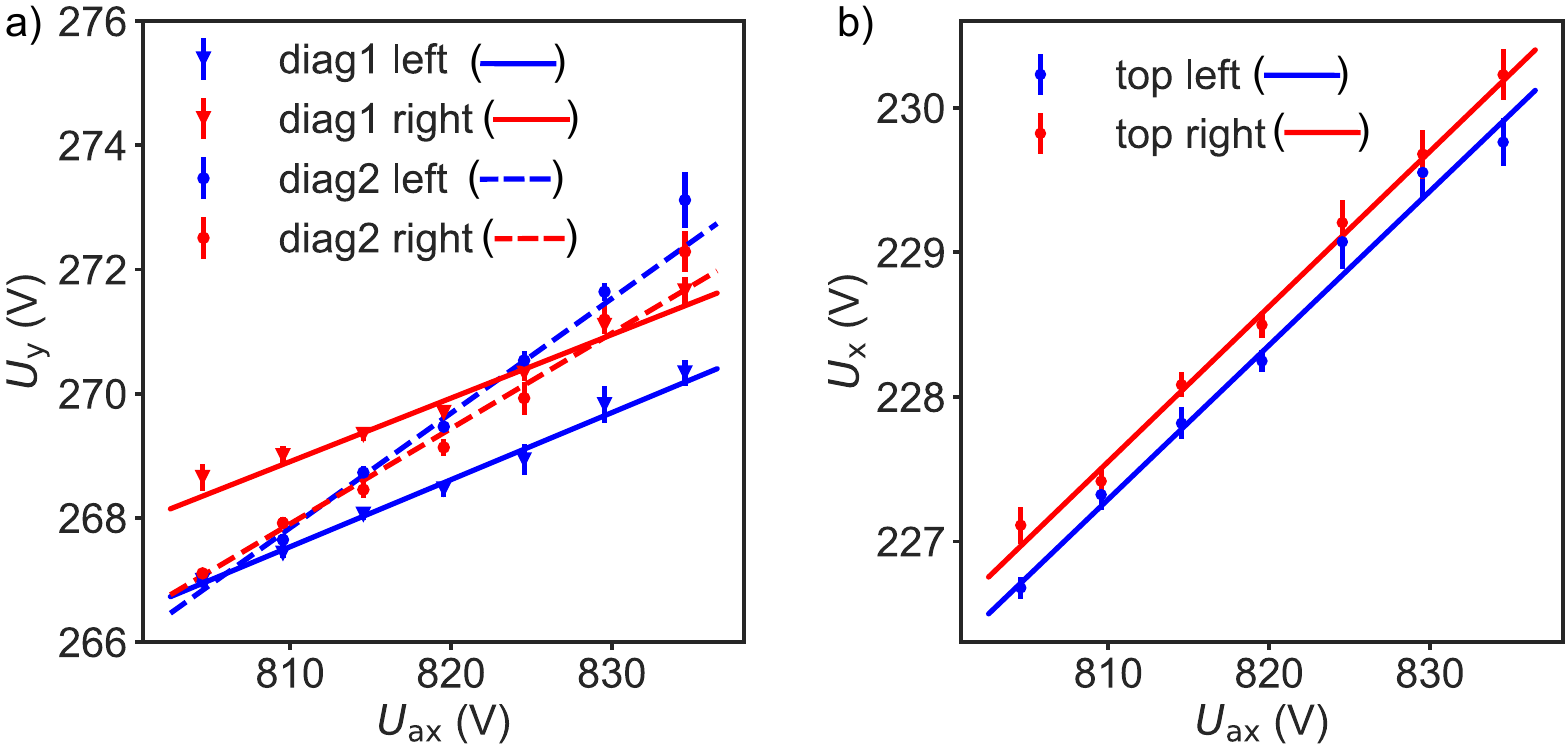}
    \caption[\Ca/\Al crystal compensation voltage for different axial voltages]{\Ca/\Al crystal compensation voltage for different axial voltages. 
    a) The first diagonal direction has two different distinct compensation values depending on the calcium position "left" or "right". This difference is not visible for the second diagonal direction. b) For the top direction the difference between the two directions is small.}
    \label{fig:CaAl_MM_compensation_point}
\end{figure}

We observe this effect experimentally by repeating the single-ion scan shown in Fig.~\ref{fig:single_Ca_MM_compensation_point} with a two-ion crystal, as shown in Fig.~\ref{fig:CaAl_MM_compensation_point}.  For the "diag1" beam, we observe that depending on the position of \Ca in the crystal the compensation voltage jumps between two values, whereas the other "top" beam direction only shows a small change in compensation voltage.

\begin{figure}[htp]
    \centering
    \includegraphics[width=0.6\linewidth]{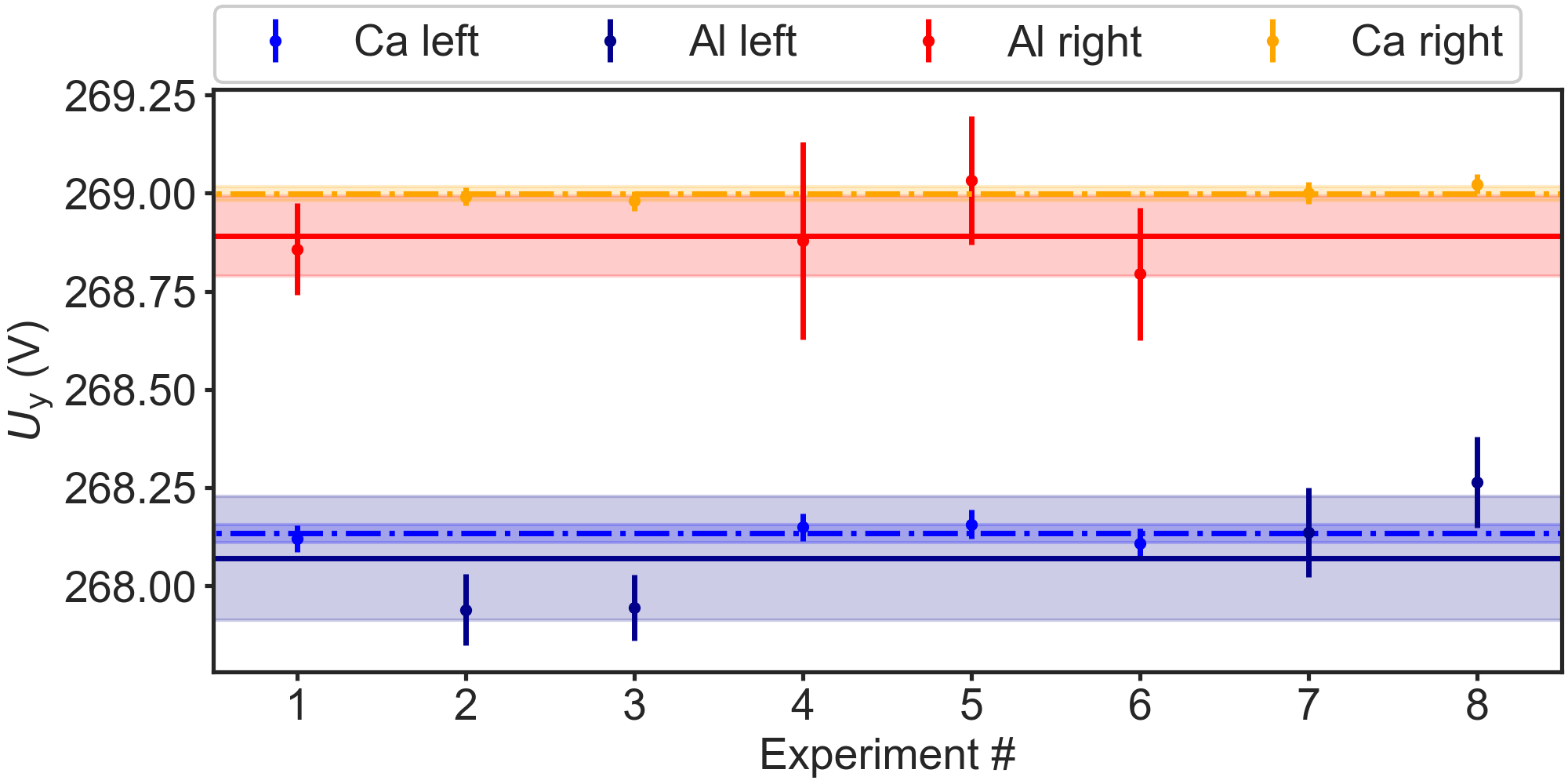}
    \caption[Compensation voltage $U_y$ for \Ca and \Al measured independently]{Comparison of the compensation voltage $U_y$ measured on \Al and \Ca for different positions. The error of the \Al measurement is larger due to its reduced sensitivity. This shows that the compensation points are not distinguishable. }
    \label{fig:CaAl_MM_compensation_point_independent}
\end{figure}

Since our observations have only been on \Ca, we have confirmed these along the $y$ direction ($U_y$). For this, we drove micromotion sidebands on the \ssztpo transition using a radial \SI{267}{\nano\meter} logic beam ($45^\circ$ angle to diag beams) over a large $U_y$ voltage range to obtain the point with the lowest modulation and therefore the lowest EMM shift for one direction. 
We scan the \SI{267}{\nano\meter} logic sideband transition and the \SI{729}{\nano\meter} "diag1" sideband transition successively. After all scans, we also scan the "diag2" transition to get a compensated point in radial direction. The results are shown in Fig.~\ref{fig:CaAl_MM_compensation_point_independent}. From the data, we see that the compensation point is the same for \Ca and \Al if their position is the same. Here we ignored errors which are caused by beam pointing.

The dependence of EMM on the crystal order has some repercussions on the measurement and automatization of the lock. One solution is to minimize micromotion for one position with \Ca, followed by a forced swap of the ions. Then the \Al would be at a compensated position. However, since no reliable method of swapping ions has been implemented yet, we adopted a different solution. We minimize the micromotion on \Ca along the "top" and "diag2" direction, which are nearly unaffected by ion swapping (see Fig.~\ref{fig:CaAl_MM_compensation_point}), thus keeping the micromotion small. The compensation point achieved by this method is not the absolute micromotion minimum, but it facilitates a stable and low micromotion over the duration of the lock. 

\begin{figure}[htp]
    \centering
    \includegraphics[width=0.6\linewidth]{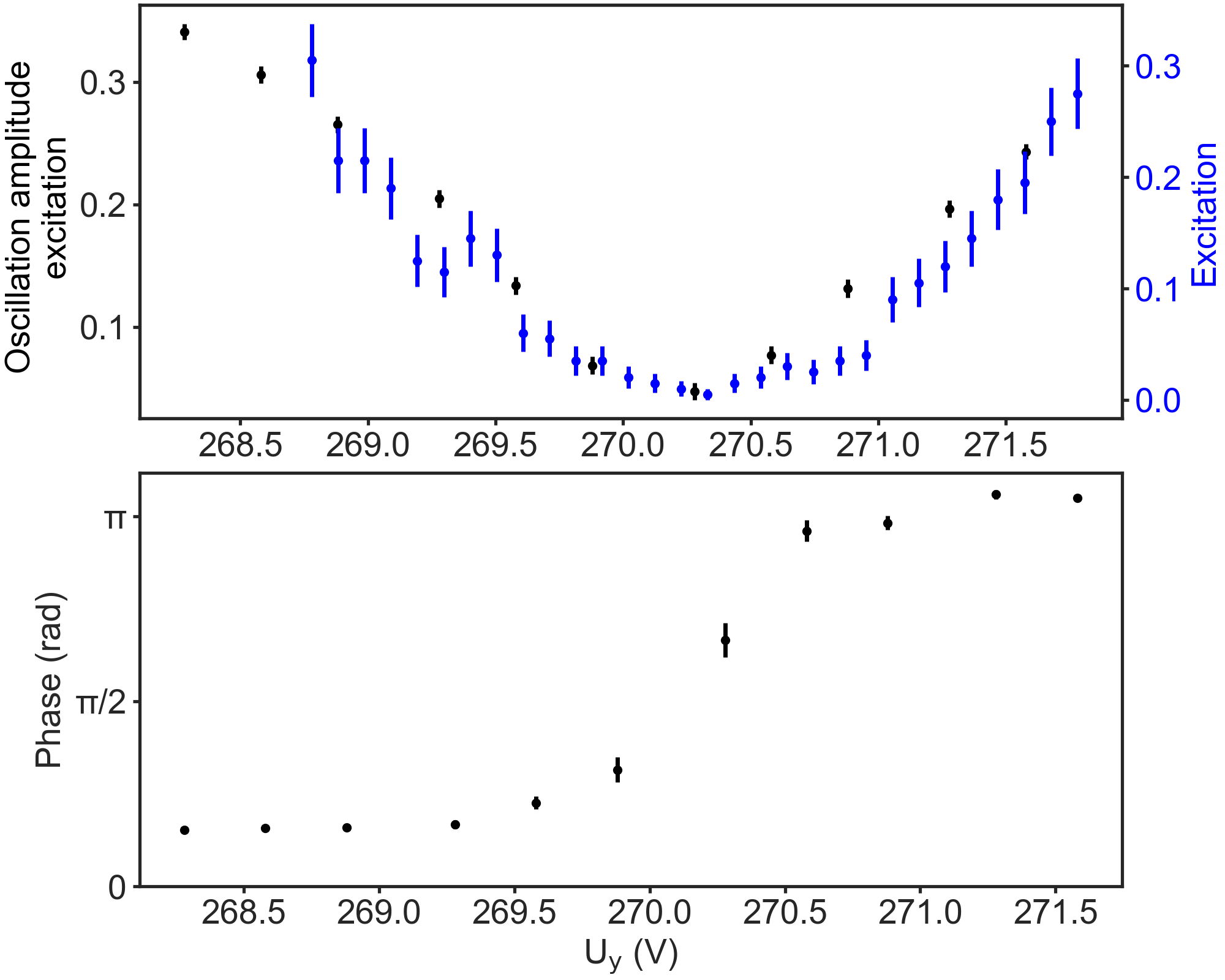}
    \caption[Comparison PMSS and micromotion sideband]{ Comparison of the sideband micromotion method (blue) with the PMSS method (black). Both show the same minima, but the PMSS method additionally adds phase sensitivity.}
    \label{fig:PMSS_comparison}
\end{figure}

For the measurement and compensation of EMM during clock operation we use the phase modulated sideband spectroscopy (PMSS) method \cite{arnold_enhanced_2024} instead of the sideband method used in Ref.~\cite{kramer_aluminum_2023}.  The sideband method and the PMSS method rely on driving the atomic micromotion sideband, and both methods result in the same minimum (see Fig.~\ref{fig:PMSS_comparison}). Compared to the sideband method \cite{keller_precise_2015}, PMSS is phase sensitive to the trap drive field, which allows to use a two-point sampling lock. 
The reason for the phase sensitivity is the phase modulation $\beta_L$ of the electric field. This modulation adds another term to Eq.~\eqref{eq:electric_field_MM_modulation}  \cite{arnold_enhanced_2024}
\begin{equation}
    E \propto \exp\left(i[\omega_L t+\beta_L \cos(\theta_L)+\beta_{mm} \cos(\theta_{mm})]\right),
\end{equation}
where $\theta_L=\wrf t+\phi$ is the phase of the laser modulation at time t and $\beta_{mm}$ is the modulation caused by EMM at a phase of $\theta_{mm}=\wrf t$. Here, we neglected terms from intrinsic micromotion of the ion. The coupling strength between the modulated laser and an atom under micromotion can be written as \cite{arnold_enhanced_2024}:
\begin{equation}
    \Omega=\frac{\Omega_c}{2}(i(\beta_Le^{i\phi}+\beta_{mm}))
    \label{eq:MM_PMSS_phase}
\end{equation}
where $\Omega_c$ is a scaling factor for the coupling strength of the laser. The effect of the phase modulation can be seen via Eq.~\eqref{eq:MM_PMSS_phase}. The change of laser modulation phase $\phi$ will lead to a different coupling between $\beta_L$ and $\beta_{mm}$, thereby enhancing or reducing the coupling. EMM can originate from a displacement from the center of the trap and a phase shift between the RF blades, resulting in modulation indices $\beta_m, \beta_p$, respectively. Since both components are $\pi/2$ out of phase, the overall micromotion can be written as $\beta_{mm}=\beta_m+i\beta_p$ \cite{arnold_enhanced_2024}. By changing the laser modulation phase one can thus distinguish between the two origins of EMM.

\begin{figure}[htp]
    \centering
    \includegraphics[width=0.6\linewidth]{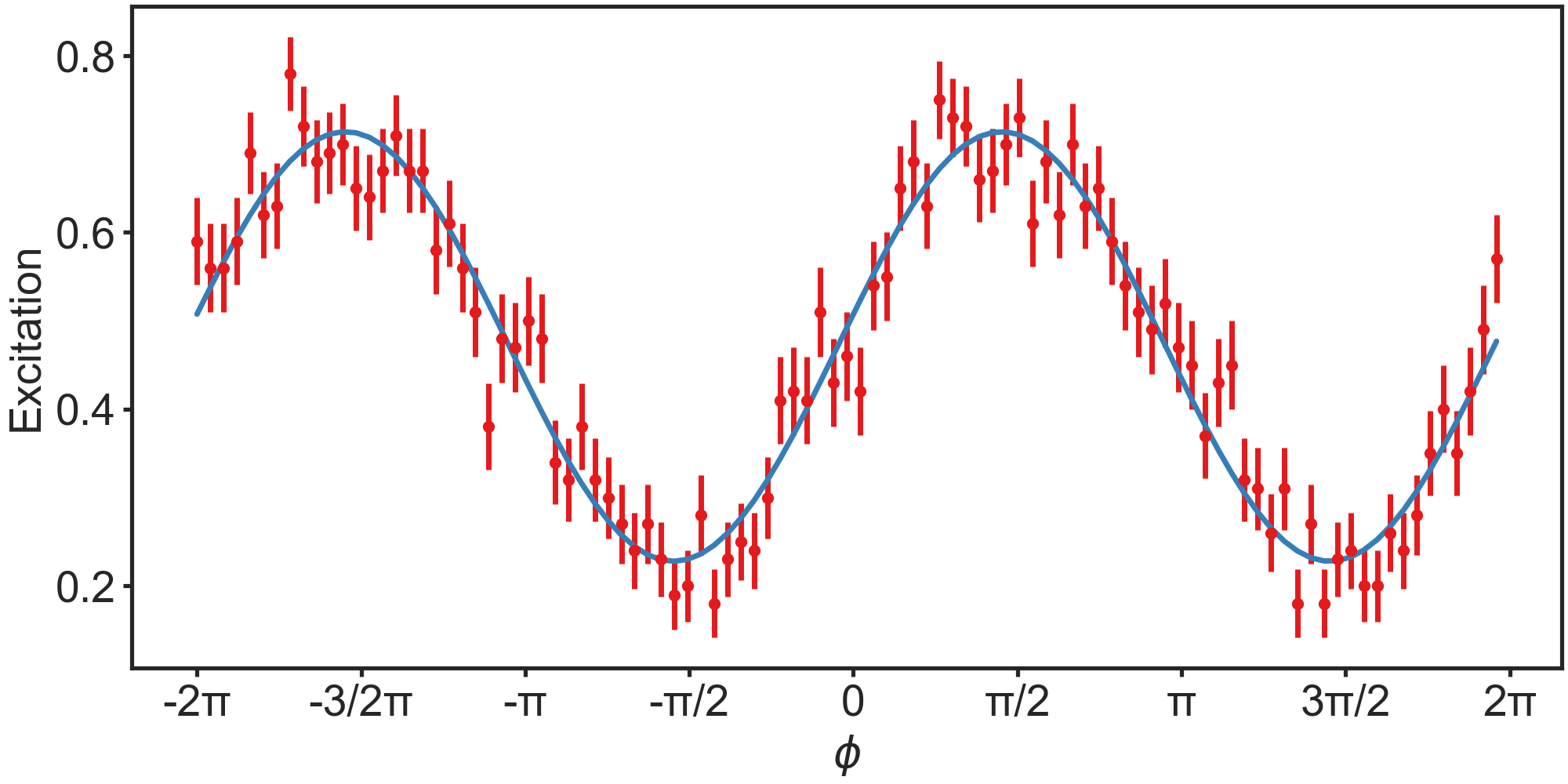}
    \caption[PMSS measurement]{PMSS measurement example. Resulting ion excitation when scanning the rf phase modulation of the laser. Here the micromotion was deliberately offset to create a large modulation.}
    \label{fig:PMSS_measurement}
\end{figure}

In the experimental sequence the laser operates at a frequency $\omega_L=\omega_0-\wrf$. Therefore, the carrier of the laser is resonant with the micromotion sideband of the atom, and the phase modulation of the laser will excite an atomic carrier transition. The probability $P$ to excite the \ddfh  state is:
\begin{equation}
    P(\Theta,R,\phi)=\frac{1}{2}[1+\cos(\Theta\sqrt{1+2R\cos(\phi+\phi_0)+R^2})],
    \label{eq:PMSS_excitation}
\end{equation}
where $\Theta=\Omega_c t \beta_L/2$ and $R=\beta_{mm}/\beta_L$. $\phi_0$ is the phase lag introduced by path length differences between the trap drive and the electro-optic modulator (EOM) modulation. The angle $\Theta$ can be calibrated by scanning the phase $\phi$ and duration of the pulse. The resulting oscillation can be employed to set $\Theta\approx\pi/2$ at a point insensitive to EMM, which provides highest sensitivity. Once $\Theta$ is set and $R$ is calibrated, Eq.~\eqref{eq:PMSS_excitation} can be used to measure the amount of micromotion. A typical measurement involving uncompensated micromotion is shown in Fig.~\ref{fig:PMSS_measurement}

\begin{figure}[htp]
    \centering
    \includegraphics[width=0.60\linewidth]{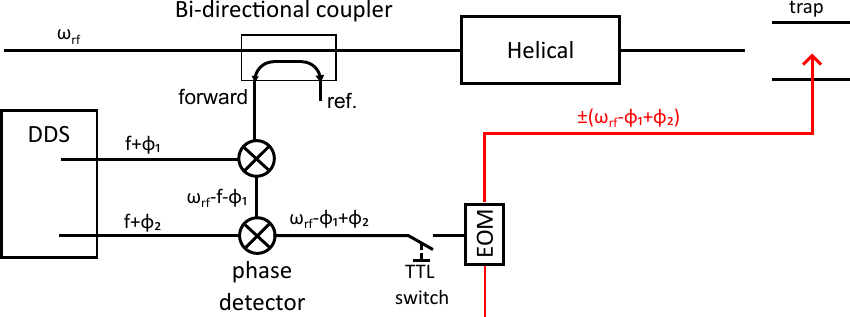}
    \caption[Setup for PMSS]{Setup used for the phase modulation of the \SI{729}{\nano\meter} laser. The red lines imply laser paths, while the black paths are electronic signals. We use the forward port of a bidirectional coupler to modulate the trap drive frequency on the laser. By mixing the signal up an down with a common DDS source we can adjust the phase of the signal without changing its frequency.}
    \label{fig:PMSS_setup}
\end{figure}

The phase modulation of the light is achieved via an EOM. By changing the phase of one DDS source, we can scan the phase of $\wrf$ (see also Fig.~\ref{fig:PMSS_setup}). 

For the lock, we calibrate the phase shift at the ion and set the lock such that we are probing at the minimum ($\phi_1+\phi_0=0$) and maximum ($\phi_2+\phi_0=\pi$) of the oscillation. The excitation difference is then fed into a servo loop for minimization. We can also use the excitation probability difference to measure the excess micromotion shift. The difference is given by

\begin{equation}
    P(\Theta,R,\phi_2)-P(\Theta,R,\phi_1)= \frac{1}{2}\cos(\Theta(1-R))-\frac{1}{2}\cos(\Theta(1+R)).
    \label{eq:MM_error_gen}
\end{equation}

We can rearrange this equation to:
\begin{equation}
    R = \frac{1}{\Theta}\arcsin\left(\frac{P(\Theta,R,\phi_2)-P(\Theta,R,\phi_1)}{\sin(\Theta)}\right).
\end{equation}
From this we can obtain $R=\beta_{mm}/\beta_L$ and determine the modulation index for a known $\beta_L$. For the calibration of $\beta_L$, we measure the carrier Rabi frequency of the $S_{1/2}(m=1/2)\leftrightarrow D_{5/2}(m=1/2)$ transition. For this the laser is detuned by \SI{26}{\mega\hertz} (not on the rf sideband) and the EOM is set to \SI{26}{\mega\hertz} to suppress excitation from other transitions. Therefore, we can drive the carrier with the modulation sideband. After this, we set the laser on resonance of the transition and measure the coupling strength of the carrier (while the EOM is still running at $26\,$MHz). The ratio of the two coupling strengths $\Omega_{sb}/\Omega_0=\beta_L/2$ is a measurement of the modulation strength of the laser.

The modulation index of the 2025 measurement is shown in Fig.~\ref{fig:MM_plot} in the main text. We lock the micromotion compensation voltages along the "diag2" and "top" directions using \Ca. 
This compensates micromotion for \Ca in the radial direction, but leaves residual micromotion on \Al (see above).

To infer the modulation index at the \Al position we must transfer it from the \Ca modulation index measurement. This is particularly important for the "diag1" direction, where we have non-zero micromotion for both positions. For this purpose, we calculate the mean difference of the modulation index between the two positions and add this on the measured \Ca value. Consequently, we expect to obtain the modulation index at the \Al position, which can then be used for calculating the frequency shift. This is illustrated in Fig.~\ref{fig:MM_plot} of the main text. 

For a full evaluation of the micromotion in every lock there is another "service" measurement which measures all three micromotion directions using the PMSS technique as described in the main part of the manuscript.

The three modulation directions $\beta_i$, are normally distributed (see also Fig.~\ref{fig:MM_plot} in the main text) and the frequency shift due to micromotion has the form:
\begin{equation}
    \frac{\Delta\nu}{\nu}=-\left(\frac{\wrf}{2 c_0 k}\right)^2 \sum_i^3 \beta_i^2.
\end{equation}
We can compare this with the random variable of a non-central chi-squared distribution $X$, where $X=\sum_{i}^{df} \beta_i^2$, where $\beta_i$ is normally distributed with a mean $\mu_i$, unit variance and $df$ degrees of freedom. Thus, the motional shift must follow a scaled non-central chi-squared distribution \cite{arnold_enhanced_2024}. 

Here, we need to scale the distribution function such that it also allows for a standard deviation $\sigma$. In Fig.~\ref{fig:MM_plot} of the main text we fit the measurement values to the probability density function with the form \cite{van_aubel_analytic_2003}:

\begin{equation}
    f_X(x,k,\lambda,\sigma)=\frac{1}{2\sigma^2}e^{-(x+\lambda)/(2\sigma^2)}\left(\frac{x}{\lambda}\right)^{(df-2)/4}I_{(df-2)/2}(\sqrt{\lambda x}/\sigma^2),
\end{equation}

where $\lambda$ is the non-locality parameter, $df$ are the degrees of freedom and $I_\gamma$ is the modified Bessel function of first-order of degree $\gamma$. From the fit, we can extract the statistical shift and uncertainty expected from the EMM, which depends on the position. For the \Al on the left position we expect a shift of \num{3.5(2.3)e-19}. For \Al on the right position we expect \num{2.9(2.1)e-19}. This indicates that the position has only a small influence on the overall frequency shift due to micromotion. For this evaluation, we excluded data points from the fit that exceed \SI{>5}{\milli\hertz} ($>5\sigma$), corresponding to faulty measurements, which would otherwise dominate the statistical evaluation. Overall, there was just one value excluded for over \num{1000} measurements in the April \num{2025} measurement.

Apart from the statistical uncertainty, as previously described, we also need to consider systematic uncertainties. As outlined above, the micromotion evaluation depends on three linearly independent measurement directions, leading to an additional uncertainty from angle errors in the laser directions. Using a Monte Carlo simulation, we simulate the frequency shift for different angles of each laser within an interval of $\pm$\SI{5}{\degree}. The modulation parameters are set to a fixed set of $\beta=[\beta_{diag_1},\beta_{diag_2}, \beta_{top}]\approx[0.010,0.001,0.001]$, which are close to the experimental results. The simulation shows an uncertainty in the frequency shift of \SI{15}{\percent} due to changing angles, like those reported by Ref.~\cite{kramer_aluminum_2023}. We therefore attribute an additional systematic fractional frequency uncertainty of \num{5e-20} from angle errors to both position values.

An additional systematic uncertainty arises from the position-dependent micromotion on \Al for the directions where we servo the \Ca micromotion to zero and the compensation voltages thus depend on the crystal ordering. For the $y$ direction, we compensate on a direction ($\beta_{\mathrm{diag}_2}$), which does not depend on the ion position or only has a very small position dependence. This is not the case for the $x$ direction. In Fig.~\ref{fig:CaAl_MM_compensation_point}, we see a small offset, which leads to an underestimation of the shift when locking it to zero micromotion. The electric field change between the two positions in $x$ is \SI{<5}{\volt\per\meter}. We also attribute a smaller effect in the $y$ "diag2" direction of \SI{<2.5}{\volt\per\meter}. The resulting shift difference for $\beta=[\beta_{diag_1},\beta_{diag_2}, \beta_{top}]\approx[0.010,0.0027,0.0043]\,\,(\Erf=[15,4.0,6.5]\,\mathrm{V}/\mathrm{m}$) corresponds to around \num{2.9e-19}.

Another systematic error is caused by cross coupling of Zeeman shifts on the micromotion sideband. At the modulation frequency $\wrf$, we can measure the ac-Zeeman shift resulting from the trap drive. From the measurement in Ref.~\cite{kramer_aluminum_2023}, we know that the modulation of the trap drive's magnetic field is primarily in the y direction (see below). The modulation index due to rf magnetic fields along the quantization field are \cite{gan_oscillating-magnetic-field_2018}:
\begin{equation}
    \beta_m=\frac{(g_Dm_D-g_Sm_S)}{\hbar\wrf}\mu_B B_{z,\mathrm{rf}} \approx 7(2) \times 10^{-4}, 
\end{equation}
where $g_D,g_S$ are the g-factors of the \Ca D and S state, $m_D,m_S$ are the magnetic quantum numbers of the D and S state, respectively, $|B_{z,\mathrm{rf}}|\approx$\SI{3.5(1.1)}{\micro\tesla} is the rf magnetic field along the quantization axis, and $\mu_B$ is the Bohr magneton. The lock will counteract any measured modulation by adjusting the compensation voltage, leading to a compensation between micromotion and rf-induced sidebands and thus non-vanishing micromotion. The modulation corresponds to an electric field of around \SI{1}{\volt\per\meter}, resulting in a relative shift of \num{0.7e-20}, which we don't correct for but take as the uncertainty.

\begin{table}[htp]
    \centering
    \caption[Errorbudget Micromotion]{Different contributions to the error of micromotion.}
    \label{tab:Error_micromotion}
    \begin{tabular}{l|cc}
    Effect & Shift (\SI{}{\milli\hertz}) & Fractional uncertainty \\
    && contribution /$10^{-19}$ \\
    \hline
    Position error   &  0.00(33) & \num{2.9} \\
    Measurement  & 0.39(26) & \num{2.3} \\
    Magnetic field modulation  &  0.00(8) & \num{0.7} \\
    Angle uncertainty    &  0.00(6) & \num{0.5} \\
    \hline
    $\Delta\nu$ &  0.39(43) & \num{3.8} \\
    \end{tabular}
\end{table}

Including all additional uncertainties (listed in Tab.~\ref{tab:Error_micromotion}), we arrive at a final EMM shift of \num{-3.5(3.8)e-19} for the right position of \Al and at \num{-2.9(3.6)e-19} for the left position. For the frequency ratio measurements we corrected for the position-dependent shifts and assumed the larger of the two uncertainties.

\section{Zeeman shift}
\label{sec:Zeeman_shift}
The linear Zeeman shift from the nuclear magnetic moment is cancelled by probing the stretched states on the clock transition as described in the main text.

One of the largest shifts for \Al clocks is the quadratic Zeeman shift which depends on the magnetic field at the position of the ion \cite{brewer_27al+_2019,rosenband_frequency_2008,marshall_high-stability_2025}. Large magnetic fields can be used to reduce effects like line pulling \cite{king_optical_2022}. We will first examine the shift caused by the static magnetic fields and in the second part we will investigate the trap drive-induced ac-Zeeman shift.

\subsection{Quadratic Zeeman shift}
\label{sec:DC_Zeeman_shift}
The Zeeman shift in an external magnetic field $B$ can be written as \cite{von_lindenfels_experimental_2013, spies_excited-state_2025}
\begin{equation}
    h \nu_Z = m_J g \mu_B \langle B\rangle + g_s^{(2)}(m_J) \frac{\mu_B^2 \langle B^2\rangle}{m_e c_0^2}+\mathcal{O}(\langle B^3\rangle)
\end{equation}
where $m_J$ is the magnetic quantum number, $g$ is the $g$-factor of the state. $g_s^{(2)}(m_J)$ is the second-order Zeeman coefficient, which is symmetric in $m_J$ and depends on the state s. For a differential shift between two states $J$ and $J'$, a common notation is that of the $C_2(m_J,m_{J'})$ coefficient, where the Zeeman shift becomes
\begin{equation}\label{eq:zeemanfreq}
    \Delta\nu=\frac{\mu_B \langle B\rangle}{h}(m_{J'}g_{J'}-m_{J}g_{J})+C_2(m_J,m_{J'}) \langle B^2\rangle.
\end{equation}
Here, the time average of the magnetic field $\langle B \rangle$ is used, which can also be written as $\langle B^2 \rangle = B^2_{dc}+\langle B_{ac}^2 \rangle$. The ac contribution to the oscillating field comes from the \SI{50}{\hertz} line noise, its higher harmonics and furthermore the \SI{28}{\mega\hertz} trap drive. These frequencies are faster than the clock interrogation cycle. Therefore, we measure their time average. Since the frequencies are far detuned from the fine structure splitting, we can neglect effects due to frequency dependencies and treat them like the dc contribution. It is important to note that
\begin{equation}
    \langle B^2 \rangle = \langle B_{dc}^2\rangle +\langle2B_{dc}B_{ac}\rangle+\langle B_{ac}^2 \rangle \approx  B_{dc}^2+\langle B_{ac}^2 \rangle,
\end{equation}
where we used that the mean of an oscillating field is zero $\langle B_{ac}\rangle = 0$. For different oscillation frequencies we can write:

\begin{equation}
    \langle B_{ac}^2 \rangle = \langle B_{ac,1}^2\rangle +\langle2B_{ac,1}B_{ac,2}\rangle+\langle B_{ac,2}^2 \rangle \approx  \langle B_{ac,1}^2 \rangle +\langle B_{ac,2}^2 \rangle,
\end{equation}

where the mixing term vanishes, as a term $\sin(f_1t)\sin(f_2t)=0.5(\cos((f_1-f_2)t)-\cos((f_1+f_2)t))$ is averaged out for long probe times if the frequency difference is large. In our case, the frequencies are at \SI{50}{\hertz} and its harmonics, as well as the trap drive frequency of \SI{28}{\mega\hertz}, validating the approximation.

\begin{figure}[htp]
    \centering
    \includegraphics[width=0.6\linewidth]{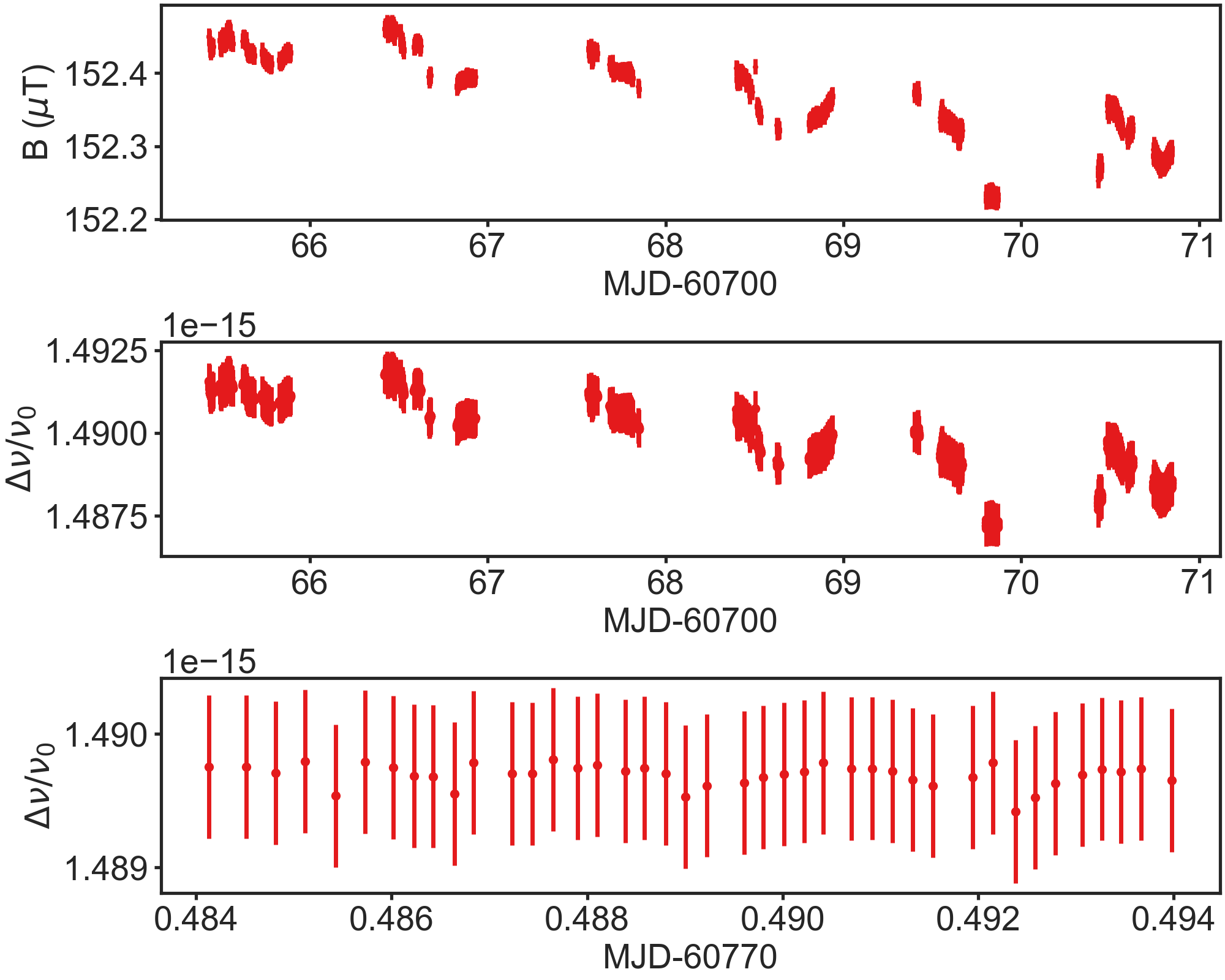}
    \caption[Magnetic field lock data]{Measurement of the magnetic field for the lock in April 2025. The upper plot shows the average magnetic field with error bars. The center plot shows the quadratic Zeeman shift with errors, and the lowest plot shows a zoom of data for one single run at the last day.}
    \label{fig:magnetic_field_lock}
\end{figure}

When running \Al as a clock, one probes the $^1$S$_0(m=+5/2)\leftrightarrow$$^3$P$_0(m=+5/2)$ and the $^1$S$_0(m=-5/2)\leftrightarrow$$^3$P$_0(m=-5/2)$ transition. The average over both transitions according to Eq.~\ref{eq:zeemanfreq} is:
\begin{equation}
    \frac{1}{2} \left(\Delta\nu(m=+5/2) + \Delta\nu(m=-5/2)\right) = C_2\langle B^2\rangle.
\end{equation}
The average of the transition frequencies of both stretched states is free of the first-order Zeeman shift. This leaves the second-order dc-Zeeman shift contribution. The $C_2$ coefficient was determined in Ref.~\cite{brewer_measurements_2019} as $C_2=$\SI{-71.944(24)}{\mega\hertz\per\tesla\squared} for the employed transitions. Theoretical calculations show that the magnetic dipole-allowed coupling between the $^3$P$_0\leftrightarrow$ $^3$P$_1$ transition has the largest contribution to the $C_2$ coefficient \cite{brewer_measurements_2019}. 
To determine the static magnetic field, we use the difference between the two measured clock transitions, as this provides us with the static magnetic field $B_{dc}$
\begin{equation}
    B_{dc} = \frac{h(\Delta\nu(m=+5/2)-\Delta\nu(m=-5/2))}{5(g_P-g_S)\mu_B}
\end{equation}
The difference of the g-factors was determined in Ref.~\cite{rosenband_observation_2007} to be $g_P-g_S=$\num{-1.18437(8)e-3}. The continuously measured magnetic field during the clock run in April 2025 is illustrated in Fig.~\ref{fig:magnetic_field_lock}. This also allows us to correct the frequency shift because of magnetic fields after each clock cycle. The daily fluctuations are mostly \SI{<0.1}{\micro\tesla}.

The uncertainty limits for one single measurement are shown in Tab.~\ref{tab:Err_buget_magnetic_field}. One can see that the dominating contribution is from the $C_2$ coefficient, while the next important uncertainty is the $g$-factor difference. 

\begin{table*}[htp]
    \centering
    \caption[Error contributions quadratic Zeeman shift]{Error contributions from the different constants and the magnetic field stability of \SI{4}{\nano\tesla} extracted from the first Allan deviation point of the first day in Fig.~\ref{fig:magnetic_field_lock}. The main limitation for the uncertainty is the $C_2$ coefficient followed by the $g$-factor difference in \Al. The actual magnetic field instability only plays a minor role. The given shift and magnetic field represent a typical measurement and are corrected for continuously during clock operation.}
    \label{tab:Err_buget_magnetic_field}
     \begin{tabular}{l|c S[table-format=5.3e2]}
        Parameter & Value & {Fractional uncertainty} \\
         &  & {contribution /\num{e-19}} \\
         \hline
         $C_2$ (\SI{}{\mega\hertz\per\tesla\squared})& \num{-71.944(24)} & 4.9 \\
         $g_P-g_S$ & \num{-1.18437(8)e-3} & 2.0 \\
         B instability (\SI{}{\micro\tesla}) & \num{152.355(4)} & 0.7 \\
         $\mu_B$ (\SI{}{\joule\per\tesla}) & \num{9.2740100657(29)e{-5}} & 9.3e-6 \\
         \hline
         Total (\SI{}{\hertz}) & \num{1.6700(6)} & 5.4
    \end{tabular}
\end{table*}

In Ref.~\cite{kramer_aluminum_2023}, it was discussed that the magnetic field can also be measured via \Ca instead of \Al. This approach has the advantage that absolute magnetic field could be measured with lower uncertainty, thanks to the lower uncertainty of the \Ca $g$-factor.  While this would reduce the overall uncertainty due to the magnetic field, it introduces a systematic effect due to the ion position dependence of the magnetic field arising from gradients. A more convenient way would be to measure the $g$-factor of \Al with the one of \Ca \cite{rehmert_lande_2026} This would lead to a lower uncertainty without adding interleaved measurements, which would be beneficial for the duty cycle of the clock.

Apart from the static field, we also need to estimate frequency shifts due to \SI{50}{\hertz} magnetic field noise. The active compensation for magnetic field noise only suppresses a certain amount of the \SI{50}{\hertz} field. Ref.~\cite{kramer_aluminum_2023} measured a \SI{50}{\hertz} magnetic field of $\sqrt{\langle B_{50\,\mathrm{Hz}}^2 \rangle}=$\SI{27.6(8)}{\nano\tesla} using the contrast loss on \Ca. This would cause an additional shift of \num{4.9(3)e-23}, which we neglect.

\subsection{Trap-induced ac-Zeeman shift} 
\label{sec:AC_zeeman shift}

In the previous section, the focus was on slow and static magnetic fields. Here, the influence of the oscillating rf magnetic field of the trap drive is evaluated. 
A Paul trap is very similar to a driven electric LC circuit, which means that the trap capacitance charges up and discharges periodically with the trap drive frequency. Hence, there will be oscillating currents which create magnetic fields. These fields can cause an ac-Zeeman shift on the ion and must be characterized. 

The trap-induced Zeeman shift was already measured and analyzed in Ref.~\cite{kramer_aluminum_2023}. Here, only the method is summarized and the results will be scaled to the current experimental settings.
The magnetic field of the trap was measured using the method described in Ref.~\cite{gan_oscillating-magnetic-field_2018} which couples the Zeeman states of the atom using the rf magnetic field of the trap drive. When the splitting between two Zeeman states is close to the trap drive frequency, one will observe an Autler-Townes splitting. The frequency difference of the Autler-Townes splitting corresponds at resonance to the ac magnetic field perpendicular to the quantization axis. Measuring the magnetic field with different magnetic field directions (quantization axes) for different trap drive powers and ion positions in the trap allows us to obtain the ac magnetic field caused by the trap drive. The resulting effective magnetic field for a single \Ca radial secular trapping frequency of $\omega_\mathrm{rad}=\SI{1.92}{\mega\hertz}$ is:
\begin{align*}
    B =& 20.11(8)\,\mu\mathrm{T} \\
    \phi_B =& 84.8(3.0)^{\circ} \\
    \vartheta_B =& 98.9(2.5)^{\circ} \\
    r( \phi_B,\vartheta_B) =& 0.429.
\end{align*}
Here, $\phi_B,\vartheta_B, r( \phi_B,\vartheta_B)$ follow the definition given in Ref.~\cite{kramer_aluminum_2023}, representing the angles of a spherical coordinate system with the z axis aligned along the trap axis, while $ r( \phi_B,\vartheta_B)$ is the correlation coefficient between both angles. The $x$ axis spans together with the $z$ axis a plane which is parallel to the optical table. The $y$ vector is pointing upwards, away from the optical table. For the given coordinates, the magnetic field mainly points along y.  

We now need to scale our results to the parameters given in Ref.~\cite{kramer_aluminum_2023}, which results in a magnetic field of $B_{\perp,\mathrm{rad}}=$\SI{22.557(4)}{\micro\tesla} for a radial trapping frequency of $\omega_\mathrm{rad}=\SI{2.041}{\mega\hertz}$ (mean value of both modes). To calculate the total $B$-field we have to account for the angle between the measured and actual field direction by using:
\begin{equation}
    B = \frac{B_{\perp,\mathrm{rad}}}{\sqrt{1-[\vec{e}(\phi_B,\vartheta_B)\cdot\vec{n}(\phi_\mathrm{rad},\vartheta_\mathrm{rad})]^2}}
\end{equation}
with $\vartheta_\mathrm{rad}=90(1)^\circ$ and $\phi_\mathrm{rad}=0(1)^\circ$ the unit direction $\vec{n}$ of the radial magnetic field measurement and $\vec{e}$ the unit vector of the magnetic field. This changes the magnetic field to $B=$\SI{22.65(11)}{\micro\tesla} limited by the uncertainty of the angles. From Ref.~\cite{kramer_aluminum_2023}, we know that due to trap drive power changes the magnetic field fluctuates. The fluctuations are on a level $u(B_{ac})=$\SI{0.17}{\micro\tesla} at \SI{21.265}{\micro\tesla} field.  If we assume a linear dependence, we can scale these power fluctuations to our value and get a total field of $B=$\SI{22.65(22)}{\micro\tesla}. 
 The magnetic field is now treated like a static magnetic field, and we get a relative frequency shift of
\begin{equation}
    \frac{\Delta\nu}{\nu} = C_2 \langle B^2_{ac,trap}\rangle = 16.46(31)\times 10^{-18},
\end{equation}
where the main limitation to this shift still remains the trap drive fluctuations, as shown in Tab.~\ref{tab:AC_zeeman_uncertainties}.

\begin{table*}[htp]
    \centering
    \caption[Uncertainty contributions ac-Zeeman shift due to the trap dirve]{Uncertainty contributions to the ac-Zeeman shift. For the fluctuation we only give the error. }
    \label{tab:AC_zeeman_uncertainties}
    \begin{tabular}{l|c S[table-format=5.3e2]}
        Parameter & Value & {Fractional uncertainty} \\
         &  & {contribution /\num{e-19}} \\
        \hline
         $\Delta$B$_{\mathrm{ac}}$ (\SI{}{\micro\tesla}) & 0.00(17) & 2.6\\
         $\vartheta_\mathrm{B}$ ($^\circ$) & 98.9(2.5) &  1.3\\
         $\phi_\mathrm{B}$ ($^\circ$) & 84.8(3) &  0.8 \\
         $\phi_\mathrm{rad}$ ($^\circ$) & 0(1) &  0.5 \\
         $\vartheta_\mathrm{rad}$ ($^\circ$) & 90(1) &  0.08 \\
         power extrapolation (dBm) & 0.964(1) &  0.06 \\
         $C_2$ (\SI{}{\mega\hertz\per\tesla\squared}) & 71.944(24) & 0.055 \\
         \hline
         frequency shift (mHz) & 18.45(35)& 3.1
    \end{tabular}
\end{table*}

A direct measurement of this shift during the clock operation is not possible with the method proposed in Ref.~\cite{gan_oscillating-magnetic-field_2018}, as it requires large magnetic fields to split the ground-state of \Ca. An alternative would be the method used by Ref.~\cite{joshi_characterization_2024} to determine the coupling strength of a $S_{1/2}\leftrightarrow D_{5/2}$ transition with $|\Delta m|=3$ using an optical and rf photon. The required high magnetic field coherence and stability to resolve this transition is not achievable with our setup, as the less magnetic field sensitive $|\Delta m|=0$ transition already has \SI{<1}{\milli\second} coherence time. 

In the future, we could employ indirect measurements, like repeated measurements of the radial trap drive or monitoring of the rf field in the trap or helical resonator, to ensure similar rf magnetic field strengths during the clock interrogation.

\section{Light shift}
\label{sec:light_shifts}

In this section, we explore the influence of electric fields on ions, including the effect of the cooling lasers mentioned earlier. The effect of the cooling laser light shift presented here has already been published in \cite{dawel_high-stability_2025_2}, and those parts are based on that publication.

\subsection{Cooling light}
\label{sec:light_shifts_cooling}
When \Ca cooling light is applied during the interrogation of \Al, a frequency shift can be observed. Overall, three beams contribute to the cooling light shift: the \SI{866}{\nano\meter} beam, the \SI{397}{\nano\meter} $\pi$ beam, and the \SI{397}{\nano\meter} $\sigma$ beam, as they are utilized for EIT cooling.

Measuring the electric field at the position of the ion is difficult because one needs to determine the correct beam size at the position of the ion, which can deviate from an elliptical Gaussian beam \cite{brewer_27al+_2019}. To circumvent this issue, we measure the electric field of each beam by measuring the frequency shift on \Ca. The polarizability of the optical dipole transitions of calcium is well-known due to experimental measurements \cite{hettrich_measurement_2015} and additional theoretical data \cite{UDportal}. Therefore, the electric field of each laser beam can be determined at the position of the \Ca ion. Since the distance between both ions is \SI{5.2}{\micro\meter} along the z axis, the electric field can be transferred from \Ca to \Al.

The frequency shift on \Al is mainly affected by the scalar shift, as the \Al clock transition is a $J=0\leftrightarrow J'=0$ transition. The hyperfine interaction induces small vector and tensor contributions similar to Sr \cite{shi_polarizabilities_2015_2}, which are in our case negligible. 

When the electric field is known on the \Ca ion, we can use this field for the \Al ion to obtain the frequency shift: 
\begin{equation}
    \Delta\nu_{Al^+} = -\frac{1}{4h}|E_0^2|(\alpha_{S,P-Al^+}-\alpha_{S,S-Al^+}) = -\frac{1}{4h}|E_0^2|\Delta\alpha_{S,Al^+}
    \label{eq:AC_stark_alu}
\end{equation}

with $\alpha_{S,P-Al^+}$ the polarizability of the $^3\mathrm{P}_0$ state, $\alpha_{S,S-Al^+}$ the polarizability of the $^1\mathrm{S}_0$ state and $\Delta\alpha_{S,Al^+}$ the differential polarizability between the states. 

\begin{figure}[htp]
    \centering
    \includegraphics[width=0.6\linewidth]{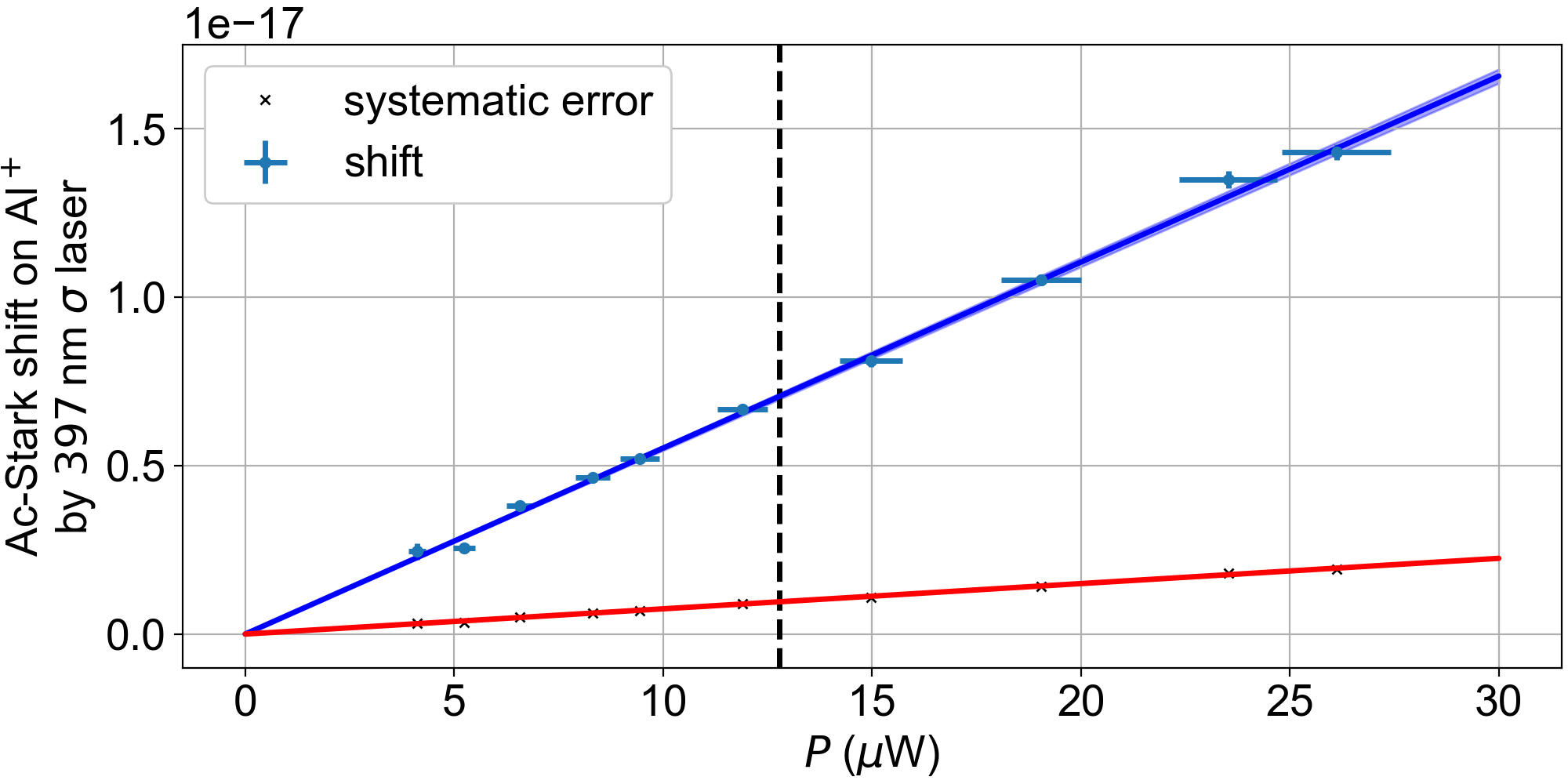}
    \caption[Ac-Stark shift \SI{397}{\nano\meter} laser]{Estimated ac-Stark shift of the \SI{397}{\nano\meter} $\sigma$ beam. The dashed line indicates the power used during the clock interrogation. }
    \label{fig:AC_stark_397_sigma}
\end{figure}

The frequency shift on \Ca was determined for multiple power values as shown in Fig.~\ref{fig:AC_stark_397_sigma} for the $\sigma^-$ \SI{397}{\nano\meter} beam. The estimated frequency shift is extrapolated from the different power measurements, resulting in an additional uncertainty contribution. The estimated frequency shift on \Al from the \SI{397}{\nano\meter} beams are $\Delta\nu_{Al^+,397\sigma}=\num{-71(10)e-19}$ and $\Delta\nu_{Al^+,397\pi}=\num{-8.4(1.2)e-19}$, where the main uncertainty contribution is due to the uncertainty in the \Al polarizability.

An additional light shift arises from the \SI{866}{\nano\meter} beam. To determine its electric field, we measure a frequency shift on the S$_{1/2}\leftrightarrow$D$_{5/2}$ transition. Since the laser is $4.9\,$THz detuned from a resonant transition coupling to the D$_{5/2}$ state, the frequency shift is small (around $100\,$Hz in our setup) compared to the once of the \SI{397}{\nano\meter} laser. To still be able to measure this frequency in the presence of magnetic field dephasing, we use the quantum-lock-in amplifier technique \cite{kotler_single-ion_2011}. 
The frequency shift on \Al is also evaluated for different laser powers. This is necessary, since the \SI{866}{\nano\meter} laser power used during EIT cooling is so low, that it is not directly measurable with our setup. 
We can extrapolate the absolute electric field at the experimental power and obtain a frequency shift on \Al of $\Delta\nu_{Al,866}=-1.37(27)\times10^{-18}$.

\begin{table*}
\caption[Uncertainty budget ac-Stark shift cooling lasers]{\label{tab:uncertainty_light_shift} Source of the uncertainties for all cooling and repumping lasers involved. For all beams, we use multiple power measurements and fit the results  to estimate the frequency shift at the power used during the experiment. This leads to an additional statistical extrapolation error. The $\alpha_\mathrm{X,Y-Ca}$ are the polarizability components of \Ca, where X is the scalar, vector and tensor component and Y indicates the state of \Ca. $I$ is the intensity illuminating each ion. This table is taken from Ref.~\cite{dawel_high-stability_2025_2}.}
\centering
\begin{tabular}{l|ccc}
Effect & Uncertainty   & Uncertainty  & Uncertainty  \\
 & \SI{397}{\nano\meter} $\sigma$ ($10^{-18}$)  & \SI{397}{\nano\meter} $\pi$  ($10^{-18}$) &  \SI{866}{\nano\meter}   ($10^{-18}$) \\
\hline
$\Delta\alpha_{\mathrm{S,Al}^+}$ & 0.89 & 0.10 & 0.27\\ 
$\alpha_{\mathrm{S,S-Ca}}$ & 0.24 & 0.03 & 0.0002\\
$\alpha_{\mathrm{V,S-Ca}}$ & 0.24 & - & -\\
$\alpha_{\mathrm{S,D-Ca}}$ & $\approx10^{-9}$& $\approx10^{-9}$ & 0.004\\ 
$\alpha_{\mathrm{V,D-Ca}}$ & $\approx10^{-10}$& - & -\\
$\alpha_{\mathrm{T,D-Ca}}$ &$\approx10^{-10}$& $\approx10^{-9}$ & 0.002\\ 
Extrapolation & 0.09& 0.008 & 0.015\\ 
$I(\mathrm{Ca})/I(\mathrm{Al})$ & 0.2& 0.12& 0.04 \\ 
\hline
Total uncertainty & 1.0 & 0.16 &  0.27 

\end{tabular}

\end{table*}

The uncertainty due to the different sources are listed in Tab.~\ref{tab:uncertainty_light_shift}. The differential polarizability of \Al is the largest uncertainty contributor for all wavelengths. An experimentally measured polarizability of \Al at both \Ca cooling laser wavelength would allow to further reduce the uncertainty of the ac-Stark shift. Overall, we end up at a light shift of $-9.3(1.1)\times10^{-18}$.

\subsection{Clock light}
\label{sec:light_shifts_clock}

In addition to the cooling lasers, the clock laser also causes a light shift due to off-resonant coupling of both clock states to other excited states. The light shift of the clock laser for a Rabi pulse was measured by Ref.~\cite{chou_frequency_2010}. They established an upper bound of $0(2)\times10^{-19}$ uncertainty for a probe time of $150\,$ms, corresponding to a $\pi$ pulse. This uncertainty applies to all experiments with the same coupling strength to the clock transition. One can convert the uncertainty for different probe times $\tau$, where $\tau$ corresponds to the time required to perform a $\pi/2$ rotation, by scaling using the formula

\begin{equation}
    \frac{\Delta\nu_{\mathrm{Stark},2\tau}}{\Delta\nu_{\mathrm{Stark},150\,\mathrm{ms}}} =\left(\frac{0.5\Delta\alpha_S E_{2\tau}^2}{0.5\Delta\alpha_S E^2} \right)=\left(\frac{\Omega_{2\tau}}{\Omega_{150\,\mathrm{ms}}}\right)^2 = \left(\frac{150\,\mathrm{ms}}{2\tau}\right)^2.
    \label{eq:Light_shift_clock_laser_Rabi}
\end{equation}
Since we are using Ramsey spectroscopy, the light shift changes depending on the pulse time $\tau$ of each Ramsey $\pi/2$ pulse and the dark time $T$. The shift of the pulses can be calculated using Eq.~\eqref{eq:Light_shift_clock_laser_Rabi}. To include the effect of the Ramsey dark time, we assume that we are close to resonance $|\Delta/\Omega_0|\ll1$ and follow the approach of Refs.~\cite{taichenachev_compensation_2010, benhelm_precision_2008}:
\begin{equation}
    \Delta\nu_{\mathrm{Stark,Ramsey}}\approx\frac{\Delta\nu_{\mathrm{Stark},2\tau}}{1+(\pi/4)(T/\tau)}=\frac{\Delta\nu_{\mathrm{Stark},2\tau}\times\frac{4\tau}{\pi}}{\frac{4\tau}{\pi}+T}
    \label{eq:ramsey_stark_shift}
\end{equation}
where $\Delta\nu_{\mathrm{Stark},2\tau}$ is the ac-Stark shift of a pulse of length $2\tau$ and $\frac{4\tau}{\pi}+T$ is the effective Ramsey time. The term $\Delta\nu_{\mathrm{Stark},\tau}\times\frac{4\tau}{\pi}$ can be interpreted as the phase shift due to an ac-Stark effect. For a $\pi/2$ pulse time of $\tau=25\,$ms and a dark time of $T=250\,$ms, we expect a clock laser light shift of $0(2)\times10^{-19}$. 

\subsection{Polarization mismatch}
\label{sec:shift_polarisationmismatch}

In Sec.~\ref{sec:light_shifts_clock}, we discussed the influence of an ac-Stark shift on the clock transition due to coupling to other levels. Now, we want to take look at the influence of the polarization of the clock laser on the light shift through coupling within the Zeeman manifold of ground and excited clock states. Such a shift has first been observed for ac-Zeeman shifts in a highly charged ion clock employing an M1 clock transition \cite{king_optical_2022}. We follow Ref.~\cite{yudin_probe-field-ellipticity-induced_2023}, which conducted a detailed evaluation of this shift. 

\begin{figure}
    \centering
    \includegraphics[width=0.3\linewidth]{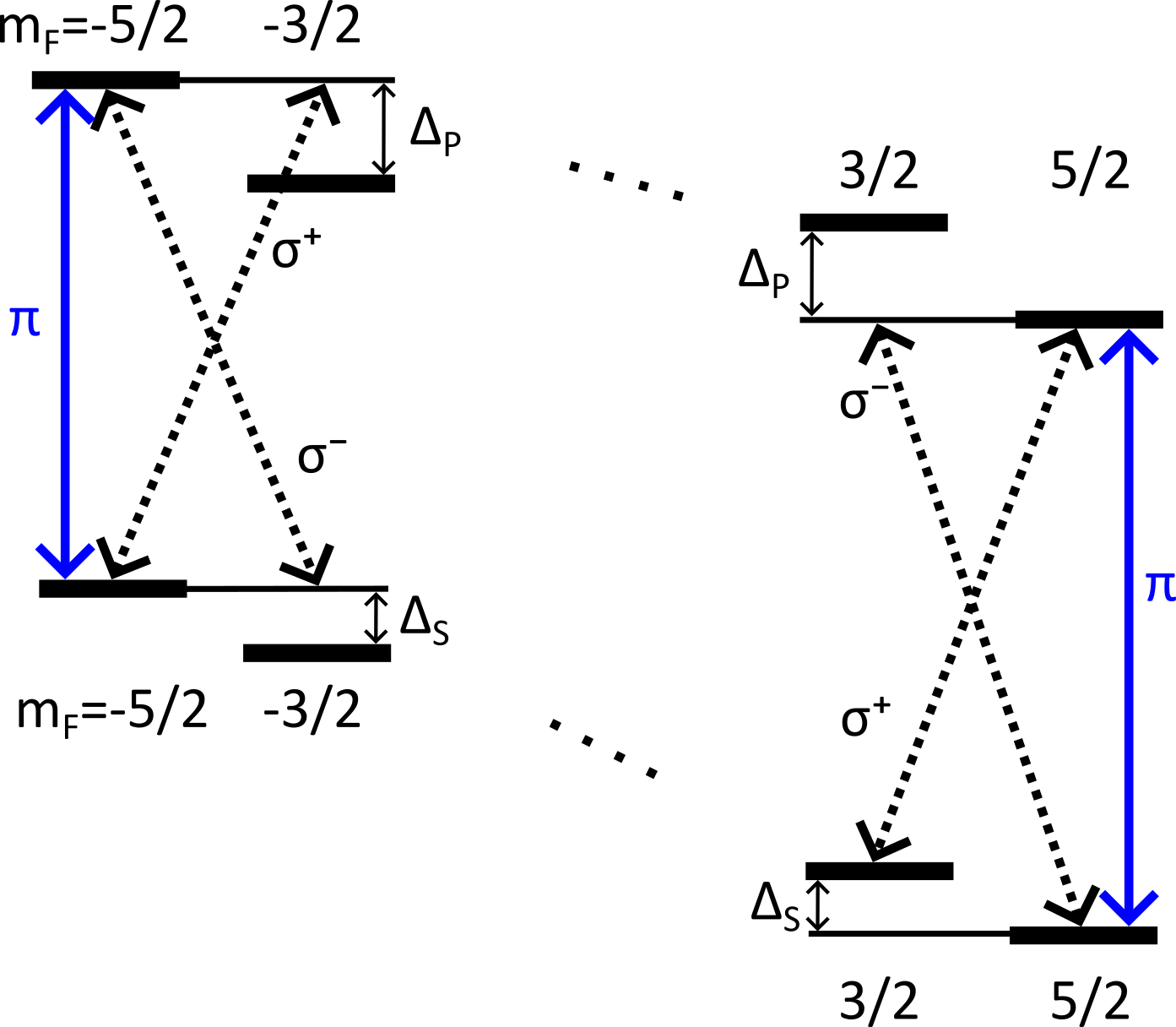}
    \caption[Clock laser polarization dependent light shift]{The light shift of the clock laser depends on the polarization. We want to drive the $\pi$ transition (blue line), but imperfect polarization will lead to additional shifts due to the coupling to other Zeeman states. This picture is adapted from \cite{yudin_probe-field-ellipticity-induced_2023} for \Al.}
    \label{fig:Pol_dep_light_shift}
\end{figure}

The clock transition ($^1$S$_0\leftrightarrow$ $^3$P$_0$) is measured on the magnetic sub states of $|\mS=-5/2\rangle \leftrightarrow |\mP=-5/2\rangle$ and $|\mS=+5/2\rangle \leftrightarrow |\mP=+5/2\rangle$, to suppress the first-order Zeeman shift (see Sec.~\ref{sec:Zeeman_shift}). With imperfect polarization, we couple the $\mS=\pm 5/2$ ground-state to neighboring Zeeman components $\mP=\pm 3/2$ in the excited state, and vice versa (see Fig.~\ref{fig:Pol_dep_light_shift}), resulting in an additional light shift.
The frequency shift $\delta_{-5/2,-5/2}$ for the $|\mS=-5/2\rangle \leftrightarrow |\mP=-5/2\rangle$ transition can be written as 
\begin{equation}
       \delta_{-5/2,-5/2} = \frac{|\Omega_+|^2}{4\DP} - \frac{|\Omega_-|^2}{4\DS}, 
\end{equation}
with $|\Omega_{\pm}|$ the Rabi frequency of the $\sigma^{\pm}$ polarized light and $\Delta_\mathrm{S,P}$ the detuning of the S and P state due to the Zeeman shift. This equation is valid as long as $|\Omega_{\pm}|\ll|\Delta_\mathrm{S,P}|$. In our case for a magnetic field of $B=\SI{0.15}{\milli\tesla}$, we have a shift of the P state of \SI{4.1}{\kilo\hertz} and of \SI{1.6}{\kilo\hertz} for the S state. Our Ramsey pulses have a Rabi frequency of $\Omega_0/2\pi \approx \SI{20}{\hertz}$.  
In analogy, we can write the shift $\delta_{5/2,5/2}$ for the plus transition as
\begin{equation}
    \delta_{5/2,5/2} = \frac{|\Omega_+|^2}{4\DS} - \frac{|\Omega_-|^2}{4\DP}.
\end{equation}
Taking the average over the two clock transitions we get:

\begin{equation}
    \nu_{\mathrm{av}} =\frac{1}{2} (\nu(m=-5/2) + \nu(m=+5/2)) = \nu_0 + \frac{\DS+\DP}{16\pi\DS\DP}(|\Omega_+|^2-|\Omega_-|^2).
\end{equation}

From this equation, we see that there is an additional shift, which is only zero if the polarization is linear ($|\Omega_+|^2=|\Omega_-|^2$) or the Zeeman splitting of the excited state is equal to the ground-state with a different sign (which is not true for \Al). The equation can be reformulated using an angular parameter $\epsilon$ to describe the ellipticity degree of polarization. The complex polarization vector $a$ is expressed as
\begin{equation}
    a=\cos(\epsilon)e'_x+i\sin(\epsilon)e'_y,
\end{equation}
where $e'_x,e'_y$ are the major and minor axis of the polarization ellipse.
For a polarization ellipse with an angle $\xi$ to the magnetic field, the shift can be represented by:
\begin{equation}
    \Delta\nu = |\Omega_{\pi}|^2\frac{\DS+\DP}{16\pi\DS\DP} \frac{\sin(2\epsilon)\sin(\xi)}{\left[\cos^2(\epsilon)\cos^2(\phi)+\sin^2(\epsilon)\sin^2(\phi)\right]\cos^2(\xi)},
\end{equation}
where $\phi$ is the angle between the projection of the magnetic field on the polarization ellipse and its major axis and $\Omega_{\pi}$ is the amount of $\pi$ polarization coupling.

Using Eq.~\eqref{eq:ramsey_stark_shift}, the frequency shift of the measured clock transition caused by imperfect polarization can be calculated. From geometrical considerations we conservatively assume an error of around $6^{\circ}$ for all angles and the ellipticity, a magnetic field of \SI{0.15}{\milli\tesla}, a $\pi/2$ pulse time of \SI{25}{\milli\second} for each pulse and a dark time of \SI{250}{\milli\second}. With that we obtain a shift and an uncertainty of \num{0(8)e-21} for Ramsey spectroscopy. For a Rabi pulse of \SI{300}{\milli\second} we would expect an uncertainty of \num{0(2)e-21}.

Ref.~\cite{yudin_probe-field-ellipticity-induced_2023} identified this effect as an additional component to the far-off resonant ac-Stark shift, which we quantified in Sec.~\ref{sec:light_shifts_clock}. We utilized Ref.~\cite{chou_frequency_2010} which provided an upper bound for the clock laser light shift for a probe time of \SI{150}{\milli\second}. For a magnetic field of \SI{0.1}{\milli\tesla}, we would obtain a light shift of \SI{0(97)e-22}, which is significantly below the given ac-Stark shift uncertainty. We can therefore assume that the main uncertainty is due to coupling to far off resonant transitions.

\subsection{Black-body radiation shift}
\label{sec:light_shifts_BBR}
The light shift caused by the black-body radiation (BBR) is one of the major frequency shifts in neutral atom clocks \cite{li_strontium_2024, koller_transportable_2017, hobson_strontium_2020, beloy_atomic_2014, aeppli_clock_2024} and plays an important role for ion clocks \cite{dube_evaluation_2013, zeng_toward_2023, lindvall_88sr_2025}. This shift depends on the temperature and the materials surrounding the atoms or ions. The influence of BBR on \Al was analyzed in Ref.~\cite{brewer_27al+_2019,chou_frequency_2010,rosenband_blackbody_2006} and specifically characterized for this setup in Ref.~\cite{dolezal_analysis_2015, kramer_aluminum_2023}. Since the last analysis, the differential polarizability of \Al was measured with lower uncertainty by Ref.~\cite{wei_improved_2024}, which leads to a minor change in the evaluation. We assume that the temperature evaluations are still valid and use the formula given in Refs.~\cite{brewer_27al+_2019,dube_high-accuracy_2014} to evaluate the frequency shift:

\begin{equation}
    \frac{\Delta\nu}{\nu}=-\frac{1}{2h\nu}\langle E^2\rangle_T\Delta\alpha_\mathrm{S,Al^+}(0)(1+\eta)=-\frac{1}{2h\nu}\frac{\pi^2 (k_B T)^4}{15\epsilon_0\hbar^3c_0^3}\Delta\alpha_\mathrm{S,Al^+}(0)(1+\eta)
    \label{eq:BBR_shift}
\end{equation}

Here, $\hbar$, $h$ are the reduced and normal Planck constants, respectively, $k_B$ is Boltzmann's constant, $\epsilon_0$ is the vacuum permittivity, $c_0$ is the speed of light in vacuum, $T$ is the temperature, $\Delta\alpha_\mathrm{S,Al^+}(0)$ is the differential static scalar polarizability of \Al, $\eta=0.00024$ is a dynamic correction factor \cite{rosenband_blackbody_2006} and $\langle E^2\rangle_{300\,\mathrm{K}}=\left(\SI{831.943}{\volt\per\meter}\right)^2$ is the quadratic electric field expected after Plank's law at \SI{300}{\kelvin} integrated over all frequencies. We use $\Delta\alpha_\mathrm{S,Al^+}(0)=\SI{6.86(23)e-42}{\coulomb^2\meter^2\per\joule}$ and a temperature of $T\approx \SI{300.5(3)}{\kelvin}$ to obtain a relative frequency shift of
\begin{equation}
    \frac{\Delta\nu}{\nu}=-\num{32.1(17)e-19}.
\end{equation}

\begin{figure}[htp]
    \centering
    \includegraphics[width=0.6\linewidth]{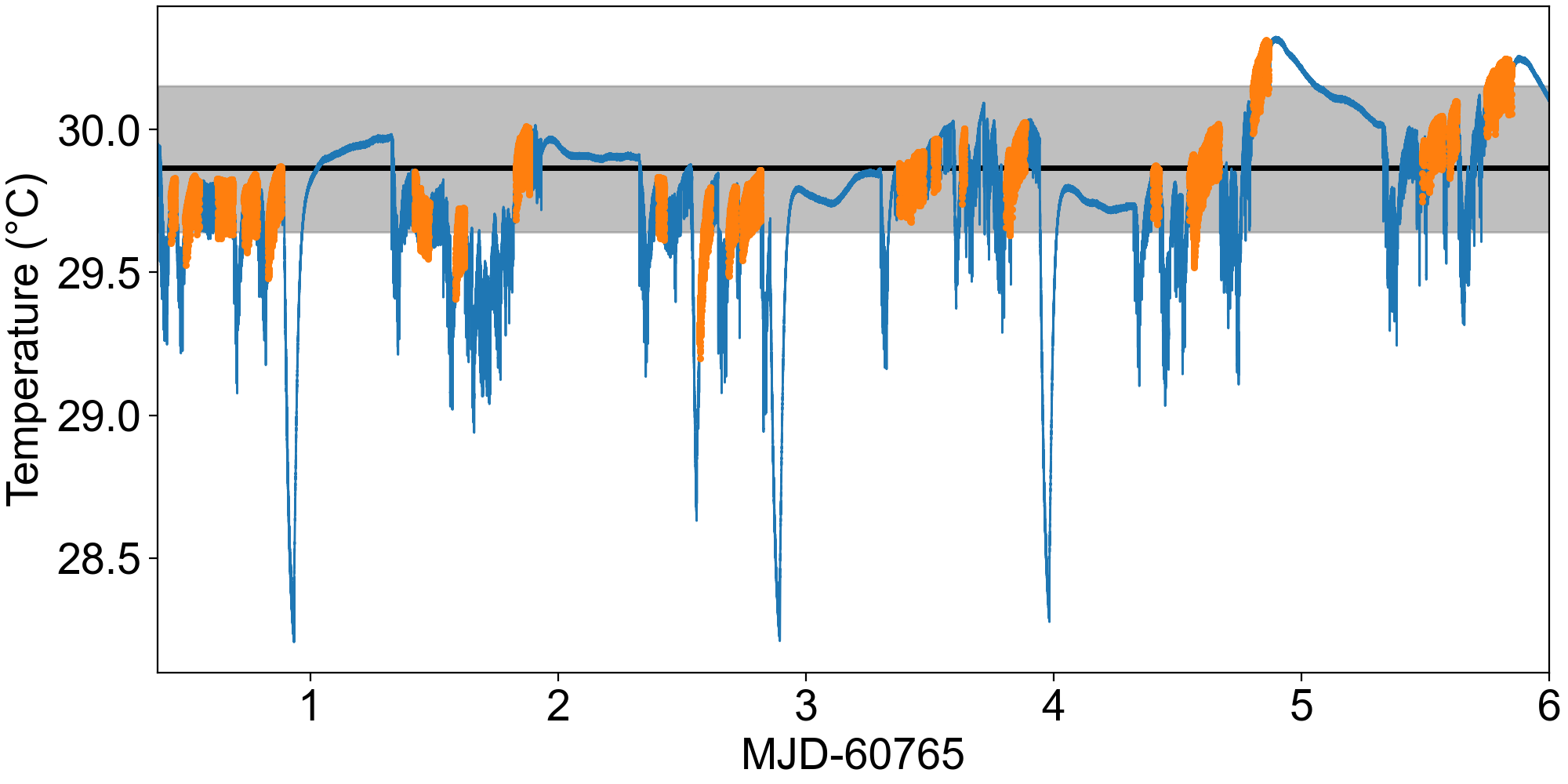}
    \caption[Temperature of the trap]{Temperature measurement with the PT100 sensor on the trap's sapphire discs. The blue line shows the entire measurement set, while the orange points are for times of active clock measurement.  The black line shows the mean and the gray area the standard deviation (29.87(26)$\,^\circ$C) of the blue data points. Loading and recrystallization of the \Al/\Ca crystal reduce the temperature, since lower rf trap powers are used.}
    \label{fig:PT_100_trap}
\end{figure}

 The different uncertainty contributions are listed in Tab.~\ref{tab:BBR_uncertainy_contributions}. Here, the main error is due to the temperature uncertainty. The temperature measured around the vacuum chamber is shown in Tab.~\ref{tab:temperature_gradient}. So far, we are only tracking the temperature of the trap by a PT100 sensor glued to one of the sapphire discs into which the blades of the trap are mounted (see Fig.~\ref{fig:PT_100_trap} for a temperature measurement during the clock run in April 2025). But this does not provide any information about the surrounding chamber. In the future, we will also track the vacuum chamber temperature at different positions to reduce the temperature uncertainty. 

\begin{table*}[htp]
    \centering
    \caption[uncertainty contributions BBR]{BBR shift uncertainty contributions. The largest uncertainty is from the temperature uncertainty.}
    \label{tab:BBR_uncertainy_contributions}
    \begin{tabular}{l|c S[table-format=5.3e2]}
        Parameter & Value & {fractional uncertainty} \\ 
        &&{contribution / $10^{-19}$} \\
        \hline
         T (K) & 300.5(3.0) & 1.3\\
         $\Delta\alpha_\mathrm{S,Al^+}(0)$ ($\frac{\mathrm{C}^2\mathrm{m}^2}{\mathrm{J}}$)& $6.86(23)\times10^{-42}$ & 1.1\\
         $\epsilon_0$ ($\frac{\mathrm{C}^2 \mathrm{s}^2}{\mathrm{kg} \mathrm{m}^3}$) & $8.854~ 187~ 8188(14)\times 10^{-12} $ &  5.1e-9 \\
         \hline
         $\Delta\nu$ (mHz) & \num{3.59(19)}& 1.7
    \end{tabular}
\end{table*}

\begin{table*}[htp]
	\caption[Temperature around the vacuum chamber]{\label{tab:temperature_gradient} Temperature distribution around the vacuum chamber when the trap is turned on. The average temperature of the lowest and highest point is \SI{27.33(25)}{\degreeCelsius}, with a difference to the highest and lowest temperature of \SI{2.57(25)}{\degreeCelsius}. Therefore, we use a conservative temperature of \SI{300.3(3.0)}{\kelvin} for the BBR shift.}
		\begin{tabular}{l|S[table-format=3.3]}
			\textrm{Where}&
			{\textrm{Temperature [\SI{}{\degreeCelsius}]}}\\
			\hline
			Below the chamber, near rf copper tubing  & 27.56(28)\\
			Octagon Chamber , near the diagonal pump port  & 25.34(25)\\
			Breadboard mounted on top of the chamber  & 24.80(40)\\
			Sapphire disc (inside chamber)  & 29.90(30)\\
		\end{tabular}
\end{table*}

\subsection{Ac-Stark shift of the trap}
\label{sec:AC_stark_trap}

\begin{figure}[htp]
    \centering
    \includegraphics[width=0.6\linewidth]{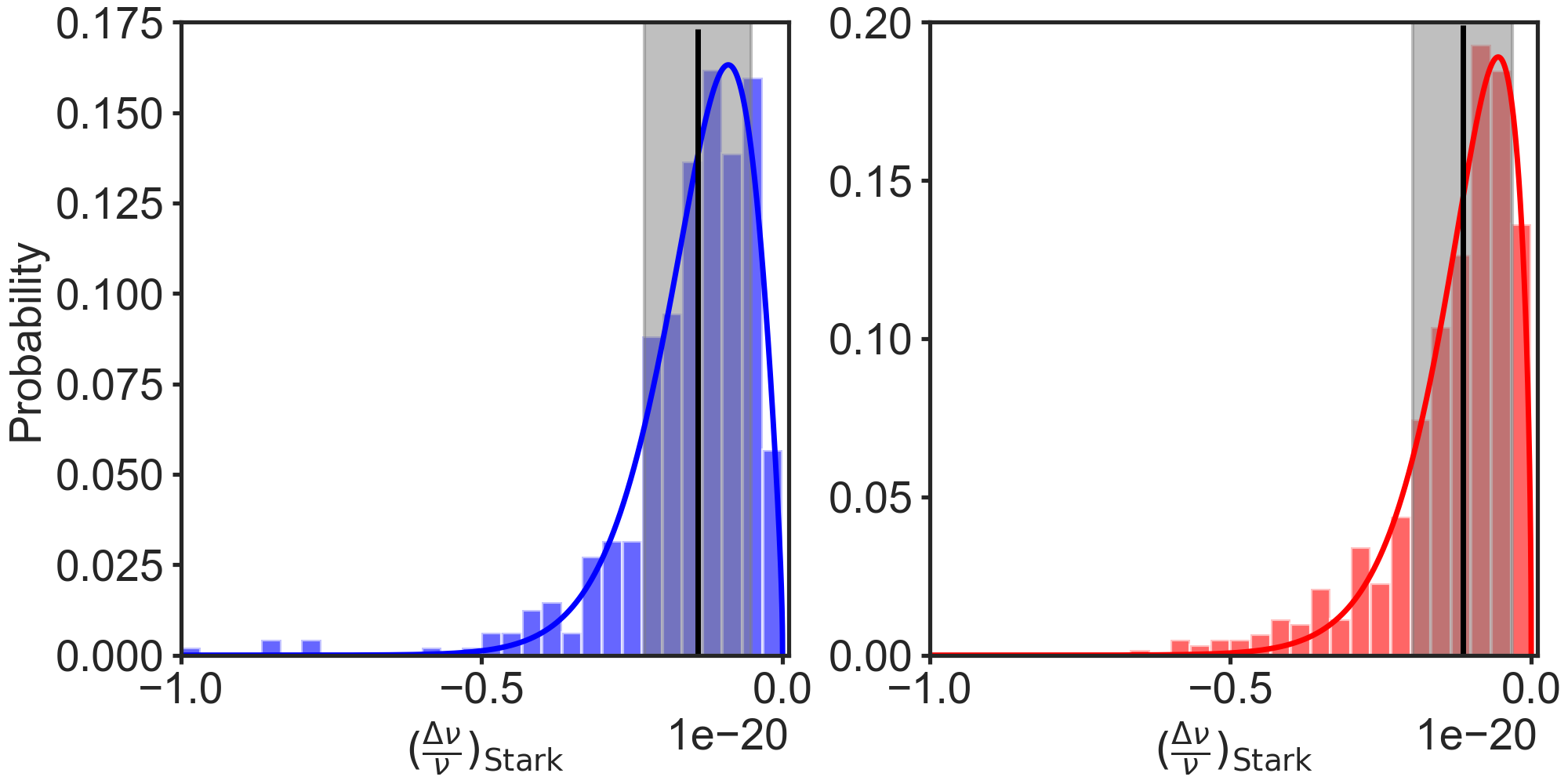}
    \caption[Ac-Stark shift of the trap on \Al]{Frequency shift caused by the electric field of the trap. Since the electric field is different for the two-ion positions we get a different shift. Overall, the shift is so small that we can neglect the impact of position changes.}
    \label{fig:Stark_shift_trap}
\end{figure}

Besides causing micromotion as discussed in Sec.~\ref{sec:excess_micromotion}, the electric field of the trap also causes an ac-Stark shift.
As for the case of micromotion, we find a component related to excess micromotion (EMM) and a thermal component arising from the spatial extent of the ion's wave function. The ac-Stark shift of the trap field is then given by \cite{keller_precise_2015}:
\begin{equation}
    \frac{\Delta\nu}{\nu} = -\frac{\Delta \alpha_\mathrm{Al^+}}{h \nu}\left(\underbrace{\frac{\Erf^2}{2}}_\textrm{EMM electric field}+\underbrace{\frac{m_\mathrm{Al^+}\wrf^2}{Q^2}k_B T_\mathrm{Al^+}}_\textrm{spatial overlap}\right) .
\end{equation}
with $\Erf$ the electric field seen due to EMM, $h$ is Planck's constant, $k_B$ Boltzmann's constant, $Q$ the charge of the ion, $T_\mathrm{Al^+}$ the temperature of the ion, $\wrf$ is the trap drive frequency, $m_\mathrm{Al^+}$ is the mass of aluminum and $\Delta\alpha_\mathrm{Al^+}$ is the static differential polarizability.
Here we estimate the temperature of \Al at $T=\SI{650(650)}{\micro\kelvin}$ as in Ref.~\cite{kramer_aluminum_2023}, which results in a negligible shift of
\begin{equation}
    \frac{\Delta\nu}{\nu} = -1.4(1.1)\times 10^{-21}.
\end{equation}
The error components are listed in Tab.~\ref{tab:Err_buget_magnetic_field}. The influence of the ion temperature is small compared to the effect caused by EMM. The main systematic uncertainty is the differential polarizability.  The position dependence of the Stark shift is plotted in Fig.~\ref{fig:Stark_shift_trap}. The shape of the distribution is the same as for the EMM shift, since they both depend on the same parameter $\Erf$. The effect of the Stark shift is significantly diminished for \Al attributable to its low differential polarizability.

\begin{table*}[htp]
    \centering
    \caption[Uncertainty contributions ac-Stark shift of the trap]{Uncertainty contributions for the ac-Stark shift of the trap. The temperature has only a very small effect on the shift, which is mainly dominated by the rf electric field. The uncertainties from physical constants can be neglected. u is the atomic mass unit taken from Ref.~\cite{mohr_codata_2025}}
    \label{tab:Err_buget_stark_shift}
    \begin{tabular}{l|c S[table-format=5.3e2] }
         Parameter & Value & {Fractional uncertainty } \\
        && {contribution /$10^{-19}$}\\
         \hline
         \Erf (V/m) & 16(14) & 0.011\\
         $\Delta\alpha$ (J$\mathrm{m}^2/\mathrm{V}^2$)& $6.87(23)\times 10^{-42}$& 3.8e-4 \\
         $T$ (\SI{}{\micro\kelvin})& 650(650)&2.5e-5 \\
         $m_{\mathrm{Al}}$ (kg) & $27 \times 1.66053906892(52)$ & 7.6e-15 \\
         \hline
         $\Delta\nu$ (\SI{}{\micro\hertz})&  1.5(1.2) & 0.011 \\
    \end{tabular}
\end{table*}

The results are consistent  with Ref.~\cite{brewer_27al+_2019}, which states that the ac-Stark shift of the trap is around 1\,\% of the excess micromotion shift. 

\section{Servo error}
\label{sec:Servo_error}

A servo loop ensures that the frequency of the oscillator follows the resonance of the atomic clock transition.
Consequently, the frequency of the laser must be updated regularly via a frequency change on an AOM. In the simplest approach, the $n$-th feedback cycle probing at frequency $\nu_n$ generates an error signal  $e_n$. This signal is weighted with a gain $g_1$ and the frequency of the next cycle $\nu_{n+1}$ is:
\begin{equation}
    \nu_{n+1} = \nu_{n}+g_1e_n.
    \label{eq:frist_order_servo}
\end{equation}
Thus, frequency changes of the oscillator or external field changes (e.g. B-field) are tracked and corrected for. However, in the presence of a linear drift of the laser
or the magnetic field, every cycle is probed with an excitation frequency which has always the same detuning from the atomic resonance frequency instead of fluctuating around it. This effect results in a systematic frequency shift that depends on the gain and the drift rate and is not compensated by the first-order servo loop given in Eq.~\eqref{eq:frist_order_servo}. To counteract this offset, we implement a second-order integrator with gain $g_2$, represented by
\begin{equation}
    \nu_{n+1} = \nu_{n}+g_1e_n+g_2 \sum_{i=0}^n 0.9^{n-i}e_i .
    \label{eq:second_order_integrator}
\end{equation}
Here, we utilize a limited second-order integrator with smaller weights on older error signal contributions. 
An unlimited integrator could also decrease the servo error to zero \cite{peik_laser_2006}. However, for the \Al clock used here this would be unfavorable. The reduction in uptime due to AlH$^+$ formation leads to repeated interruptions, and changes of the drift rate during these interruptions could impair the effectiveness of the second-order integrator. 
Furthermore, the limited integrator allows for faster settling of the lock after a restart of the clock. 

To understand the impact of the servo on the clock, we performed a realistic simulation of the clock cycle and feedback assuming a drift rate of \SI{-100}{\micro\hertz\per\second} at \SI{1068}{\nano\meter}. From the simulation we extracted the frequency offset for each gain setting used in the experiment. For simplicity, we omitted laser, magnetic field, and quantum projection noise, since these noise sources only introduce random fluctuations without impacting the offset \cite{yuan_suppression_2021_2}. An overview of the different shifts for different gain parameters is shown in Tab.~\ref{tab:servo_error_for_different_servos}. Through these simulations, we can attribute a suppression factor $s$ for the offset frequency to each gain setting, which is independent of the drift rate \cite{yuan_suppression_2021_2}. 

\begin{table}[htp]
    \centering
    \caption[Servo error shifts]{Servo error shifts for different locking parameters for a fixed drift rate of \SI{-100}{\micro\hertz\per\second} at \SI{1068}{\nano\meter}. 
    The values below are valid for the measurements in 2025. The amount of shift factor $s$ is given by the ratio of the shift $\Delta\nu$ and the laser's drift $d$ ($s=\Delta\nu/d$). The bottom line represents optimal lock parameters for which the servo error vanishes independent of the laser's drift rate. 
    The lock is simulated multiple times, and the uncertainty is the standard deviation of the different simulation runs.}
    \label{tab:servo_error_for_different_servos}
    \begin{tabular}{cc S[table-format=5.3] ccc ccc}
         $g_1$& $g_2$ & {Shift $\Delta\nu$ (mHz)}  & shift factor s & \multicolumn{3}{c}{employed during MJD} \\
         \hline
         0.25&0.005&-34(3)& $3.4(3) \times 10^2$&  60765.00&-&60768.60\\  
         0.35&0.01&-19(2)& $1.9(2) \times 10^2$& 60768.75&-&60769.47\\    
         0.7&0.01&-6.3(6)& $63(6)$& 60769.48&-&60769.77\\  
         0.9&0.01&-3.1(3)& $31(3)$ & 60769.78&-&60770.00\\ 
         0.46&0.0825&0.0(1)&  0(1)
    \end{tabular}
\end{table}

\begin{figure}[htp]
    \centering
    \includegraphics[width=0.6\linewidth]{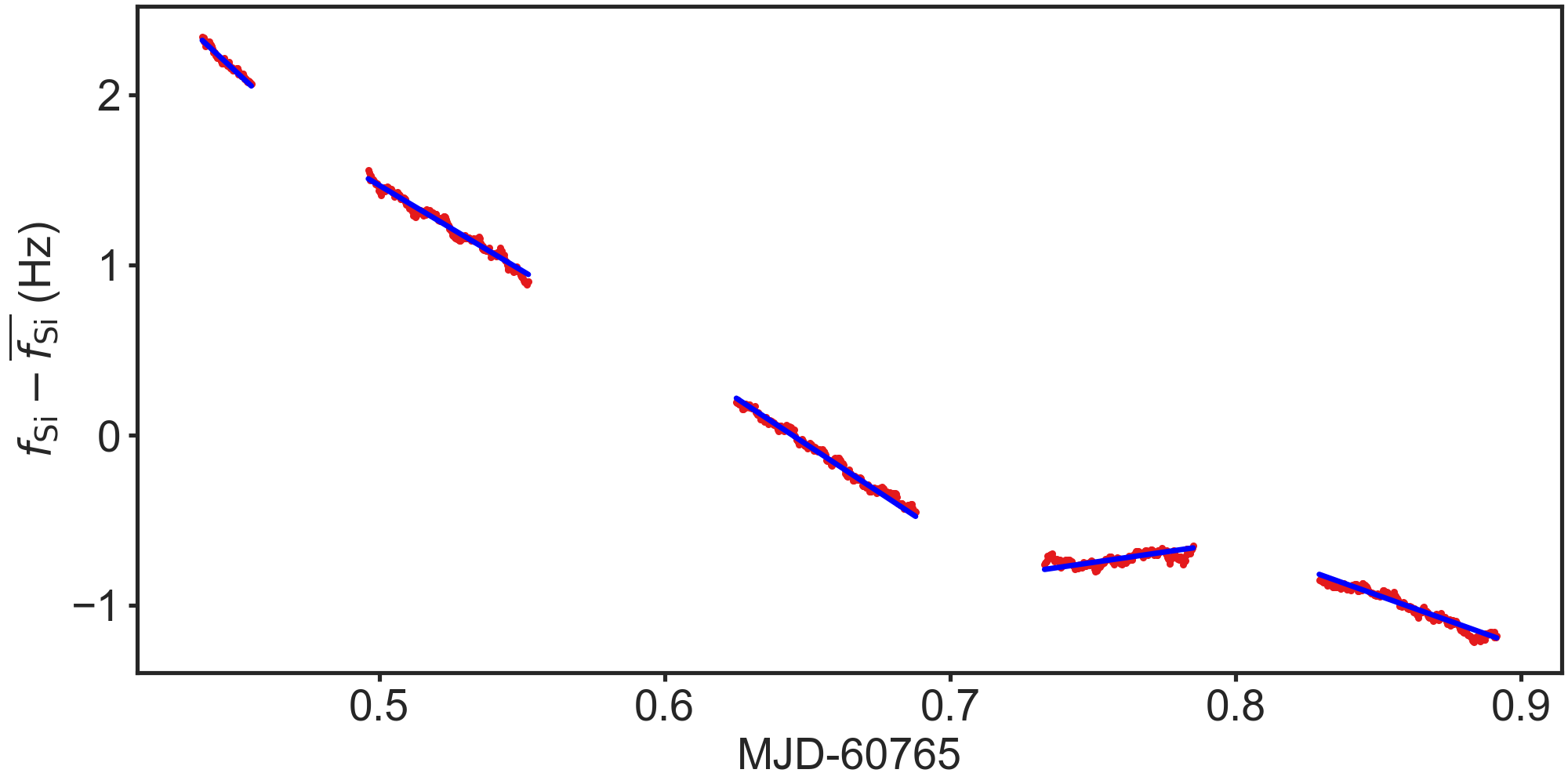}
    \caption[Laser drift at 1068\,nm]{Laser drift seen by the lock. The feedback enables us to calculate the drift of the laser without feedback. We see that at different lock times the drift fluctuates between $-180\,\mu$Hz/s and $+28\,\mu$Hz/s.}
    \label{fig:Si_cavity_drift_from_lock}
\end{figure}

We can utilize the simulated shift factor to determine the frequency offset (including the uncertainty from the simulated shift factor and the fitted laser drift) of each measurement that has been performed with non-optimum gain factor settings by extracting the frequency drift of the Si cavity-stabilized laser from the measurements (see Fig.~\ref{fig:Si_cavity_drift_from_lock}).

From the simulations we obtained an optimal parameter set of $g_1=0.46$ and $g_2=0.0825$ for which the servo error vanishes within the uncertainty of the simulations, i.e. $0(1)\times10^{-19}$. However, we conservatively estimate an uncertainty of $2\times10^{-19}$ to account for potential additional errors due to laser noise fluctuation \cite{lindvall_noise-induced_2023} or quadratic laser frequency changes, which are not included in the simulation.

\section{Phase chirp}
\label{sec:Phase_chirp}
\begin{figure}[htp]
    \centering
    \includegraphics[width=0.45\linewidth]{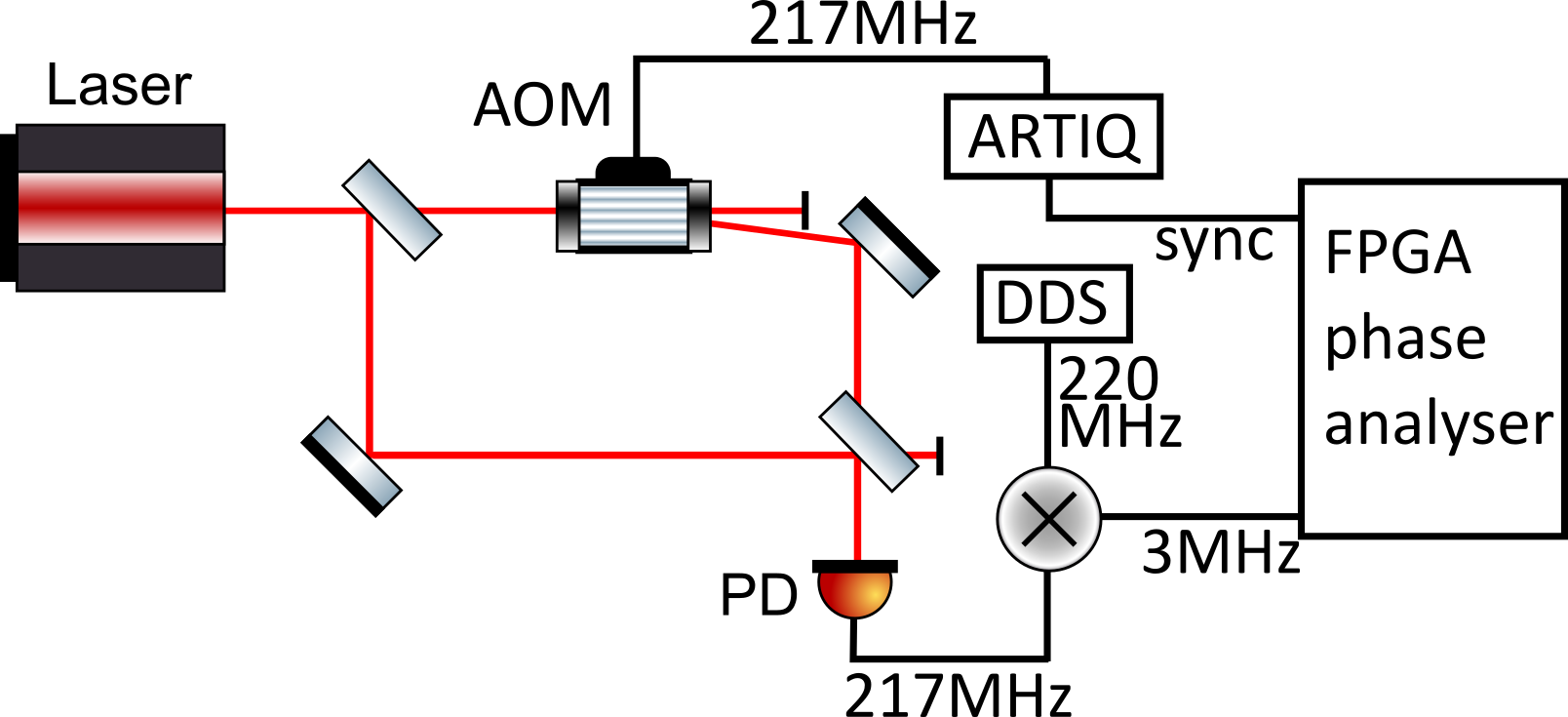}
    \caption[Phase chirp measurement scheme]{Measurement scheme for the phase chirp caused by an AOM.}
    \label{fig:Phase_chirp_scheme}
\end{figure}

AOMs are used in various spectroscopy setups as switches, primarily to control the light pulse duration on the atoms as well as adjusting the laser frequency and intensity. The duration of the light pulse and the driving rf power used in the AOM might result in a phase chirp. For instance, a fast pulse can become convoluted by the response of the electronic circuit of the AOM \cite{degenhardt_calcium_2005, rosenband_frequency_2008,wan_precision_2014}. Additionally, high rf power pulses can cause the AOM to heat, thereby altering the optical path length of the light inside the AOM crystal. Therefore, it is crucial to assess whether the AOM introduces a phase chirp on the clock laser light. To determine the resulting frequency shift seen by the atom, we measure the phase fluctuations of the first-order diffracted beam against a known reference (see Fig.~\ref{fig:Phase_chirp_scheme}).

We employ the setup described in Ref.~\cite{kazda_phase_2016}, which uses an FPGA unit to compare the measured phase of the resulting beat note on a photodiode with a common reference phase. All devices are referenced to the same \SI{10}{\mega\hertz} reference signal and the phase analyzer is triggered by the experimental control system. This allows coherent averaging of the phase measurement. 

\begin{figure}[htp]
    \centering
    \includegraphics[width=0.6\linewidth]{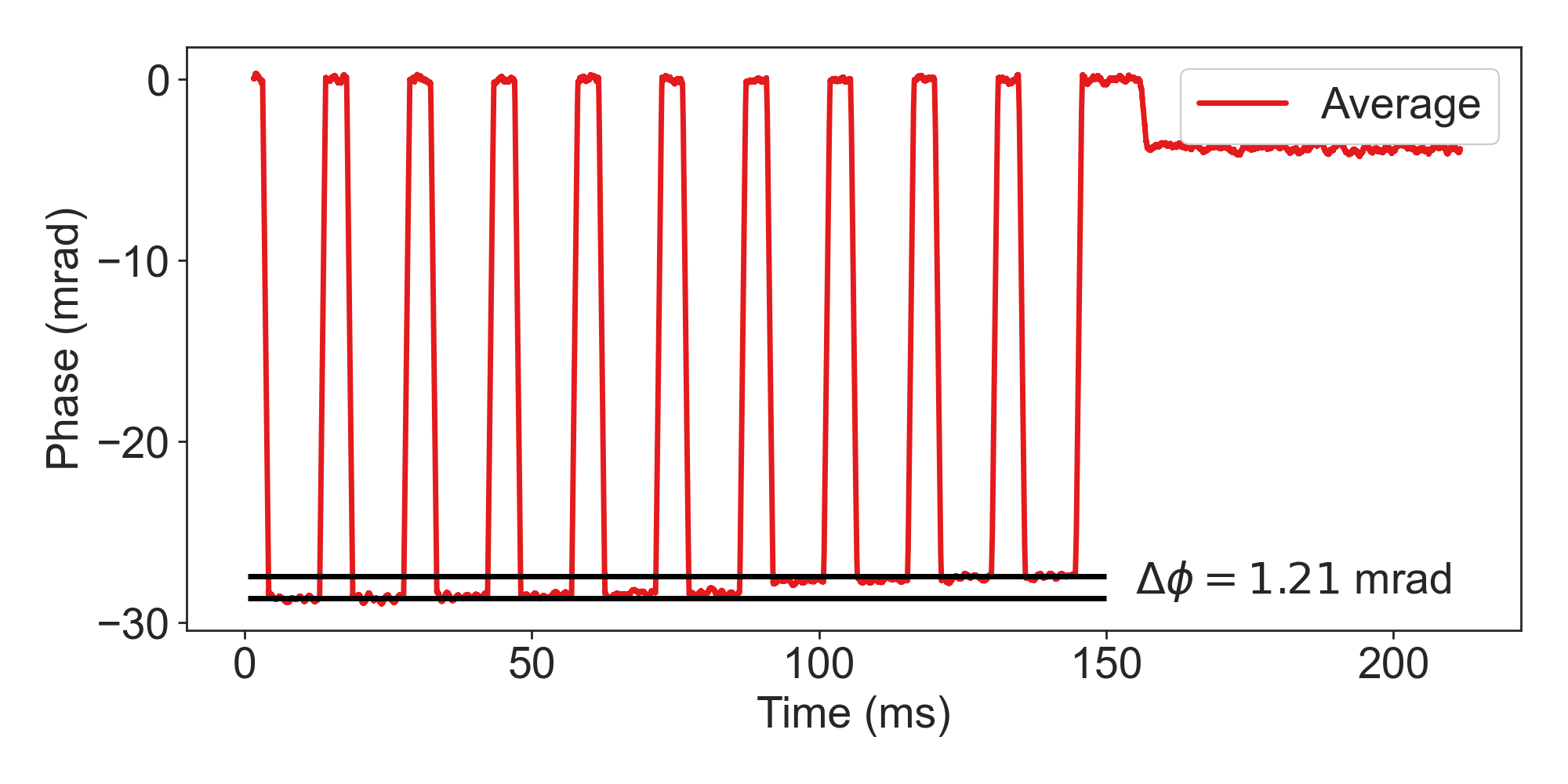}
    \caption[Phase chirp measurement]{Measurement of the phase chirp of a Rabi clock cycle. Here, only the averaged data is displayed. The relative phase difference between the ten pulses can be determined, which shows a slow drift and a clear jump between the \nth{6} and the \nth{7} pulse. This behavior is visible for all measurements made. 
    }
    \label{fig:Phase_chirp_data}
\end{figure}

The phase fluctuations measured with the device can now be utilized to estimate the frequency shift. We mimic the clock sequence, consisting of ten pulses at high rf power (\SI{10}{\milli\watt}) to transfer the excited clock state into the ground-state, followed by a single (instead of ten in the actual clock sequence) weak clock pulse at low rf power (\SI{31.6}{\micro\watt}). Only the strong rf pulses are visible in the measurement shown in Fig.~\ref{fig:Phase_chirp_data}, since the clock pulse signal is too weak to be detected by the phase analyzer.
Due to the low power (\SI{31.6}{\micro\watt}) of the clock pulse compared to the (\SI{10}{\milli\watt}) of the pumping pulses, we anticipate a thermal decay over the clock pulse. Consequently, it is crucial to understand how sensitive the clock excitation probability is towards phase variations of the light. For this purpose, we use the sensitivity function $g(t)$ \cite{dick_local_1990,falke_delivering_2012}. Considering the scenario of a phase step $\phi(t)=\epsilon H(t-t')$ ($H(x)$ is the Heaviside function) at the most phase sensitive point of the Rabi oscillation $t'$ (worst case), we can derive a frequency shift of $d\phi(t)/dt=2\pi \Delta\nu(t)=\epsilon\delta(t-t')$ ($\delta(x)$ is the delta function)\cite{dick_local_1990, falke_delivering_2012}. We can then calculate the effect on the excitation using the formula:
\begin{equation}
    dP=\frac{1}{2}\int_0^{t_i}\epsilon\delta(t-t')g(t)dt=\frac{\epsilon}{2}g(t'),
\end{equation}
which means that a phase step of $\epsilon=\SI{1.23(31)}{\milli\radian}$ results in an excitation change of \num{0.61(16)e-3}. Using $d\nu=dP/(g_{Rabi} t_i\pi)$, with $t_i=\SI{0.3}{\second}$ the frequency shift would be \SI{1.07(28)}{\milli\hertz} (relative: \SI{0.96(25)e-18}). Within a Rabi clock sequence, we cycle through ten Rabi pulses before we switch to another stretched state. Therefore, we would expect an additional suppression of ten, because the remaining nine pulses do not encounter a phase shift. Thus, for a Rabi sequence with no additional phase changes, we predict a relative shift of \num{0.96(25)e-19}.

Ramsey spectroscopy is used for interrogation, utilizing two AOMs to probe from opposite sides. The rf power for Ramsey is higher than for Rabi pulses, leading to an additional frequency shift by the clock pulses. We use pulses of \SI{25}{\milli\second} length with a power of \SI{2.2}{\milli\watt}, in comparison to the pulses measured for the phase chirp with \SI{10}{\milli\second} length at a power of \SI{10}{\milli\watt}.  Although no new measurements were taken with the AOMs, one is reused, while the other, being of the same brand and type, is expected to exhibit a similar behaviour. A phase error of $\epsilon=\SI{1.23(31)}{\milli\radian}$ would result in a frequency shift of \SI{0.69(18)}{\milli\hertz}. In a scenario where after ten pulses (as in the case before), the laser phase jumps by $\epsilon$, we expect a shift and uncertainty of \SI{0.139(36)}{\milli\hertz} for 20 pulses in ten cycles. For a conservative estimate, we assume the full shift as uncertainty. Thereby, we arrive at a phase chirp of \SI{0.0(2)}{\milli\hertz} (\num{0(2)e-19}) for Ramsey interrogation. Additionally, the effect of the de-excitation pulses add an uncertainty of $<\num{1e-19}$. Therefore, the relative frequency uncertainty for the phase chirp is at \num{0(3)e-19}.

\section{Background-gas collisions}
\label{sec:BGC}

background-gas collisions can shift the frequency through various processes, as described in Ref.~\cite{hankin_systematic_2019, barrett_analysis_2025_2}. An elastic transition (glancing collision) transfers small amounts of momentum to the ion, whereas a spiraling collision (Langevin collision) can transfer a lot of momentum \cite{meir_dynamics_2016}.  Both collision types between the ion and the atom/molecule will therefore affect the motional state distribution of the ion and thus cause and additional time dilation shift (TDS). Each collision will also polarize the colliding atom/molecule leading to an attractive interaction potential of $V(r)=-C_4/2R^4$, influenced by the atom-ion distance $R$ and the interaction strength $C_4$. For Langevin collisions the distance between the collision partners can become small, resulting in an additional differential energy (and thus phase) shift between the ground and excited clock state.

\begin{figure}[htp]
    \centering
    \includegraphics[width=0.6\linewidth]{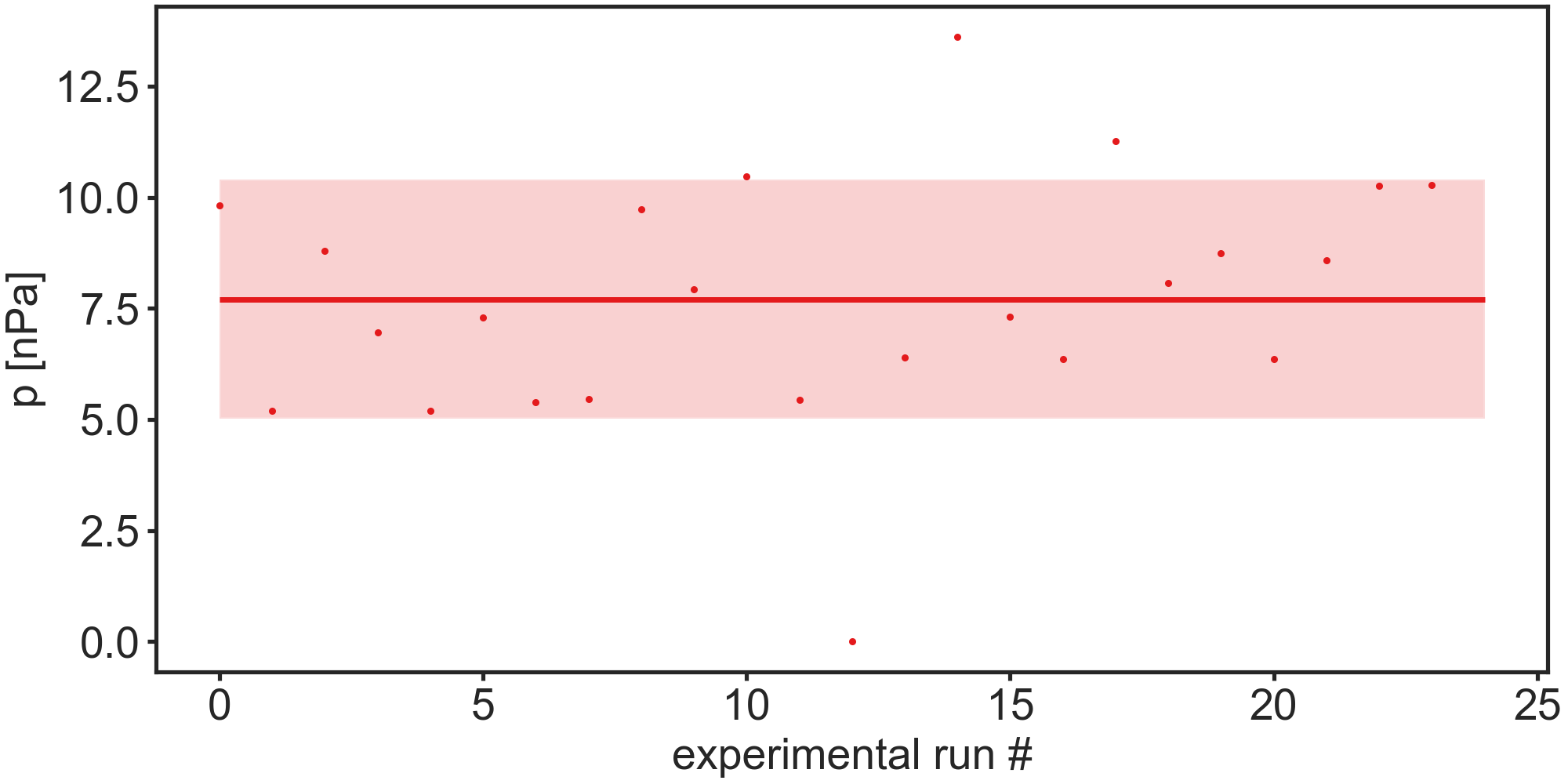}
    \caption[Pressure measurement during clock runs]{Estimated pressure at the ion positions measured via position changes during the clock measurement due to position swaps.}
    \label{fig:pressure_swaps}
\end{figure}

For our experiment we follow the procedures described in Ref.~\cite{kramer_aluminum_2023} for the estimation of the systematic phase shift caused by collisions. The pressure can be determined from swap rate measurements of a \Ca/\Al crystal \cite{hankin_systematic_2019}. By detecting the position of the ion during the clock operation, we determine the number of swaps during the clock run, assuming a \SI{50}{\percent} probability for a position swap after a decrystallization event. This is illustrated in Fig.~\ref{fig:pressure_swaps}, showing an average pressure of $p=\SI{8.1(2.5)}{\nano\pascal}$. 

For the calculation of the fractional frequency shift due to phase shifts caused by collisions we apply the formula provided in Ref.~\cite{davis_improved_2019}. We assume that the background-gas is primarily hydrogen H$_2$,  as it is expected to dominate residual background-gas in a stainless steel vacuum chamber. It also has the largest uncertainty and thus establishes the most conservative bound. This results at $300(5)\,$K in a shift of 
\begin{equation}
    \frac{\Delta\nu}{\nu}=\num{0.2(28)e-19}.
\end{equation}
The main source of uncertainty is attributed to the model, as illustrated in Tab.~\ref{tab:error_phase_collisions}. 

\begin{table}[htp]
    \centering
    \caption[Errors sources of the collisional phase shift]{Different error contributions to the collisional phase shift with $n_{bg}$ is the number density of the background-gas in [cm$^{-3}$]. The model error for H$_2$ dominates the error estimation.}
    \label{tab:error_phase_collisions}
    \begin{tabular}{l|c S[table-format=5.3]}
        Parameter & Value & {Fractional uncertainty } \\
        && {contribution /$10^{-20}$}\\
        \hline
        $\Delta\nu/n_{bg}$ (\SI{}{\pico\hertz \per \centi\meter\cubed}) & 12(161) & 28 \\
         
         p (\SI{}{\nano\pascal}) & 8.1(2.5) & 0.64 \\
         T (\SI{}{\kelvin}) & 300(5) & 0.035 \\
         \hline
         $\Delta\nu$ (mHz)& 0.02(31) &  28
    \end{tabular}
\end{table}

Apart from the phase shift there is also a motional shift from the energy transfer of the background-gas particle to the \Al ion, which is usually not covered by sideband thermometry because this method underestimates non-thermal state distributions \cite{chen_sympathetic_2017, rasmusson_optimized_2021}. While this shift can be large, it is suppressed by the reduced interaction of the probe light with a motionally excited clock ion \cite{barrett_analysis_2025_2}: the Rabi frequency for probing the clock transition is reduced by the Debye-Waller factor  $e^{-\eta^2(\bar{n}+1/2)}$ \cite{wineland_experimental_1998}, which depends on the mean motional state occupation $\bar{n}$. 
In case of a Ramsey sequence, where collisions are likely to occur during the dark time, the interaction of the second pulse is reduced. For vanishing coupling of the second Ramsey pulse, this results in a purely statistical measurement outcome with no frequency bias, effectively suppressing a frequency shift. 
The suppression of this effect is simulated in Ref.~\cite{hankin_systematic_2019}, showing a suppression of $\xi\approx3.67\times10^3$, if no cooling is applied. We can use the same factor in our experiment if no cooling is used during the interrogation.

In the case of cooling during the interrogation, the motional state decreases back to its steady state after the collision, depending on the cooling rate. This can lead to dynamic changes in the time-dilation shift during the dark time, while the second pulse is interacting with the ion at the motional ground-state, thus eliminating the suppression effect discussed above. Ref.~\cite{marshall_high-stability_2025} describes the use of Doppler cooling during the interrogation. As Doppler cooling reduces a motional state of $n=1400$ in around \SI{2.7}{\milli\second} to its steady-state value, the unsuppressed shift is reduced by \SI{2.7}{\milli\second}/$T$ to account for the decrease of the mean motional state by cooling over the entire probe time of $T=\SI{1}{\second}$.

We use EIT cooling to keep the ions near the motional ground-state during the interrogation. Our Lamb-Dicke factor for the radial IP mode is $\eta_{397}\approx0.01$, which is a factor of 5 lower than the largest Lamb-Dicke factor for the slowest cooled mode in Ref.~\cite{marshall_high-stability_2025}. Consequently, the ion will remain longer in a higher motional state and our cooling rate is slower by a factor of $5^2$, resulting in a cooling time of $\Delta t_\mathrm{cool}=\SI{2.7}{\milli\second}\cdot5^2=\SI{62.5}{\milli\second}$. Importantly, we must note that EIT cooling is different to Doppler cooling. We observe in the experiment that ions do not recrystallize after large momentum transfers using only EIT cooling. This means that collisions with large momentum transfer are still suppressed, while collisions with small momentum transfer contribute a motional shift, which is reduced by cooling. Therefore, instead of the $\xi\approx3.67\times10^3$ given in Ref.~\cite{hankin_systematic_2019}, we conservatively attribute a suppression factor of $\xi=37$, which is 100 times smaller due to the presence of EIT cooling. This approach allows us to estimate the motional shift due to the collisions through scaling of the simulation result of Ref.~\cite{hankin_systematic_2019}, where the motional shift was simulated at $p_\mathrm{Han}=\SI{38}{\nano\pascal}$ with a probe time of $T_{\mathrm{Han}}=\SI{150}{\milli\second}$, resulting in a time dilation frequency change of $ \Delta\nu_{\mathrm{Han}}=2200(1100)\times 10^{-19}$.

We use (see also \cite{marshall_high-stability_2025})

\begin{equation}
    \Delta\nu = \Delta\nu_{\mathrm{Han}}\frac{p}{p_\mathrm{Han}}\frac{T}{T_{\mathrm{Han}}}\frac{\Delta t_\mathrm{cool}}{T}\frac{1}{\xi} = \num{-5.7(4.3)e-19},
\end{equation}
and scale this shift with our probe time of $T=300\,$ms as it increases linearly with longer interrogations. Here, we take the complete shift as uncertainty, as the suppression factor is just an estimation. Therefore, the total uncertainty of the collisional shift is \num{-5.4(6.5)e-19}. Without EIT cooling, we expect a shift of \num{0.2(28)e-19}, where the motional shift due to collisions is \num{0.3(0.3)e-19} as the suppression factor is much larger in the absence of cooling. 

\section{Electric Quadrupole shift}
\label{sec:elec_quadrupole_shift}
The electric quadrupole shift was previously assessed in Ref.~\cite{kramer_aluminum_2023}. It depends on the quadrupole moment, which is small for both clock states in \Al \cite{beloy_hyperfine-mediated_2017}. Following the derivation of Ref.~\cite{kramer_aluminum_2023}, we can write the shift as:
\begin{equation}
    \frac{\Delta\nu}{\nu} = \frac{1}{2h} \frac{\partial^2 \Phi}{\partial z^2} \frac{3m_F^2-F(F+1)}{F(2F-1)}\Theta_P,
    \label{eq:quadrupol_shift}
\end{equation}
where F is the total angular momentum, $m_F$ is the magnetic quantum number, $\frac{\partial^2 \Phi}{\partial z^2}$ is the gradient of the electric field, and $\Theta_P$ is the quadrupole moment of the clock state \tpz ($\Theta_P=$\SI{8(3)e-46}{\ampere\second\meter\squared}),
while the quadrupole moment of the \ssz state $\Theta_S<$\SI{4e-48}{\ampere\second\meter\squared}) is negligible \cite{beloy_hyperfine-mediated_2017}). 
The electric potential of the trap as a function of cartesian coordinates with $Z$ aligned along the symmetry axis of the linear Paul trap can be written as \cite{wubbena_sympathetic_2012}:
\begin{equation}
    \Phi = \frac{\tilde{U}}{2}\cos(\wrf t) \frac{X^2-Y^2}{R^2} + U \frac{Z^2-\alpha X^2-(1-\alpha)Y^2}{d^2}.
\end{equation}
Here, $\tilde{U}$ and $\wrf$ are the voltage and angular frequency at the rf electrodes, respectively. $U$ is the voltage for the axial trapping potential. The quantities $d, R$ are characteristic dimensions of the trap and $\alpha$ is a factor for the radial asymmetry of the trap. For the case that the trap axis does not align with the quantization axis and dropping the rf term, the curvature of the field takes the form \cite{beloy_hyperfine-mediated_2017}:
\begin{equation}
    \frac{\partial^2 \Phi}{\partial z^2}=\frac{2 U}{d^2}\left[\frac{3\cos^2(\theta)-1}{2}-(\alpha-\frac{1}{2})\sin^2(\theta)\cos(2\phi)\right],
\end{equation}
where $\theta$ and $\phi$ are the spherical coordinates between the trap and the quantization axis.
In addition to the trap potential, the \Al ion will be influenced by the electric field curvature introduced by the \Ca logic ion which is expressed by:
\begin{equation}
    \frac{\partial^2 \Phi}{\partial z^2}=\frac{2 U}{d^2}\left(\frac{3\cos^2(\theta)-1}{2}\right).
\end{equation}
Both curvatures must be added together for the calculation of the quadrupole shift in Eq.\eqref{eq:quadrupol_shift}.
The ratio between the voltage of the endcaps and their characteristic length can be calculated via the secular frequency by
\begin{equation}
\frac{U}{d^2} = \frac{m_{Ca} \omega_{z,IP}^2}{2 e} \frac{\mu}{1+\mu-\sqrt{1-\mu+\mu^2}} = \frac{m_{Ca} \omega_{z,Ca}^2}{2 e},
\end{equation}
where $\omega_{z,IP}$ is the axial IP mode, $m_{Ca}$($m_{Al}$) is the mass of \Ca(\Al), and $\mu=m_{Al}/m_{Ca}$ is the mass ratio between the ions. For the case that the quantization axis is aligned with the trap axis ($\theta\approx0$, $\phi\approx 0$, $\omega_{z,\mathrm{IP}}/2\pi=\SI{1.255(10)}{\mega\hertz}$) the shift is $-2.3(8) \times 10^{-20}$. 

\begin{table*}[htp]
    \centering
    \caption[uncertainty contributions quadrupole shift]{Uncertainty contributions of different factors to the quadrupole shift. The quadrupole moment is the largest uncertainty source.}
    \label{tab:Quadrupol_shift_uncertainy_contributions}
    \begin{tabular}{l|c S[table-format=7.3e1]}
        Parameter & Value & {Fractional uncertainty} \\
        && {contribution / $10^{-19}$} \\
        \hline
        $\Theta_P$(\SI{}{\ampere\second\meter\squared}) &  \num{8(3)e-46} & 0.08 \\
        $\omega_{z,\mathrm{IP}}/2\pi$ (\SI{}{\mega\hertz}) & 1.255(10) & 0.0036 \\
        $\theta$ (°) & 1(5) & 0.0015\\
        $\phi$ (°) & 1(5) & 2.5e-7\\
         u (kg) & $1.660~539~0689(5)\times 10^{-27}$ &  7e-11 \\
         $a_0$ (m) & $5.291~772~1054(8)\times10^{-11}$ & 7e-11 \\
         \hline
         $\Delta\nu$ (mHz) & -0.026(7) & 0.08
    \end{tabular}
\end{table*}

\section{Gravitational redshift}
\label{sec:grav_red_shift}

\begin{figure}
    \centering
    \includegraphics[width=0.45\linewidth]{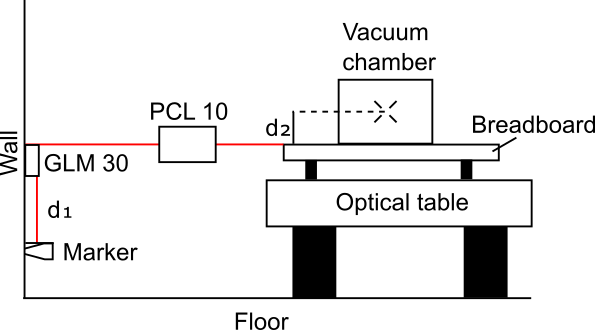}
    \caption[Height determination of the experiment]{Height determination of the experiment to the reference marker. GLM30 is a digital laser measure and the PCL is a crossline laser.}
    \label{fig:Height_measurment}
\end{figure}

The frequency of clocks in comparisons needs to be corrected for the gravitational redshift difference between the clocks \cite{petit_relativistic_2005}, requiring a precise height difference determination.
Various heights of height markers were measured at PTB in a campaign detailed in Ref.~\cite{denker_geodetic_2018, denker2014geodetic}. Additionally, the local gravitational acceleration was measured using a gravimeter. This allows for a precise determination of the gravitational redshift which is defined as \cite{mehlstaubler_atomic_2018}:
\begin{equation}
    \frac{\Delta\nu}{\nu_0} = \frac{\Delta W_{gp}}{c_0^2}=\frac{g(H)\Delta H}{c_0^2}.
\end{equation}
Here, $\Delta W_{gp}$ is the difference in the gravity potential, including the Newtonian gravitational potential and the centrifugal potential, $\Delta H$ is the height difference between the clocks, and $g(H)$ is the local acceleration depending on the height $H$ of the clock. 

 The height difference between the ion and the reference marker was measured as shown in Fig.~\ref{fig:Height_measurment}. The length $d_1$ was determined using a crossline laser and a digital laser measure. The height $d_2$ of the ion was assessed using the mean height of the axial \SI{397}{\nano\meter} $\sigma$ beam, both before and behind the vacuum chamber. The total height of the ion to the reference marker is \SI{775(4)}{\milli\meter}, where the main uncertainties are from the crossline laser's angle errors and the distance uncertainty of the laser measure.
 
 The relative shift of \Al to the different PTB clocks is given in Tab.~\ref{tab:Gravitational_red_shift}. The uncertainty is primarily limited by the uncertainty of the height measurement.

\begin{table*}[htp]
    \centering
    \caption[Table with local gravitational redshifts]{Relative gravitational redshift between \Al and other PTB clocks. The \Al clock is located on the \nth{5} floor, resulting in a significantly larger gravitational redshift than that experienced by the other clocks at PTB.}
    \label{tab:Gravitational_red_shift}
    \begin{tabular}{l|c c}
        clock & relative gravitational redshift ($10^{-19}$) & local relative gravitational  redshift uncertainty ($10^{-19}$)\\
        \hline
        Sr3  & -13761.9 & 6.3 \\
        CSF1 & -13948.7 & 7.7 \\
        CSF2 & -14064.4 & 7.7
    \end{tabular}
\end{table*}

\section{Systematic frequency uncertainty budget} 
\label{sec:error_budget}

The total relative systematic frequency uncertainty budget for operation of the \Al/\Ca clock with EIT cooling during clock interrogation is given in Tab.~\ref{tab:Error_buget_complete}. The differences for the uncertainty budget without cooling are listed in Tab.~\ref{tab:Error_buget_complete_without_cooling}.
The main difference between the two tables is the motional shift at \SI{285}{\milli\second}, the light shift of the cooling lasers, and the collisional shift. While cooling during the interrogation does not look favorable right now, future measurements of the differential polarizability will reduce the light shift, making cooling during interrogation the preferred option.
Compared to the systematic frequency uncertainty budget of Ref.~\cite{kramer_aluminum_2023}, we also include the servo error, phase chirp, first-order Doppler shift, and a reevaluation of the excess micromotion shift.

When we look at the overall uncertainty, the three largest uncertainties are the cooling laser light shift, the quadratic Zeeman shift, and collisions. While the first two can be reduced by measuring the respective atomic constants more accurately and/or reducing the magnetic field for the Zeeman shift, reducing the collisional shift will only be possible with a lower background-gas pressure or a more advanced collisional model.

\begin{table*}[htp]
    \centering
    \caption[Uncertainty budget of the \Al clock]{Uncertainty budget contributions of the \Al clock sorted in order of their uncertainty contribution. All relative shifts $\Delta\nu$ and uncertainties $u(\Delta\nu)$ are scaled by $10^{-19}$.}
    \label{tab:Error_buget_complete}
    \begin{tabular}{ c |S[table-format=7.3] S[table-format=7.3] c c}
         Effect & {$\Delta\nu$}  & {$u(\Delta\nu)$}  & Ref. of used constants & Limitation \\
         \hline
         Cooling laser light shift & -93 & 11 & $\Delta\alpha$:\cite{dawel_high-stability_2025_2} & $\Delta\alpha_{Al^+}$ \\
         Collisions & -5.4 & 6.5 & phase shift \cite{davis_improved_2019} & model and vacuum \\
         Quadratic Zeeman - dc & -14900.3 & 5.4 & $C_2$:\cite{brewer_27al+_2019}, $g_P-g_S$: \cite{rosenband_observation_2007} & $C_2$ \\
         
         Excess micromotion & -3.5 & 3.8 & & measurement uncertainty \\
         Quadratic Zeeman - ac & -164.6 & 3.1 & $C_2$:\cite{brewer_27al+_2019},$g_S(Ca)$: \cite{tommaseo_g_j-factor_2003} & trap drive stability \\
         Phase chirp & 0.0 & 3.0 & & measurement uncertainty \\
         First-order Doppler shift & 0.0& <2.3 & & measurement uncertainty \\
         Clock laser light shift & 0.0 & 2.0 & $\Delta\nu_\mathrm{ac-Stark} $ \cite{chou_frequency_2010} & $\Delta\nu_\mathrm{ac-Stark}$ \\
         Second-order Doppler shift & -16.9 & 2.0 & & measurement $\bar{n}$ \\
         Servo error   & 0 & <2 & & uncertainty of the model \\
         Black-body radiation & -32.2 & 1.7 & $\Delta\alpha_{Al^+}$:\cite{wei_improved_2024} & temperature uncertainty \\
         Electric quadrupole shift & -0.23 & 0.08 & $\Theta_P$: \cite{beloy_hyperfine-mediated_2017} & uncertainty of the constant \\
         ac-Stark shift trap & -0.013 & 0.010 & $\Delta\alpha_{Al^+}$:\cite{wei_improved_2024} & position and measurement uncertainty \\
         \hline
         \textbf{Total} & -15216 & 16 & &\\
    \end{tabular}
\end{table*}

\begin{table*}[htp]
    \centering
    \caption[Uncertainty budget of the \Al clock without cooling]{Uncertainty budget of the \Al clock without cooling. All shifts and uncertainties are given in $10^{-19}$. Only differences to Tab.~\ref{tab:Error_buget_complete} are listed. The interrogation time is fixed to \SI{300}{\milli\second}. }
    \label{tab:Error_buget_complete_without_cooling}
    \begin{tabular}{c| S[table-format=7.3] S[table-format=7.3]c c}
         Effect & {$\Delta\nu$}  & {$u(\Delta\nu)$}  & Ref. of used constants & Limitation \\
         \hline
         Second-order Doppler shift without cooling & -71.2 & 7.7 & & measurement $\bar{n}$ and $\dot{\bar{n}}$ \\
         Collisions without cooling & 0.2 & 2.8 & phase shift \cite{davis_improved_2019} & model and vacuum \\
         Cooling laser light shift & 0.0 & 0.0 \\
         other shifts (see Tab.~\ref{tab:Error_buget_complete}) & -15100 &  1.0 \\
         \hline
         \textbf{Total without cooling} & -15171 & 12
    \end{tabular}
\end{table*}

\section{Frequency measurement against CSF1 and CSF2}

The absolute frequency of \Al was measured against both caesium fountain clocks as PTB \cite{weyers_advances_2018}.
The measurements were performed in the intervals between the Modified Julian Days (MJDs) 60528 through 60531, 60534 through 60538 and 60765 through 60770 with average uptimes of \SI{17}{\percent}, \SI{15}{\percent} and \SI{22}{\percent}, respectively.
Tab.\ \ref{tab:absolute-frequencies} summarizes the results.
For this we used the method described in Refs.~\cite{schwarz_long_2020, grebing_realization_2016}, where the maser serves as a flywheel oscillator to bridge gaps in overlapping uptime of the two clocks and thus increase the averaging time. The noise model of the maser (H9) $S_y=\sum_{\alpha=-1}^1h_\alpha f^\alpha$ was extracted from data against the \Al clock. We used the following parameters:
\begin{itemize}
    \item $h_1=$\num{3.1e-26}; $\sigma_{y,\mathrm{FPN}}=$\num{5.9e-14}
    \item $h_0=$\num{2.3e-27}; $\sigma_{y,\mathrm{WFN}}=$\num{3.4e-14}
    \item $h_{-1}=$\num{6.1e-33}; $\sigma_{y,\mathrm{FFN}}=$\num{0.9e-16}
\end{itemize}
which are similar to the ones reported for the same maser in Ref.~\cite{schwarz_long_2020}, but can change over time. Here we used a cutoff frequency of \SI{0.5}{\hertz}.
The maser is used as a flywheel to bridge gaps between intervals when the \Al clock was running but notably not to extend the measurement beyond the start of the first interval or the end of the last one. The maser drifts over the measurement periods with a rate of \SI{0(2)e-17}{\per\day}, which can be neglected.
Overall we accumulated a measurement time of \SI{310}{\hour}.
The conversion between the microwave and optical domains was performed on an optical frequency comb in the building that houses the maser.

\begin{figure}
    \centering
    \includegraphics[width=0.6\linewidth]{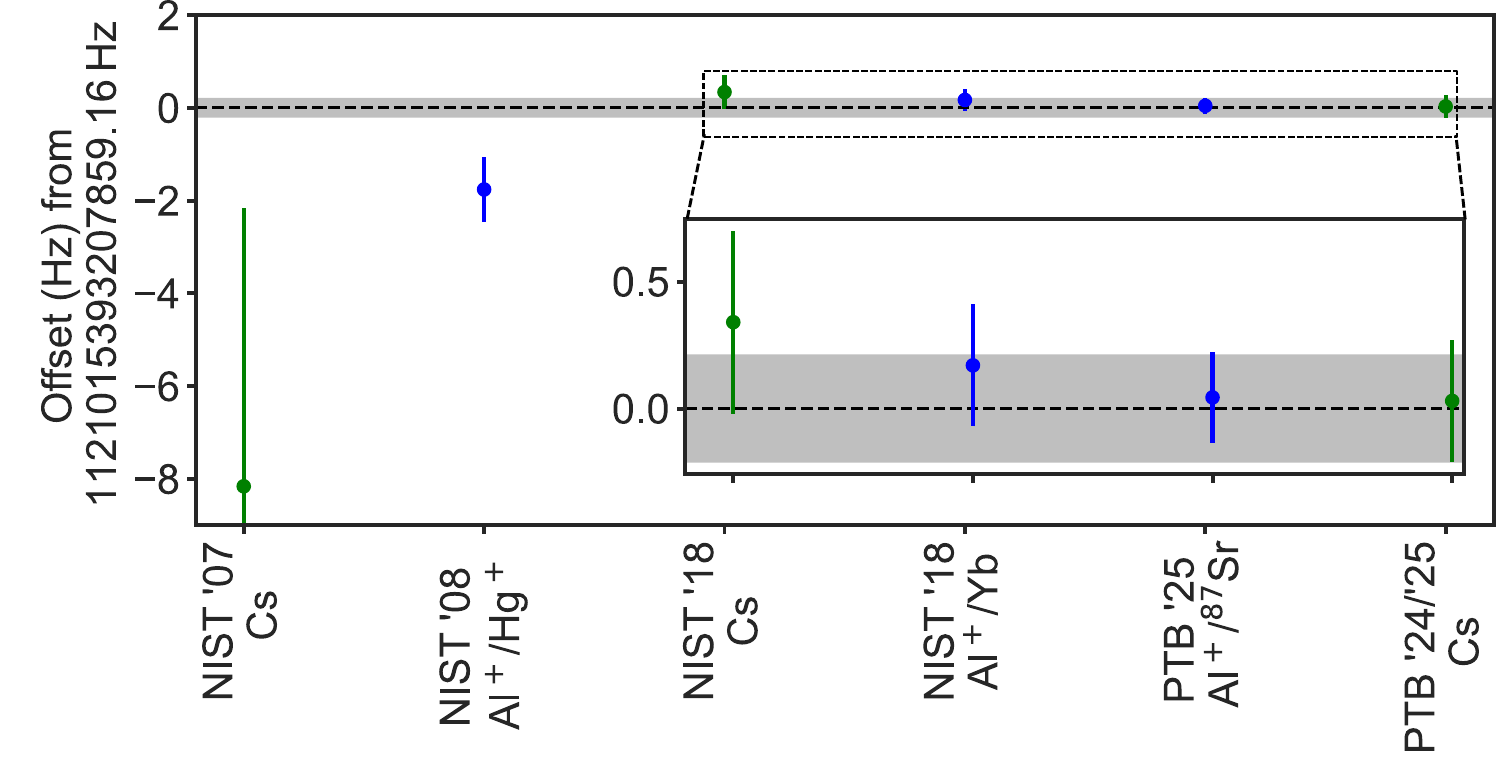}
    \caption{Comparison of absolute frequency measurements of the \Al \ssztpz measurements. Blue point mark measurements, which are transferred from ratios, while green points show direct measurements against caesium. The NIST measurement are extracted from Refs.~\cite{leopardi_measurement_2021,rosenband_observation_2007,rosenband_frequency_2008}}
    \label{fig:Absolute_frequency_comparison_Al}
\end{figure}

We use the procedure described in Refs.\ \cite{nosske_transportable_2025_2, klo26} to determine the average values of the absolute frequency with respect to one or both caesium clock as well as their relevant correlations.
The procedure treats the systematic uncertainties $u_{b, i}$ as correlated between measurements, the statistical uncertainties $u_{a,i}$ as uncorrelated between measurements, and the extrapolation uncertainties $u_\mathrm{ext}$ as correlated for measurements during the same interval and uncorrelated otherwise; the weights of individual measurements in the average values are adjusted to minimize the uncertainties of the latter \cite{nosske_transportable_2025_2}.
The weights and correlations of all three averages are listed in Tabs.\ \ref{tab:absolute-frequencies-weights} and \ref{tab:absolute-frequencies-correlations}, respectively.
The average values with respect to the individual caesium clocks are listed in Tab.\ \ref{tab:absolute-frequencies}.
The overall absolute frequency with respect to both caesium clocks is $\nu_\Al = \SI{1121015393207859.19(24)}{\hertz}$ with a fractional uncertainty of \num{2.1e-16}.
This agrees with and slightly improves over a previous measurement stating $\nu_\Al=\SI{1121015393207859.50(36)}{\hertz}$ \cite{leopardi_measurement_2021}. A comparison of different measurements is shown in Fig.~\ref{fig:Absolute_frequency_comparison_Al}. 

\setlength{\tabcolsep}{5pt}
\begin{table}[]
    \caption{All measurements of \Al against CSF1 and CSF2 around the Modified Julian Date (MJD). The frequency difference $\Delta\nu$ is added onto $\nu=\SI{1 121 015 393 207 000}{\hertz}$ for the absolute frequency of the \Al clock transition. The statistical uncertainty of \Al is below $u_\mathrm{A,\Al}<$\num{5e-18} and has a negligible influence on the overall uncertainty of each measurement. The $T_{j}$ are the total measurement times of clock j in the given interval, while the $u_\mathrm{A,j},u_\mathrm{B,j}$ are the statistical and systematic frequency uncertainty of each clock, respectively. The $u_{\mathrm{ext}}$ is the additional noise caused by the extrapolation with maser as a flywheel. }
    \label{tab:absolute-frequencies}
    \centering
    \begin{tabular}{cccccccccccccc}
    \toprule
         									& & &	&		\multicolumn{5}{c}{CSF1}		&			\multicolumn{5}{c}{CSF2}			\\		
                                            \cmidrule(r){5-9}
                                            \cmidrule(r){10-14}
     &     $T_\mathrm{\Al}$	&	$u_{B,\Al}$ &		$u_\mathrm{ext}$ &		$T_\mathrm{Cs}$		& $u_\mathrm{A,Cs}$ 		& $u_\mathrm{B,Cs}$ & $\Delta\nu$	 & u &     $T_\mathrm{Cs}$			&$u_\mathrm{A,Cs}$ 		& $u_\mathrm{B,Cs}$ & $\Delta\nu$	& u \\  
     MJD  &     (s)	&	($10^{-18}$) &	($10^{-16}$) &		(days)		&   	\multicolumn{2}{c}{($10^{-16}$)}	& (Hz) 	 & (Hz) &     (days)		&   	\multicolumn{2}{c}{($10^{-16}$)}	& (Hz) 	 & (Hz) \\  
\midrule
60530	&	46754	&	10.0	&	1.5	&	3.14	&	7.8	&	1.9	&	858.97	&	0.92	&	3.13	&	3.1	&	1.7	&	859.26	&	0.43	\\
60537	&	56178	&	10.0	&	1.4			&	4.59	&	6.3	&	1.6	&	858.87	&	0.75	&	4.57	&	2.7	&	1.7	&	859.71	&	0.39	\\
60768	&	106300	&	3.5	&	1.0			&	5.41	&	5.9	&	2.0	&	858.80	&	0.71	&	5.41	&	2.5	&	1.7	&	858.98	&	0.36	\\
 \colrule																												
\multicolumn{2}{l}{\textbf{Average}}			&			&		&		&		&		&	\textbf{858.87}	&	\textbf{0.48}	&	&			&		&	\textbf{859.29}	&	\textbf{0.27}	\\

\bottomrule

    \end{tabular}

\end{table}

\begin{table}[ht]
    \caption{
        Optimized weights of the individual measurements shown in Tab.\ \ref{tab:absolute-frequencies} in the average values with respect to one or both caesium clocks, determined according to Ref.\ \cite{nosske_transportable_2025_2}.
    }
    \label{tab:absolute-frequencies-weights}
	\begin{tabular}{lrrrr}
		\toprule
		\multicolumn{1}{c}{\bfseries MJD}    & \multicolumn{4}{c}{\bfseries Weight} \\ \cmidrule(r){2-5}
		\multicolumn{1}{c}{\bfseries (days)} & \multicolumn{1}{c}{\bfseries CSF1 average} & \multicolumn{1}{c}{\bfseries CSF2 average} & \multicolumn{2}{c}{\bfseries CSF1+CSF2 average}                              \\ \cmidrule(r){4-5}
		                                      &                                             &                                             & \multicolumn{1}{c}{\bfseries (CSF1)} & \multicolumn{1}{c}{\bfseries (CSF2)} \\
		\colrule
		60530 & 0.232 & 0.254 & 0.054 & 0.194\\
		60537 & 0.364 & 0.326 & 0.083 & 0.245\\
		60768 & 0.404 & 0.421 & 0.097 & 0.326\\
		\bottomrule
	\end{tabular}
\end{table}

\begin{table}[ht]
    \caption{
        Correlation coefficients $r$ of the average \Al transition frequencies measured using either caesium fountain clock, $\bar{\nu} (\Al|i)$, or both, $\bar{\nu}(\Al)$, with respect to each other and to the atomic clocks' systematic relative errors, $\Delta_{\mathrm{B},j}$, computed according to Ref.\ \cite{nosske_transportable_2025_2}.
    }
    \label{tab:absolute-frequencies-correlations}
	\begin{tabular}{l|l|d}
		\hline
		$q_i$ & $q_j$ & \multicolumn{1}{l}{$r(q_i, q_j)$} \\
		\hline
		\colrule
		$\bar{\nu} (\mathrm{Al\textsuperscript{+}}|\mathrm{CSF1})$ & $\Delta_{\mathrm{B}, \mathrm{Al\textsuperscript{+}}}  $    &  0.017 \\
		$\bar{\nu} (\mathrm{Al\textsuperscript{+}}|\mathrm{CSF1})$ & $\Delta_{\mathrm{B}, \mathrm{CSF1}}$    & -0.430 \\
		$\bar{\nu} (\mathrm{Al\textsuperscript{+}}|\mathrm{CSF2})$ & $\Delta_{\mathrm{B}, \mathrm{Al\textsuperscript{+}}}  $    &  0.030 \\
		$\bar{\nu} (\mathrm{Al\textsuperscript{+}}|\mathrm{CSF2})$ & $\Delta_{\mathrm{B}, \mathrm{CSF2}}$    & -0.699 \\
		$\bar{\nu} (\mathrm{Al\textsuperscript{+}})$               & $\Delta_{\mathrm{B}, \mathrm{Al\textsuperscript{+}}}  $    &  0.034 \\
		$\bar{\nu} (\mathrm{Al\textsuperscript{+}})$               & $\Delta_{\mathrm{B}, \mathrm{CSF1}}$    & -0.200 \\
		$\bar{\nu} (\mathrm{Al\textsuperscript{+}})$               & $\Delta_{\mathrm{B}, \mathrm{CSF2}}$    & -0.604 \\
		$\bar{\nu} (\mathrm{Al\textsuperscript{+}})$               & $\bar{\nu} (\mathrm{Al\textsuperscript{+}}|\mathrm{CSF1})$ &  0.507 \\
		$\bar{\nu} (\mathrm{Al\textsuperscript{+}})$               & $\bar{\nu} (\mathrm{Al\textsuperscript{+}}|\mathrm{CSF2})$ &  0.887 \\
		\hline
	\end{tabular}
\end{table}
\end{document}